\documentclass[acmsmall]{acmart}

\AtBeginDocument{%
  \providecommand\BibTeX{{%
    \normalfont B\kern-0.5em{\scshape i\kern-0.25em b}\kern-0.8em\TeX}}}

\setcopyright{acmlicensed}
\acmJournal{CSUR}
\acmYear{2022} \acmVolume{1} \acmNumber{1} \acmArticle{1} \acmMonth{1} \acmPrice{15.00}\acmDOI{10.1145/3510409}

\acmJournal{CSUR}

\usepackage{xspace}
\usepackage{hyperref}
\usepackage{subfigure}
\usepackage{diagbox, eqparbox, hhline}
\usepackage{tabularx, colortbl, booktabs}
\usepackage{filecontents}
\usepackage{multirow}
\usepackage{xspace}
\usepackage{pifont}
\usepackage{spverbatim}
\usepackage{tikz}
\usepackage{pgfplots} \pgfplotsset{compat=1.12}
\usepackage{pgfplotstable}

\newif\ifloadtables
\loadtablesfalse

\ifloadtables
\usepgfplotslibrary{external}
\tikzset{external/force remake}
\fi

\iffalse
\newcommand{\ReviewOne}[1]{\textcolor{blue}{#1}}
\else
\newcommand{\ReviewOne}[1]{\textcolor{black}{#1}}
\fi
\iffalse
\newcommand{\ReviewTwo}[1]{\textcolor{green}{#1}}
\else
\newcommand{\ReviewTwo}[1]{\textcolor{black}{#1}}
\fi

\usepackage{array}
\newcolumntype{?}{!{\vrule width 1pt}}

\newcommand{\cmark}{\ding{51}}% for a valid tick
\newcommand{\xmark}{\ding{55}}% for a crossmark tick
\newcommand{\qmark}{{\fontfamily{ugq}\selectfont \textbf{?}}}

\newcommand{\USG}{USG\xspace}
\newcommand{\LARS}{LARS\xspace}
\newcommand{\UPOIMINE}{UPOI-Mine\xspace}

\newcommand{\GTBNMF}{GT-BNMF\xspace}
\newcommand{\FPMCLR}{FPMC-LR\xspace}
\newcommand{\LRT}{LRT\xspace}

\newcommand{\IGSLR}{iGSLR\xspace}
\newcommand{\UTE}{UTE\xspace}
\newcommand{\SE}{SE\xspace}
\newcommand{\UTESE}{UTE+SE\xspace}
\newcommand{\LARSS}{LARS*\xspace}
\newcommand{\UPOIWALK}{UPOI-Walk\xspace}
\newcommand{\GTAG}{GTAG\xspace}
\newcommand{\GEOMF}{GeoMF\xspace}
\newcommand{\IRENMF}{IRenMF\xspace}
\newcommand{\LORE}{LORE\xspace}

\newcommand{\CORE}{CoRe\xspace}

\newcommand{\RANKGEOFM}{RankGeoFM\xspace}
\newcommand{\GEOSOCA}{GeoSoCa\xspace}
\newcommand{\PRMEG}{PRME-G\xspace}
\newcommand{\CAPRF}{CAPRF\xspace}

\newcommand{\GE}{GE\xspace}

\newcommand{\ASMF}{ASMF\xspace}
\newcommand{\STELLAR}{STELLAR\xspace}
\newcommand{\WWO}{WWO\xspace}

\newcommand{\GEOTEASER}{Geo-Teaser\xspace}

\newcommand{\PACE}{PACE\xspace}

\newcommand{\CTFARA}{CTF-ARA\xspace}

\newcommand{\TGSCPMF}{TGSC-PMF\xspace}

\newcommand{\LBPR}{LBPR\xspace}

\newcommand{\VPOI}{VPOI\xspace}

\newcommand{\SAENAD}{SAE-NAD\xspace}

\newcommand{\CARA}{CARA\xspace}

\newcommand{\TENMF}{TenMF\xspace}
\newcommand{\GEOEISO}{GeoEISo\xspace}

\newcommand{\GEOIE}{GeoIE\xspace}

\newcommand{\MEAPT}{MEAP-T\xspace}

\newcommand{\MLR}{MLR\xspace}

\newcommand{\APRASA}{APRA-SA\xspace}

\newcommand{\STA}{STA\xspace}

\newcommand{\STGN}{STGN\xspace}
\newcommand{\HILDA}{HI-LDA\xspace}
\newcommand{\GAIMC}{GAIMC\xspace}
\newcommand{\SPR}{SPR\xspace}
\newcommand{\MMBE}{MMBE\xspace}
\newcommand{\TECF}{TECF\xspace}

\newcommand{\NA}{(N.A.)\xspace}
\newcommand{\unk}{N.A.\xspace}
\newcommand{\KNN}{$k$-NN\xspace}
\newcommand{\SIMILARITY}{CF sims\xspace}

\newcommand{\GraphLink}{Graph/Link\xspace}
\newcommand{\nfold}{$n$-fold\xspace}
\newcommand{\Fix}{Per User\xspace}
\newcommand{\CC}{System\xspace}

\newcommand{\PRECISION}{P\xspace}
\newcommand{\RECALL}{R\xspace}
\newcommand{\NDCG}{NDCG\xspace}

\newcommand{\checkin}{check-in\xspace}
\newcommand{\checkins}{check-ins\xspace}

\newcommand{\UB}{UB\xspace}
\newcommand{\IB}{IB\xspace}
\newcommand{\MODS}{Mods\xspace} %mods are modifications of the algorithm
\newcommand{\POPULARITY}{Pop\xspace}
\newcommand{\RANDOM}{Rnd\xspace}
\newcommand{\MF}{MF\xspace}
\newcommand{\RANDOMWALK}{RW\xspace}
\newcommand{\MAE}{MAE\xspace}
\newcommand{\RMSE}{RMSE\xspace}

\begin{document}

\ifloadtables
\begin{figure}
\centering
\scalebox{0.8}{
\begin{tikzpicture}
\begin{axis}[
    ybar stacked,
	bar width=15pt,
	nodes near coords,
    enlargelimits=0.1,
    ymin=0,
    legend style={at={(1,1)},anchor=north west, draw=none, font=\small},
    ylabel={Number of papers},
    xlabel={Years},
    symbolic x coords={2011, 2012, 2013, 2014,
		2015, 2016, 2017, 2018, 2019, 2020},
    xtick=data,
    x tick label style={rotate=45,anchor=east},
    ]
\addplot+[ybar] plot coordinates {(2011, 1) [1] (2012, 4)
  (2013,16) (2014, 12) [12] (2015, 22) (2016, 39) (2017, 26) (2018, 26) (2019, 31) (2020, 25)};
\addplot+[ybar] plot coordinates {(2011, 0) [0] (2012, 0)
  (2013,0) (2014, 2) [2] (2015, 10) (2016, 10) (2017, 13) (2018, 16) (2019, 29) (2020, 28)};
\legend{\strut conferences, \strut journals}
\end{axis}
\end{tikzpicture}
}
\caption{
Number of papers considered in this study based on their publication venue.
}
\label{fig:number_confjour}
%update figure may 2021
\end{figure}
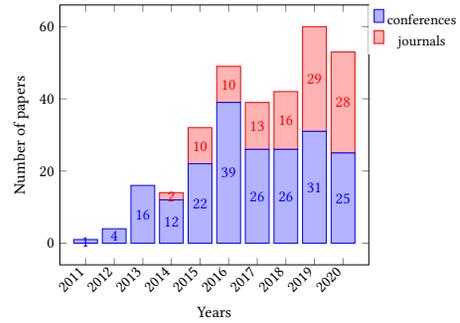
\else
\fi

%%
%% The "title" command has an optional parameter,
%% allowing the author to define a "short title" to be used in page headers.
\title[\ReviewTwo{POI RS based on LBSNs: A Survey from an Experimental Perspective}]{%
\ReviewTwo{Point-of-Interest Recommender Systems based on Location-Based Social Networks: A Survey from an Experimental Perspective}
}

\author{Pablo S\'{a}nchez}
\authornote{Both authors contributed equally to this research.}
\orcid{0000-0003-1792-1706}
\affiliation{%
	\department{Information Retrieval Group}
	\institution{Universidad Aut\'onoma de Madrid}
	\city{Madrid}
	\postcode{28049}
	\country{Spain}
}
\email{pablo.sanchezp@uam.es}
\author{Alejandro Bellog\'{i}n}
\orcid{0000-0001-6368-2510}
\affiliation{%
	\department{Information Retrieval Group}
	\institution{Universidad Aut\'onoma de Madrid}
	\city{Madrid}
	\postcode{28049}
	\country{Spain}
}
\email{alejandro.bellogin@uam.es}

\begin{abstract}
Point-of-Interest \ReviewOne{recommendation} is an increasing research and developing area within the widely adopted technologies known as Recommender Systems.
Among them, those that exploit information coming from Location-Based Social Networks (LBSNs) are very popular nowadays and could work with different information sources, which pose several challenges and research questions to the community as a whole.
We present a systematic review focused on the research done in the last $10$ years about this topic. We discuss and categorize the algorithms and evaluation methodologies used in these works and point out the opportunities and challenges that remain open in the field\ReviewOne{. More specifically, we report the leading recommendation techniques and information sources} that have been exploited more often (such as the geographical signal and deep learning approaches) while \ReviewOne{we also alert about the} lack of reproducibility in the field that may hinder real performance improvements.
\end{abstract}

%%
%% The code below is generated by the tool at http://dl.acm.org/ccs.cfm.
%% Please copy and paste the code instead of the example below.
%%



\maketitle

\section{Introduction}

\ReviewOne{%
Recommender Systems (RSs) have risen as technological solutions to the information overload, as they help users to filter the most interesting items (in whatever domain the RS is being deployed) according to their preferences.
Moreover, in the Internet era, they have become indispensable due to their ability to process large amounts of information and make personalized recommendations to users by learning their interests and tastes~\cite{DBLP:reference/sp/RicciRS15}.
However, they serve other purposes as well.
They are particularly useful to aggregate user behavior, which is pervasive nowadays, very common and easier to obtain thanks to the Internet and the increasing and diversity of social networks dealing with different domains. 
This is in fact related to the universal applicability of general RSs, since classic RSs have been oriented towards recommending music or movies, but for some years now they have been applied to other areas such as news, e-commerce, social contacts, healthcare, and tourism~\cite{DBLP:journals/ipm/KarimiJJ18,DBLP:conf/kdd/LiuNZZYCWZC16,DBLP:reference/sp/KoprinskaY15,DBLP:journals/fgcs/KaurKB18,DBLP:conf/icc/AujlaJC0VSO19,DBLP:journals/jitt/TrattnerOMP18}.
}

\ReviewOne{%
In particular, Location-Based Social Networks (LBSNs) are a special kind of social Web systems where it is possible for users to register whenever they visit a specific Point-of-Interest (POI) through the so-called \checkins, or to establish social links with other users in the system~\cite{DBLP:journals/csur/DingLJZ18}.
In retrospective, they represent the digital versions of historical catalogs such as the Zagat survey or the Michelin guides, which aimed at summarizing and synthesizing ratings and reviews provided by amateur (since the beginning in Zagat around 1980s) or expert (since the 1930s in Michelin) food reviewers.
These systems, as modern RSs based on LBSN data, had the same goal: reducing the choice overload of users, while providing a subjective measurement of the POI (for these two examples, restricted to restaurants) quality.
Since then, location-based services that deliver information according to the location and context of the user and her device play a key role~\cite{DBLP:journals/jlbs/HuangGKRW18}.
They appeared in the early 1990s, but thanks to the evolution of the technology (mobile devices, availability of GPS and navigation systems) a wide range of applications have emerged, not limited to LBSNs, but for gaming, health, fitness, and assistive technology. We refer the reader to~\cite{DBLP:journals/jlbs/HuangGKRW18} for a review on the research trends on that topic.
}

\ReviewOne{%
A popular demanded service in these LBSNs is POI recommendation.
In general, these RS techniques aim at recommending users new places to visit when they arrive to a city or region;
however, this problem is inherently multi-faceted and, hence, the following related problems are typically studied~\cite{DBLP:series/lncs/BothorelLPN18}:
suggesting interesting previously unvisited places to a target user,
recommending the next place to go, 
recommending events to attend and neighborhoods to explore in a urban setting, 
and discovering places in a city with respect to an input query and the user previous interests.
Naturally, these recommendations are contextualized for a specific type of geographical region -- such as a country, city, or town --, either implicitly (inferred from previous user history)~\cite{DBLP:conf/sigir/ZhangC15} or explicitly (requested by the system itself)~\cite{DBLP:journals/umuai/SanchezB20}.
Additionally, a number of constraints could be incorporated into the model, such as type of trip (leasure or work), price, schedule, or weather~\cite{DBLP:journals/heuristics/GavalasKMP14,DBLP:journals/jitt/TrattnerOMP18}.
In this regard, commercial systems such as the early Triplehop's TripMatcher try to replicate the interactivity observed in typical sessions with travel agents~\cite{DBLP:journals/expert/StaabWRZGFPK02}, whereas recent platforms have created new services, such as exchange or sharing tourism-related products (Airbnb, Uber), integrating users in a community (TripAdvisor, Foursquare), searching and comparing (Trivago, Skyscanner), or booking and travel support (Expedia, Booking).
}\ReviewTwo{We would also like to mention that, even though some of these companies have not yet fully exploited the potential of recommender systems -- since they are more focused on tuning their filters according to the collected interactions~\cite{DBLP:conf/recsys/BernardiEEO20} --, many researchers do make use of data from these companies to perform different types of recommendations. For example, it is well-known that hotel and/or tourist attraction recommendations can be performed using data from TripAdvisor or Booking~\cite{electronics10161920, DBLP:journals/ijon/ShenDG16}). At the same time, another company from this domain, Expedia, organized in 2013 a contest\footnote{Expedia contest: \url{https://www.kaggle.com/c/expedia-personalized-sort}} to find the best recommender for their website. However, the challenge was focused on data from search logs, so important characteristics from LBSNs (as we shall point out later) were ignored.}

\ReviewOne{Finally, it is important to highlight that} this domain exhibits other differences with respect to classical recommendation and how it has been modeled in the past, such as not being limited to a preference or rating matrix, or incorporating additional information such as geographical, social, or temporal signals to better adapt to the users' interests~\cite{DBLP:journals/pvldb/LiuPCY17}.
To obtain this type of information, researchers normally resort to exploiting LBSNs, \ReviewOne{as they include most of these attributes}.
It is worth mentioning that, even though the area of POI recommendation is of great interest to researchers because it allows to study the behavior and movement patterns of users, it is also appealing for companies and businesses in the tourism, leisure, and e-commerce domains, as they seek to attract and maintain customers by becoming popular and receiving good reviews \ReviewTwo{-- this is evidenced by the increasing number of companies working on related problems, as discussed before}.
As a consequence, a large number of articles have been published in recent years where different algorithms to recommend POIs were proposed by exploiting the information available in LBSNs.
For that reason, we believe it is necessary to analyze the current proposals in the area for this type of recommendation, with special emphasis on the different types of implemented models and algorithms, information used, and evaluation methodologies followed in these works.
This is because we consider all these pieces critical to produce real advances in current state-of-the-art, which might be hindered by reproducibility or evaluation issues~\cite{DBLP:conf/cikm/ArmstrongMWZ09,DBLP:conf/recsys/SaidB14}.
We present such an analysis in this systematic review, together with a careful and detailed discussion of the current problems that this area is facing at the moment, as well as potential future lines of research.

\subsection{What are the differences between this survey and former ones?}
\label{ss:diffs}
Due to the growing interest in the general recommendation area on this domain, there is a considerable number of surveys related to POI recommendation and its different ramifications that complement our work.
On the one hand, we have the works \cite{DBLP:series/sbece/SymeonidisNM14,yu2015survey,DBLP:journals/geoinformatica/0003ZWM15} which cannot be considered to be up to date anymore, since they were published $5$ years ago, thus our survey should provide a novel overview of the works developed in this time.
On the other hand, \citeauthor{DBLP:journals/heuristics/GavalasKMP14} present in~\cite{DBLP:journals/heuristics/GavalasKMP14} an overview of optimization approaches that aim to solve a problem with applications on related tasks: % to the general POI recommendation:
the Tourist Trip Design Problem (TTDP); this can be applicable to route recommendation, which, as we specify in the next section, is not completely in the scope of this survey.

Additionally, we found some surveys that were too focused on specific subproblems.
For instance, in~\cite{DBLP:journals/tkde/ZhengHS18} \citeauthor{DBLP:journals/tkde/ZhengHS18} consider the problem of location prediction but only based on Twitter information.
Another example is~\cite{DBLP:conf/inista/ChristoforidisK19}, where \citeauthor{DBLP:conf/inista/ChristoforidisK19} focus on deep learning techniques while neglecting the other types of recommendation algorithms.

\iffalse %% to make paper shorter
However, even after filtering those works, there are still several reviews dedicated to the problem of POI recommendation based on LBSN data, as we address herein \ReviewTwo{(works not dealing with data coming from a LBSN are out of the scope of our survey)}.
Our main difference with the following works is the comprehensive analysis that we present, involving more than \ReviewOne{$300$} papers from $10$ years categorized according to their algorithmic and evaluation models.
%
For instance, in~\cite{DBLP:journals/csur/DingLJZ18}, \citeauthor{DBLP:journals/csur/DingLJZ18} present a survey where they group the articles according to different recommendation objectives, such as location, trip, or activity recommendation, among others.
%
In~\cite{DBLP:series/sbcs/ZhaoLK18}, \citeauthor{DBLP:series/sbcs/ZhaoLK18} provide an overview of the problem from the geographical and temporal perspectives, but then they focus on two specific algorithms.
%
\fi

\begin{table}[tb]
	\caption{Queries issued to the three digital libraries considered. For ScienceDirect the query is used in the field ``Title, abstract or author-specified keywords'', indicating 2011-\ReviewOne{2020} in the field ``Years''.}
	\label{t:queries}
	%\small
	\footnotesize
	\begin{tabular}{lp{0.8\textwidth}}
	\toprule
	\textbf{Source} & \textbf{Query} \\
	\midrule
	Scopus          &
	( ( TITLE ( point-of-interest )  OR  TITLE ( venue )  OR  TITLE ( poi )  OR  TITLE ( location ) )  AND  ( TITLE ( recommendation )  OR  TITLE ( recommender ) )  AND  ( TITLE-ABS-KEY ( lbsn )  OR  TITLE-ABS-KEY ( "location-based social network" ) )  AND  ( PUBYEAR  >  2010 ) )  AND  ( PUBYEAR  <  \ReviewOne{2021} )  AND NOT  TITLE ( survey )  AND  ( LIMIT-TO ( LANGUAGE ,  "English" ) )
	\\
	ScienceDirect   &
	((lbsn) OR ("location-based social network")) AND -survey AND ("point of interest" OR venue OR location OR poi) AND (recommendation OR recommender)
	\\
	ACM             &
	[[Publication Title: "point-of-interest"] OR [Publication Title: "point of interest"] OR [Publication Title: poi] OR [Publication Title: venue] OR [Publication Title: location]] AND [[Publication Title: recommendation] OR [Publication Title: recommender]] AND [[Abstract: "location-based social network"] OR [Abstract: "lbsn"]] AND [Publication Date: (01/01/2011 TO \ReviewOne{12/31/2020})]
	\\
	\bottomrule
	\end{tabular}
	\end{table}

Finally, since our analysis is also tailored towards the evaluation aspects of the works, it is worth mentioning those reviews where this aspect has been considered.
However, we must acknowledge that we could not find any survey that focused on this particular aspect; because of that, we consider our survey is very valuable in this domain at the moment.
To somehow overcome this shortcoming, 
we believe it is important to mention the experimental comparison presented by \citeauthor{DBLP:journals/pvldb/LiuPCY17} in~\cite{DBLP:journals/pvldb/LiuPCY17},
where they compared $12$ recommendation models under different evaluation protocols and using three datasets, which could help to analyze the behavior of those methods under the same and different conditions.

\begin{center}
\begin{minipage}[t]{\textwidth}
  \begin{minipage}[t]{0.49\textwidth}
    \centering
    \vspace{10pt}
    \captionof{table}{Papers retrieved and final papers processed from the three digital libraries considered.}
    \label{tab:PapersProcessed}
    \footnotesize
    \begin{tabular}{lrr}
    \toprule
	\textbf{Source} & \textbf{Papers retrieved} & \textbf{Valid papers} \\
	\midrule
	Scopus          &      \ReviewOne{404}    &    \ReviewOne{302}     \\
	ScienceDirect   &      \ReviewOne{50}     &    \ReviewOne{30}      \\
	ACM             &      \ReviewOne{71}     &    \ReviewOne{43}       \\
	\midrule
	Unique papers	&	   \ReviewOne{\textbf{431}}	  &	   \ReviewOne{\textbf{310}}		\\
	\bottomrule
    \end{tabular}
    \end{minipage}
  \hfill
  \begin{minipage}[t]{0.49\textwidth}
    \vspace{10pt}
    \centering
    \scalebox{0.6}{
    \includegraphics[]{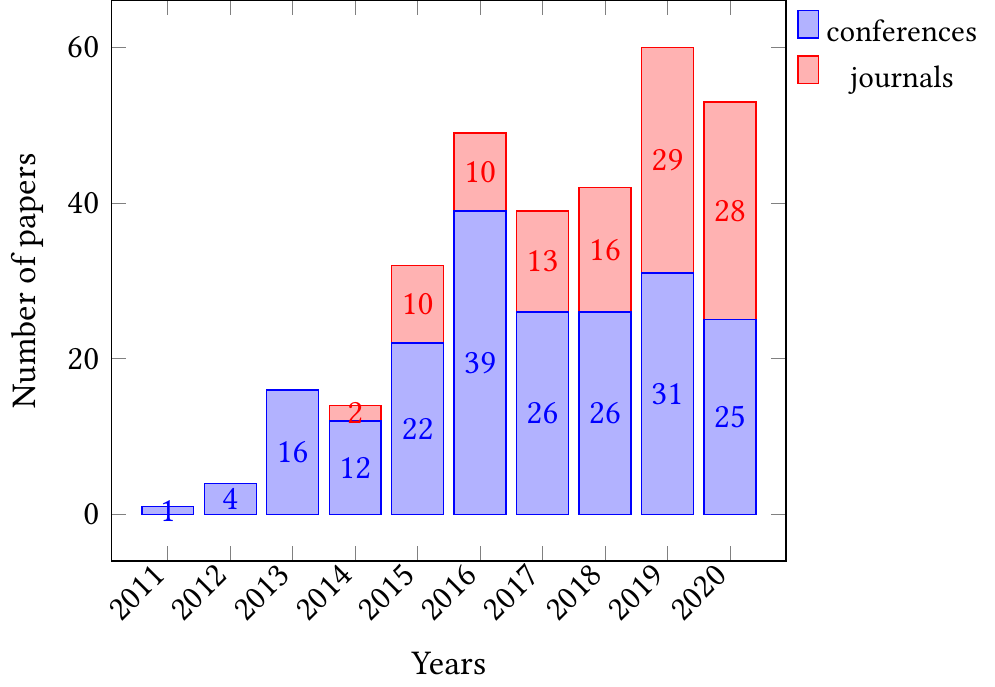}
    }
    \captionof{figure}{Number of papers considered in this study based on their publication venue by year.}
    \label{fig:number_confjour}
  \end{minipage}
  \end{minipage}
\end{center}

\subsection{How do we collect the papers?}
\label{ss:how}
In order to select the papers that we have analyzed in this survey, we have searched in three digital libraries: Scopus\footnote{\url{https://www.scopus.com/}}, ScienceDirect\footnote{\url{https://www.sciencedirect.com/}}, and ACM Digital Library\footnote{\url{https://dl.acm.org/}}.
As each library has a different query language to use within its search system, three different queries were needed to be defined and executed, however, they were designed to be as equivalent as possible\footnote{The queries were issued last time on 
\ReviewOne{April 2021}
so some of the results may have changed.}.

The main characteristics these queries should satisfy are:

\begin{itemize}
  \item Focus on articles between 2011 and \ReviewOne{2020} (both included).
  \item Each publication should include in the title: ``Point of interest recommendation'' or (similar texts such as ``POI \ReviewOne{recommender}'').
  \item Each publication should also include somewhere in the title, abstract, or keywords the terms ``location-based social network'' (or ``LBSN''), since this survey is oriented to models using data coming from these systems, 
  % as a fundamental part of the survey is to analyze
  \ReviewOne{together with an analysis on the different evaluation methodologies that are being applied using datasets generated from LBSNs.}
\end{itemize}

Thus, the final queries issued to each source are shown in Table~\ref{t:queries}.
Based on this,
Table~\ref{tab:PapersProcessed} shows the number of papers we initially obtained with each query, as well as the actual number we finally analyzed.
The difference was mostly caused to some papers not being available, some of them appeared in more than one source, and some had to be filtered out because they address a different task to the one we want to focus on this paper.
\iffalse %% 
In the same way, we have found some ``repeated'' articles that corresponded to improved versions of other works from previous years, conducted by the same authors; in those cases, we have only taken into account the oldest article. For example,
the approach \LARSS presented in~\cite{DBLP:journals/tkde/SarwatLEM14} in the year 2014 is an extension of \LARS, proposed in 2012 by the same authors; to avoid overrepresenting the same methods, we decided to ignore the second paper and only consider the oldest one (as it is the original formulation of the model).
\fi
%
%
We also decided to keep only those papers whose final goal is to recommend a list of POIs to each user; this includes related tasks such as next-POI recommendation as long as no trajectory or route recommendation is performed (as in~\cite{DBLP:conf/cikm/ZhangW15}), but discards tasks such as route, category, or friend recommendation~\cite{DBLP:conf/cikm/KurashimaIIF10,DBLP:conf/aaai/ChenZCQXD15,DBLP:conf/gis/SymeonidisPMST11}. 
\ReviewOne{Hence, the only task we aim to cover with this review is POI recommendation (see later in Section~\ref{sss:def} a formal definition of this problem, and in Section~\ref{s:RecommendationTasks} other related tasks not covered herein).}
\iffalse %%
As a final note, we have not applied additional filters in the paper collection phase such as selection by conferences or journals, to not incur in any subjective bias, although we had to remove few papers where the proposal was not presented in a clear way.
\fi

Figure~\ref{fig:number_confjour} shows the number of articles we include in this review according to their publication venue (conference or journal).
We observe that the number of publications has increased steadily since 2014; although initially most of the papers were published in conferences, over the years there has been a growing interest in publishing in journals.
This figure shows that the problem of POI recommendation is still relevant today.
All the works included in our analysis are available as supplementary information\footnote{Available here: 
\url{https://abellogin.github.io/poi_survey/}.}.

\subsection{Contributions of this survey}
The purpose of this systematic literature review is to identify the current state-of-the-art in POI recommendation \ReviewTwo{based on LBSNs} and to analyze the techniques used, the experimental protocols used to validate them, and the related research challenges.
For these reasons, we define the following research questions:
\begin{description}
  \item[RQ1] What is the state-of-the-art in POI recommendation \ReviewTwo{based on LBSNs}? To answer this question we survey the literature in terms of algorithms, information sources, and evaluation methodologies. %current
  \item[RQ2] Which are the most relevant works? We want to analyze with more detail those works that have had more impact in the community, and extract possible reasons for that, exposing these characteristics so that future researchers focus on them in their research.
  \item[RQ3] How are these recommenders evaluated? Which protocols/metrics/datasets are used? As a specific goal of this review, we want to dig in the specific evaluation methodologies followed in the POI recommendation literature, since this is a potential source of misbehavior that could limit overall improvements in the field.
  \item[RQ4] What are the most important issues to be addressed in the future? Based on the answers to the other research questions, we summarize and present the most important topics that should be considered by the researchers dealing with the POI recommendation problem \ReviewTwo{when using data from LBSNs}.
\end{description}

Therefore, our main goal with this survey is to provide a complete review of the works from the last $10$ years in the field of POI recommendation \ReviewTwo{based on data coming from LBSNs}.
As already mentioned before, this is not the first survey that has been done on this subject, however we believe we are the first ones -- to the best of our knowledge -- that have also considered and, hence, classified the articles by the evaluation protocols followed.
The key contributions, thus, of this work are:
\begin{itemize}
	\item A thorough review of state-of-the-art POI recommendation models \ReviewTwo{based on LBSNs} between $2011$ and \ReviewOne{$2020$}.
	\item A proposal to classify the algorithmic methodologies used in those works, together with the contextual information handled by the models, and the evaluation methodologies employed to evaluate their performance.
	\item A list of challenges and open issues in the field, in combination with potential future directions, to help other researchers and practitioners focus their work and resources on the problems that this area needs to fix as soon as possible.
\end{itemize}

In the next section we present a background on classical recommendation methods and their evaluation, in both cases independent of the domain.
Then, in Section~\ref{ss:poi} we contextualize these concepts to the problem of POI recommendation.
Sections~\ref{s:soa_algo}, \ref{s:soa_eval}, and~\ref{s:soa_data} present the main outcomes from our systematic review, first regarding the state-of-the-art algorithms, later about the evaluation methodologies, and finally focusing on the datasets reported in the experiments.
We conclude the paper in Section~\ref{s:future} with the most important future research directions and open issues identified after our systematic review.

%%%%
\section{Background on classical recommendation}
\label{ss:background}

\subsection{Problem definition}
\label{s:problemDefinition}
The main purpose of Recommender Systems (RSs) is to suggest hypothetically relevant items to users.
When needed, we denote with $\mathcal{U}$ the set of users in the system and $\mathcal{I}$ the set of items, with $u$, $v \in \mathcal{U}$ and $i$, $j \in \mathcal{I}$.
Furthermore, since the most typical type of interaction between users and items are ratings, we use $\mathcal{R}$ for the interactions, as it is standard in the area, although other types of interactions exist, such as clicks, buys, watchings, or listenings, depending on the domain~\cite{DBLP:reference/sp/RicciRS15}.

Normally, these algorithms exploit the interactions of the users available in the system to build a model from the data and generate recommendations.
Traditionally, the recommendation problem has been defined as an optimization problem~\cite{DBLP:journals/tkde/AdomaviciusT05}:
\begin{equation}
\label{f:optProblem}
	i^*(u) = \arg\max_{i \in \mathcal{I}}{g(u,i)}
\end{equation}
where $i^*$ is the optimal item that maximizes the relevance or utility for user $u$ on any item $i$ among those in $\mathcal{I}$, where such utility function is represented by $g$.
Depending on the domain, items may have different nature, either movies, books, electronic products, or touristic venues, as the focus of this work.
At the same time, while the final objective for any of these systems is the same in any case, we classify the most common algorithms used depending on how they work with the data, collaborative filtering and content-based being the two most popular and well-known categories, but other types such as demographic or knowledge-based exist and are applied in the community~\cite{DBLP:reference/sp/RicciRS15}.
In the next subsections, we introduce these two classes of algorithms, together with the most common ways to combine these methods as hybrid approaches;
after that, we present basic information regarding how to evaluate recommender systems.

\subsection{Content-based filtering (CB)}
Content-based recommender systems analyze the items and/or user features (content) and use them to create user and item profiles to recommend items to the target user that are similar to the ones she liked previously. 
\ReviewOne{In order to make recommendations, this type of system uses three main components~\cite{DBLP:reference/sp/GemmisLMNS15}: the \textit{content analyzer} that pre-processes the information available of the items in order to extract keywords, concepts, or other information, the \textit{profile learner} that, using the content information of the items, builds a profile for every user in the system, and, finally, the \textit{filtering component} that matches the user profile against the items in the system.}

\ReviewOne{
For modeling the items features from text many content-based algorithms use simple Information Retrieval (IR) models such as the Vector Space Model (VSM)~\cite{DBLP:books/aw/Baeza-YatesR2011}, where an $n$-size vocabulary in the form of keywords or terms is obtained from documents, and then this vocabulary is used to represent those documents in an $n$-dimensional space. To build the vectors, a common approach is using schemes based on Term Frequency (TF), Inverse Document Frequency (IDF), and combinations thereof (such as the well-known approaches of TF-IDF or BM25)~\cite{DBLP:conf/recsys/CantadorBV10}. Once we have transformed all the items into vectors, a similarity metric (such as cosine similarity) can be applied to obtain a ranking of similar items with respect to others the user has previously consumed. Even though modeling this problem with a VSM is still popular nowadays, the use of embeddings has increased lately to exploit possible latent relationships between documents and associated terms~\cite{DBLP:journals/umuai/LopsJMBK19}.}

\ReviewOne{
For modeling the users profiles several techniques have also been proposed, including \textit{probabilistic models} (e.g., Na\"ive Bayes) that will estimate for a target user the probability to classify a document $d$ into class $c$, that is $P(c|d)$ (e.g., the user likes it or dislikes it, or even one class for each possible rating value), \textit{relevance feedback} that refines the user profile by taking into account their opinion of the previous suggested items %, 
and \textit{neighborhood-based} algorithms, where it is common to use a similarity function computed on the VSM representation of the items and then select the class for the unclassified item taking into account the classes of the nearest neighbor items~\cite{DBLP:reference/sp/GemmisLMNS15}.
}

\subsection{Collaborative filtering}
Collaborative Filtering (CF) techniques analyze the interactions between users and items to establish patterns between them when making recommendations.
These techniques are normally divided into two groups: memory-based
that perform the recommendations using the interactions (usually represented as a user-item matrix) in a direct way by computing similarities between users and/or items~\cite{DBLP:reference/sp/NingDK15}, and model-based algorithms
that build a predictive model by approximating the information stored in the preference or interaction matrix~\cite{DBLP:reference/sp/KorenB15}.
We now explain some of the fundamental concepts related to these two families of CF algorithms.

\subsubsection{Memory-based methods}
\label{ss:knn}
Memory-based methods (also called nearest neighbors or \KNN) are one of the most well-known and implemented strategies in traditional recommendation due to its ease of programming and the great interpretability of the recommendations obtained~\cite{DBLP:reference/sp/NingDK15}.

The idea behind these algorithms is to recommend to the target user the most appropriate items by exploiting similarities between the rest of the users/items in the system.
For this, they build neighborhoods -- by considering those users/items with the highest similarities -- and predict the score for new items based on those similarities and the scores provided by such neighbors~\cite{DBLP:reference/sp/NingDK15}.

Obviously, the similarity function is the most critical component in this type of algorithms, since it is used to select the neighbors and to weight each of them for the final score.
Classical similarity metrics exploit trends in ratings such as Pearson correlation or cosine similarity, but recent approaches less focused on the rating prediction problem directly exploit how many items in common are recorded between user/item interactions, by means of variations of overlap measurements such as the Jaccard index~\cite{DBLP:reference/sp/NingDK15}.

\subsubsection{Model-based methods}
\label{ss:factorization}
Model-based
algorithms represent the other major family of CF methods, enjoying great popularity because they generally perform better than neighborhood-based models and because of their importance on the Netflix Prize~\cite{DBLP:journals/sigkdd/BellK07}.
These models approximate the user-item matrix by transforming both users and items into a latent factor space of low dimensionality so that the user-item interactions can be explained (or recovered) by applying dot products in that space~\cite{DBLP:reference/sp/KorenB15}.

The most popular method in the area is the standard Matrix Factorization (MF), where the latent space is learned either by applying Stochastic Gradient Descent (SGD) or Alternating Least Squares (ALS) optimization techniques, depending on the domain characteristics and efficiency constraints~\cite{DBLP:reference/sp/KorenB15}.
However, many other approaches such as PMF (Probabilistic MF), LDA (Latent Dirichlet Allocation), and even the embeddings learned in Neural Networks fit under this family.

Beyond the matrix completion paradigm, several approaches have been proposed to extend this basic formulation to include additional biases and contextual information -- like time, sequentiality, and seasonality --, or tags, including well-known techniques like Markov Chains and deep learning techniques~\cite{DBLP:journals/jmlr/ShaniHB05,DBLP:journals/csur/ZhangYST19}.

\subsection{Hybrid recommenders}
\label{ss:Hybrids}
Individually, each recommendation algorithm may have some disadvantages in certain situations. For this reason, it is common to combine several models in order to alleviate such problems.
For example, CF approaches cannot recommend items to users with very few ratings, while social models need a mechanism to recommend items to those users who have not indicated a social link in the system.
There are many ways to make these combinations (we refer the reader to the work~\cite{DBLP:conf/adaptive/Burke07} for a complete survey about hybrid approaches),
%%%% 
%
although it is usually understood that a hybrid is any algorithm that combines different sources of information, either explicitly (social and collaborative) or implicitly (two data models generated by different recommendation methods).

Initially, since the most widespread algorithms were CF and CB approaches, hybrid methods combining these two systems proliferated, as in the case of the collaborative via content technique proposed in~\cite{DBLP:journals/cacm/BalabanovicS97}.
However, due to the great expansion of recommender systems other techniques have emerged combining several instances of the same type of recommendation model, like Fossil~\cite{DBLP:conf/icdm/HeM16} that combines Markov Chains with similarities models or FPMC~\cite{DBLP:conf/www/RendleFS10} that combines Markov Chains with Matrix Factorization.

\subsection{Evaluation of recommender systems}
\label{ss:classicEv}
The aim of RS evaluation is to determine which recommenders (or configurations of recommenders) are better than others based on the results obtained in certain metrics under a specific evaluation methodology.
In fact, among the different types of experiments that can be performed with users of a particular RS -- that is, offline, online, and user studies~\cite{DBLP:reference/sp/GunawardanaS15} --, 
\ReviewOne{the RS community has been mostly focused on offline evaluation, since it is the most comparable across different settings and the one typically used in the literature}. 
\ReviewTwo{In this survey, as we focus on Location-Based Social Networks, offline evaluation will also prevail over online evaluation methodologies since, in most cases, data from LBSNs are readily available and, hence, online studies are not necessary to collect user behavior.} \ReviewOne{However, throughout the rest of this review we will not limit our analysis to this setting, even though it is the most popular one (as we will show later)}.

For this type of evaluation, the first step is to divide the available data into different sets, so that part of the data is used to build (or train) the recommendation model, while the rest is used to evaluate it (either to validate and test the model in different stages, or using a single withheld subset of the data).
The simplest way to do this division is through random partitioning, where a percentage of the interactions is considered for training and the rest for testing (a typical value is 80\% for training, thus, leaving 20\% for testing).
A more elaborate -- but quite common -- way of doing this partitioning is by $n$-fold Cross-Validation (CV), where the data is divided into $n$-disjoint sets, in such a way that $n-1$ sets are used to build the training set and the remaining one for testing, and this process is repeated $n$ times, so that each set is used once as a test set.

However, these popular random partitioning protocols ignore the temporal component of the interactions, which might be problematic due to the unrealistic evaluation setting~\cite{DBLP:journals/umuai/CamposDC14}.
As we shall see in Section~\ref{ss:evaluationProtocols}, time-aware splitting protocols are more prevalent in POI recommendation than in classical recommendation, so we defer the explanation of these protocols to that section.

Regardless of how the data is partitioned, we need a way to analyze the performance of the recommenders.
Originally, the recommendation quality was equated to how close the recommender was able to predict the rating provided by the user.
Hence, error metrics like Mean Absolute Error (MAE) and Root Mean Squared Error (RMSE) were used, however, since these metrics only account for the observed items, they do not reflect well real-world problems nor the perceived user experience~\cite{DBLP:conf/chi/McNeeRK06}.
Because of this, IR ranking metrics like Precision, Recall, or nDCG (normalized discounted cumulative gain) were used to measure how many relevant items were included in the ranking generated by the RS~\cite{DBLP:journals/ir/BelloginCC17}.

Moreover, despite the importance of relevance in recommendations, there has been a growing awareness on measuring other evaluation dimensions like novelty (as opposed to recommending popular items) and diversity (accounting to how many items with different features are recommended), as sometimes producing only accurate recommendations may not surprise or discover new items to the user~\cite{DBLP:reference/sp/CastellsHV15}.
Nonetheless, it should be noted that, even though it is advisable to have good recommendations in all these evaluation dimensions (i.e., novelty, diversity, accuracy, etc.), it is in general difficult to find an algorithm that outperforms any other method in all possible situations~\cite{DBLP:conf/recsys/SaidB14}.

\begin{figure}[t]
\centering
\scalebox{0.6}{
\includegraphics[trim=1cm 11cm 3cm 8cm, clip]{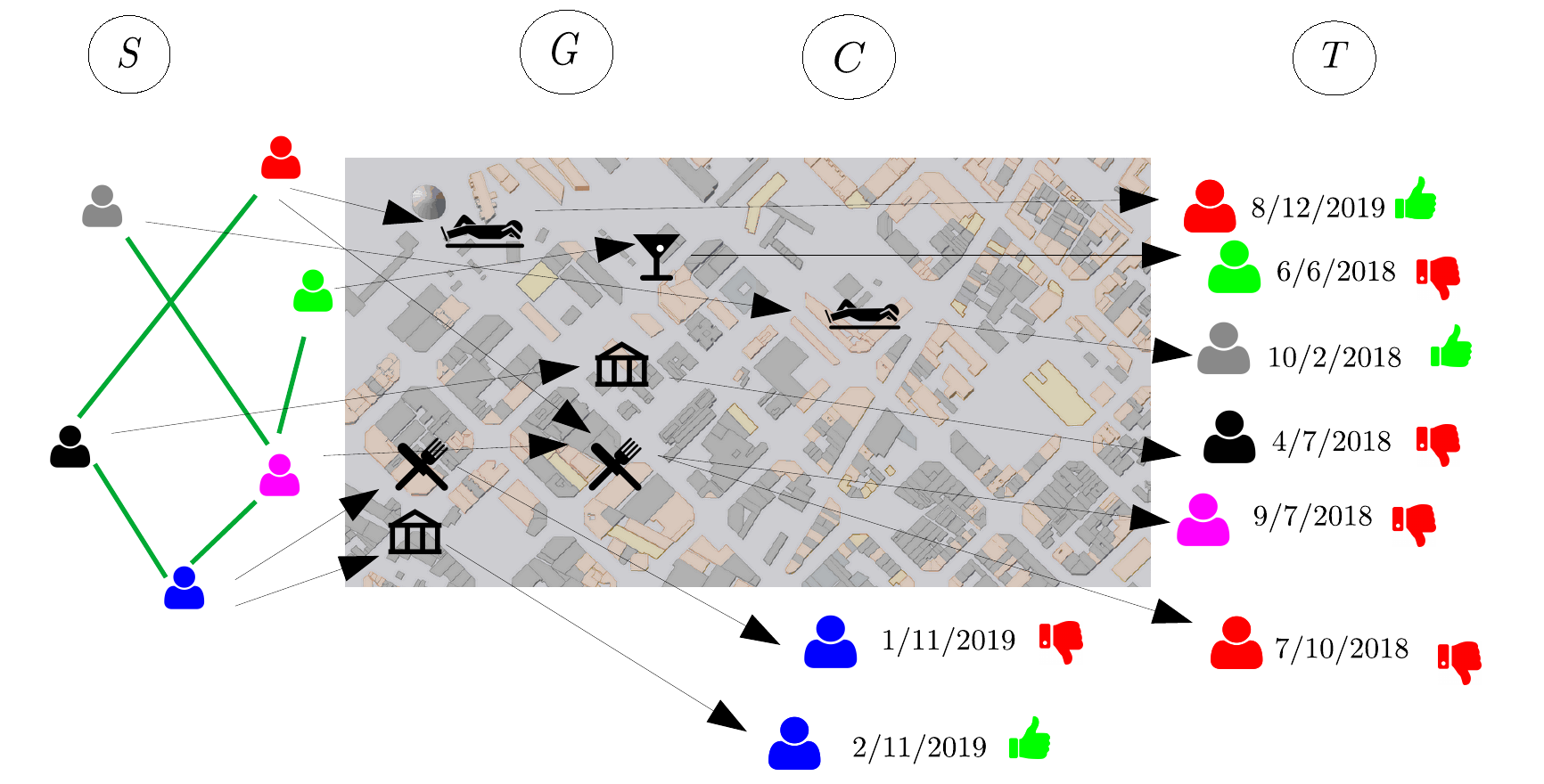}}
\caption{Graphical representation of the data found in a LBSN. Each letter represents one typical information source available in such data, with S showing the social relationships between the users, G and C show the geographical (physical coordinates in the map) and categorical (the type of POI) influences of the POIs, and T shows the moment in time when the user visited the POI and her opinion in the POI (temporal and textual/rating information).
}
\label{fig:LBSN}
\end{figure}

\section{Point-of-Interest recommendation}
\label{ss:poi}

\subsection{Problem definition}
\label{sss:def}
The key concept in POI recommendation is to suggest users new places to visit \ReviewOne{when they arrive to a city or region}, like museums, restaurants, or hotels.
Location-Based Social Networks (LBSNs) shape the data used by most of the literature devoted to this problem, and in particular, by those works analyzed here.
In these social networks users may establish social links with other users in the system, share information, and record \checkins to the specific venues they visit when located in a city.

Figure~\ref{fig:LBSN} depicts the different types of information that can be stored in and collected from these LBSNs.
Due to the great wealth of information available on these social networks, several recommendation objectives have been defined, including recommending locations, trips, activities, or friends.
As we have already introduced previously, in this review our focus is on the problem of POI or venue recommendation, for a review oriented on the rest of  recommendation objectives, we refer the reader to the survey of \cite{DBLP:journals/csur/DingLJZ18} and those discussed in Section~\ref{ss:diffs}.

Let us formalize the problem of POI recommendation.
In order to help the reader throughout the rest of the document, we will adapt the notation used in Section~\ref{s:problemDefinition} as follows.
Since in this case the items are POIs (locations) and the ratings are \checkins, we will use the letter $\mathcal{L}$ to denote the POIs and letter $\mathcal{C}$ to denote the \checkins 
as in~\cite{DBLP:journals/pvldb/LiuPCY17}. 
Moreover, even though the POI recommendation task is similar to the classical recommendation problem, it has some particularities that differ from the traditional recommendation. These include but are not limited to:

\begin{itemize}
	\item Sparsity: normally, the sparsity
  (the ratio between observed and potential preferences)
  is very high. For example, the \ReviewOne{density} of the Netflix and Movielens20M datasets (used in classical recommendation) are approximately 1.77\% and 0.537\% respectively, while the \ReviewOne{density} of datasets from Foursquare \cite{DBLP:journals/tist/YangZQ16} and Gowalla \cite{DBLP:conf/kdd/ChoML11} are 0.0034\% and 0.0047\% respectively.

	\item External influences: while in classic recommendation the only information usually exploited is the user-item matrix (user, item, score, and sometimes the timestamp associated), POI recommendation is highly affected by geographical (coordinates of the visited venues), social (friendship relationships between users), and temporal (specific moment in time when the user visited the venue) influences. Even if the use of all these influences is useful in this type of context due to the high sparsity, the geographical influence is possibly the most important aspect to consider. As the Tobler's first law of geography states~\cite{MHJ2004}: ``Everything is related to everything else, but near things are more related than distant things''. These influences are not only important to improve the performance of the algorithms, sometimes it is mandatory to take them into account because they impose certain restrictions on the recommendations. For example, some POIs such as shops, restaurants, museums, etc. are only open for a certain period and users cannot make visits to POIs that are too distant from each other.

	\item Implicit information: in classic recommendation the information encoded in the user-item matrix has been traditionally modeled using ratings. However, in most POI recommendation datasets (e.g., Brightkite, Gowalla, or Foursquare), we only have the specific moment in time when a user visited a POI. In fact, the users may have checked-in more than once in the same POI (something that it is not possible in classic recommendation).
In order to model these repeated preferences, researchers build frequency matrices in which each entry represents the number of times a user checked-in in a venue.
\end{itemize}

	\ReviewOne{Considering these features, for POI recommendation, Equation~\ref{f:optProblem}  should be replaced by a more appropriate one as follows:}
\begin{equation}
\label{eq:RecProblemPOI}
	\ReviewOne{l^*(u) = \arg\max_{l \in \mathcal{L}}{g(u,l,\theta)}}
\end{equation}
\ReviewOne{where in this case $\theta$ represents a contextual variable (e.g., the geographical information of the POIs and users, temporal influence, social context, etc.).}

In the next subsection (Section~\ref{ss:informationSource}), we present in more detail the different information sources that are usually exploited in POI recommendation.
Then, in Sections~\ref{ss:models} and~\ref{ss:evaluationProtocols} we characterize the particularities of the algorithms and evaluation methodologies, respectively, when applied to this problem.
Finally, Section~\ref{s:RecommendationTasks} presents the relation between this and other recommendation tasks.

\subsection{Alternative information sources}
\label{ss:informationSource}
As presented before, the density of most POI recommendation datasets is very low. For this reason, the vast majority of the analyzed POI recommendation approaches use more than one source of information. These include: %

\subsubsection{Interaction types}
\label{sss:int_types}
Although we have equated \checkins in LBSNs with the main interaction between users and items (as ratings in classical recommender systems),
this is not the only type of interaction recorded in this type of systems.
Other LBSNs -- such as Yelp -- allow users to perform reviews of the POIs they visit and, in some cases, rate the \ReviewOne{venue}; through these reviews we determine whether the user liked the POI or not, either by directly considering the rating or by analyzing the sentiment of the text in the review.
Other works obtain the items to be processed from the photos that users take and upload to other applications such as Flickr or Instagram~\cite{DBLP:journals/mta/NieLZS16, DBLP:conf/www/WangWTSRL17}, which may include GPS coordinates as their metadata along with visual information, so that the path followed by the users could be recovered.
Similarly, user generated content tagged with GPS coordinates -- such as tweets from Twitter or the traces left by mobile apps -- can potentially be used in POI recommendation applications.

\subsubsection{Rich side information of items}
The items in this type of systems, Points-Of-Interest, can be associated with a richer kind of information than in other domains.
First, each POI has a geographic location associated, although this information is not always available in the datasets.
This source of knowledge is especially relevant because people tend to go to places that are close to each other.
Sometimes this information is exploited to calculate centroids or clusters of activity for either users and items in order to make recommendations~\cite{DBLP:journals/kbs/SiZL19, DBLP:conf/cikm/LiuWSM14}.
We consider that an algorithm uses this kind of information if it uses the user/POIs coordinates somewhere in the proposed model (e.g., when computing distances, creating clusters, building distributions based on proximity, and so on).

Second, and more similar to the traditional recommendation situation where items usually have associated characteristics -- such as genres in the movie, book, or music domains --, in POI recommendation the venues are frequently linked to a specific POI category (for instance: restaurants, hotels, parks, museums, etc.), which may have different levels (thus, building a category hierarchy) depending on how specific the category is -- for instance, \ReviewOne{a venue} could be labeled as a \textit{Vietnamese restaurant}, an \textit{Asian restaurant}, or simply as \textit{Food}.
This information is very useful and, as we will show later, exploited in many works~\cite{DBLP:conf/kdd/YingLKT12, DBLP:conf/sigir/ZhangC15}, since some users may be more interested in visiting only certain types of POIs while, at the same time, it is not very common for a user to visit very similar POIs all the time, affecting the recommendations.

On top of this, we may find approaches that make use of the opening and closing times or the time windows or prices of the POIs, since these are particularly important characteristics when creating practical recommendations for users of real systems.
However, it should be noted that this type of information is generally considered in works that are evaluated with user studies or mobile apps, or that try to solve a different problem where constraints on the recommendations need to be taken into account (for example, trajectory instead of POI recommendation), and hence, they are less represented in this review because some of those approaches are out of its scope.

\subsubsection{Textual reviews}
In some LBSNs, users can not only register their \checkins, but also write reviews about the POIs they have visited and exchange this information with other users of the system, either as long, more elaborated texts (as in Yelp) or as short, concise texts (as the so-called \textit{tips} in Foursquare).
This type of textual information can be exploited by recommendation approaches and structure this information using topic modeling techniques like Latent Dirichlet Allocation (LDA) or Latent Semantic Analysis (LSA)~\cite{DBLP:journals/ijon/RenSES17}.
This textual information may provide more useful and high-quality information about the users' interests since, in combination with \checkin data, it is possible to capture when the user visited a venue and whether she liked it, together with the reasons about such opinion.

It is important to note that, \ReviewOne{as} the textual information available from reviews is different than the aforementioned POI features (since such features are intrinsic and static to the items, they do not change, while the reviews represent a subjective opinion from the user perspective), in our classification we will make a distinction between these two types of information, counting \ReviewOne{differently} those works that exploit textual reviews or POIs features. \ReviewOne{However, it must be taken into account that both textual and content information are related, since in some cases the researchers work with content information obtained from the text, as in~\cite{DBLP:conf/cikm/ZhangCZ15, DBLP:conf/icwe/GuoSZT19, DBLP:journals/tweb/MazumdarPB20}.}

\subsubsection{Social links}
As we already know from other domains, users tend to be more interested in a product when their friends have some opinion about it; in the same way, this type of information may influence users when receiving POI recommendations.
Because of this, some approaches exploit such social links when predicting the user preferences, for instance, by replacing the collaborative neighborhood in classical CF methods with those users who have some social relation with the user~\cite{DBLP:conf/sigir/YeYLL11, DBLP:conf/icdm/ChengC13}, or by building social graphs between the users in the system~\cite{DBLP:conf/gis/WangTM13}.

It should be considered, however, that the social links that exist in LBSNs, even though they are usually denoted as ``friends'', because of the nature of these networks, it is very likely that they do not correspond to friends outside of the system, but similar-minded people or with close tastes, interested in following their opinions or controling the places they visit.
In fact, some datasets that include this information it is actually extracted from a different social network (for instance, the global-scale dataset from Foursquare reports friends from Twitter~\cite{DBLP:journals/tist/YangZQ16}), so these social links should be exploited with great care.

\subsubsection{Sequential and temporal information}
As discussed before, the temporal dimension is \ReviewOne{essential} in the domain of POI recommendation, mainly because it affects \ReviewOne{significantly} the type of venues that can be visited, but also because users tend to diversify when deciding the next place to visit. \ReviewOne{Hence,} it becomes paramount to know, and to consider in the recommendation process, \ReviewOne{the users'} previous visits.
Similarly, since the user interactions usually have a timestamp associated, it is possible to exploit this data to know the evolution of users' tastes over time; it can also be used to detect the periods of time where some POIs have more activity than others (e.g., bars and restaurants from midday onwards).

In this survey we consider that an algorithm uses sequential information if it processes or analyzes the different events when they occur immediately one after the other or if they exploit successive visits to different POIs.
At the same time, we assume that a model uses temporal information if they work with the different timestamps of the \checkins or if they use time schedules of the POIs.

It is important to note the distinction between temporal and sequential information. While these are clearly related terms, they are not completely equivalent: not every sequential event need to be temporal and vice versa.
For example, once we know a user visited three venues at 4PM, 8PM, and 10AM, we might be tempted to create a sequence of length $3$, however, it is very likely that the user stopped to rest during the night, so the sequence should be splitted; the inverse case is more obvious: if we know the sequence followed by a user, it is impossible to recover the exact timestamps unless we know information about the initial time, and the time involved to go from each venue to the next, together with how much time was spent in each of them.

\subsection{Characterization of POI recommender systems}
\label{ss:models}
In this section, we classify existing research works according to six main classes of algorithms, based on the most frequent approaches we have identified in our analysis: based on similarities, factorization models, probabilistic approaches, deep learning techniques, graph- or link-based methods, and hybrid models.
These categories may or may not use more than one information source among those presented in the previous section, as we shall discuss in detail in Section~\ref{s:soa_algo}.
In the following, we describe these categories together with some representative methods from the state-of-the-art reviewed in this survey.

\subsubsection{Based on similarities}
These algorithms correspond to the classic \KNN approaches explained in Section~\ref{ss:knn}, where researchers use similarities between users or items like the well-known cosine similarity.
\ReviewOne{The pure user-based CF approach is defined as follows: }
\begin{equation}
	%\ReviewOne{\hat{s}_{ul} = \frac{\sum_{v \in \mathcal{N}_{l}(u)}^{}c_{vi}w_{uv}}{\sum_{v \in \mathcal{N}_{l}(u)}w_{uv}}}
	\ReviewOne{\hat{g}(u, l) \propto \sum_{v \in \mathcal{N}_{l}(u)}^{}sim(u,v) c_{vl}}
\end{equation}
\ReviewOne{where $\hat{g}(u, l)$ represents the predicted score for the user-location pair (as in Equation~\ref{eq:RecProblemPOI}), $c_{vl}$ indicates the influence of venue $l$ on user $v$ (usually as a function of the \checkin frequency), $\mathcal{N}_{l}(u)$ denotes the neighbors of user $u$ that have also visited location $l$, and $sim(u,v)$ represents the similarity between users $u$ and $v$.}

Due to the additional information available in this domain, some authors incorporate a temporal decay in the formulation or even use similarities based on the geographic distance between items. 
For example, the \UTESE approach from~\cite{DBLP:conf/sigir/YuanCMSM13} divides the check-in matrix in different time slots and uses it in a user neighborhood CF model; however, since this increases the data sparsity, the authors add a term in the prediction score to account for the similarity between time slots.

However, as social links between users are often available, instead of calculating similarities between users, in some works, they use the friends of the target user as ``nearest neighbors'' as they assume that friends in this type of networks may have common interests, \ReviewOne{as done in the \MLR model from~\cite{DBLP:journals/isci/GengJGLW19}}.

It should be noted that in the examined works, neighbor-based models are usually an intermediate phase of a more complex algorithm.
That is why we decided to extend the category beyond \KNN approaches to consider any proposal that use similarities between items/users and/or use these similarities to establish relationships between them.
As a particular example, the \LARS approach proposed in~\cite{DBLP:conf/icde/LevandoskiSEM12} would fit in this category, since it takes into account two different similarity spaces: preference locality (users in the same region tend to have similar tastes) and travel locality (users tend to travel short distances when visiting the venues of a region).

\subsubsection{Factorization}
The basic premise of this family of algorithms is to decompose the \checkin matrix $\mathcal{C}$ $\in \mathbb{R}^{|\mathcal{U}| \times |\mathcal{L}|}$ into two matrices, one for users $\mathcal{U}$ $\in \mathbb{R}^{|\mathcal{U}| \times|K|}$ and one for POIs $\mathcal{L}$ $\in \mathbb{R}^{|\mathcal{L}| \times |K|}$, with $K$ being the number of latent factors, \ReviewOne{Formally, these models try to optimize the following function:}
\begin{equation}
\ReviewOne{\min_{U,L} ||C-UL^T||^2_F + \lambda_1||U||_F^2 + \lambda_2||L||_F^2}
\end{equation}

\ReviewOne{However, in most models, the previous formulation is augmented by incorporating additional influences such as geographical or temporal ones}. We want to note we did not name this class of techniques as the most frequent name \textit{matrix factorization}, because algorithms using tensor factorization (where the additional dimension is used to model time or geographical information) or other types of latent factor models also fit in this category. 

Recommendation approaches that belong to this type include \GTBNMF, proposed in~\cite{DBLP:conf/kdd/LiuFYX13}, which is a geographical probabilistic factor analysis framework that takes into account the geographical influence and the textual information of the POIs, to avoid limitations from pure collaborative information such as the cold-start problem.
In fact, factorization approaches that exploit the geographical information are very frequent in this domain, as this is a critical information source.
The following three models have become state-of-the-art baselines because of their popularity in the area.
First, \GEOMF, a weighted factorization model proposed in~\cite{DBLP:conf/kdd/LianZXSCR14} that divides the full geographical space into different grids to model the following influences: user activity areas and POI influence areas.
Second, \IRENMF from~\cite{DBLP:conf/cikm/LiuWSM14} incorporates geographical information in the form of neighboring POIs of the target item by exploiting two types of influences: the instance level influence (assuming users tend to visit neighboring locations) and region level influence (to capture user preferences that are shared in the same geographical region). 
Third, in \RANKGEOFM from~\cite{DBLP:conf/sigir/LiCLPK15} the authors propose a geographical factorization method that incorporates the influence of the neighboring POIs of the target item by including a distance weight in the optimization formula. 

Other methods like \GEOIE proposed in~\cite{DBLP:conf/ijcai/WangSOC18} also incorporate geographical influence, but in this case a power-law distribution is used to consider that POIs that are far from other POIs in the system are less likely to be selected.
A tensor model is introduced in~\cite{DBLP:conf/aaai/HeLLSC16}, where the authors apply factorization techniques to transition tensors so that transitions between consecutive POIs are modeled, together with a geographical preference term so that far away POIs are less likely to be selected, just as in the previous approach.
\ReviewOne{Finally, \SPR from \cite{DBLP:journals/kbs/ZhaoLQH20} also fit in this category. In this model, the authors incorporate the geographical influence (distance between POIs and users) and sentiment similarity between POIs extracted from micro blogs into a classical MF approach.}

The temporal dimension is exploited in~\cite{DBLP:conf/recsys/GaoTHL13}, where the authors propose \LRT, a matrix factorization model that incorporates the temporal effects of the POIs by considering two properties: non-uniformness (the users have different preferences during the day) and consecutiveness (users tend to have similar preferences in consecutive hours).
\STELLAR, the model proposed in~\cite{DBLP:conf/aaai/ZhaoZYLK16}, is a time-aware successive POI recommendation model by using a four-tuple tensor while adapting the Bayesian Personalized Ranking (BPR) optimization criteria from~\cite{DBLP:conf/uai/RendleFGS09}.
\GEOTEASER as proposed in~\cite{DBLP:conf/www/ZhaoZKL17}, on the other hand, combines two different models: a temporal POI embedding for sequential influence that differentiates between weekday and weekends, and a hierarchical pairwise preference ranking model based on BPR that discriminates POIs based on the distance between them.

Social information has also been used in factorization techniques.
For instance, \TENMF from \cite{DBLP:journals/toit/YaoSWZQ17} is a tensor factorization approach (integrating users, venues, and time frames) that incorporates spatial and social influences in the regularization terms.
\GEOEISO is an MF approach based on the SVD++ model proposed in~\cite{DBLP:journals/ijon/GaoLLSZ18} that incorporates both geographical and social influence (in particular, the trust relationships between the users).
The model \TGSCPMF proposed in~\cite{DBLP:journals/ijon/RenSES17} also combines different information sources, since its probabilistic matrix factorization component exploits categorical and textual information by using an LDA technique, a kernel density estimation uses geographical information, and social information is combined through a power-law distribution.

Categorical or content information, as in the last method described, is easy to be integrated in factorization methods.
For instance, \CAPRF is proposed in~\cite{DBLP:conf/aaai/GaoTHL15} where besides the user and POI latent matrices, it incorporates the content and sentiment analysis obtained from the user tips.
In a more complex method, \ASMF merges social, geographical, and categorical influences, by exploiting \checkins of social, location, and neighboring friends in order to learn the potential locations to recommend, and using an additional score based on a distance distribution between the users's home and their actual \checkins to model the geographical influence, while the categorical information is considered through an additional weight in the recommendation score.

\subsubsection{Probabilistic}
Probabilistic approaches typically consider several random variables that might be related according to some laws or formulations, which in recommendation usually involve users, items, and the potential interaction between the former and the latter.
Probabilistic graphical models are one of the most useful frameworks that allow to encode these probability distributions over arbitrary domains, however it is possible to also define simple probability models just by applying Na\"ive Bayes or other simple approximations with strong (and probably not too realistic) independence assumptions.
Besides those techniques that match these formulations, we also extend our categorization as probabilistic to any model that uses some kind of probabilistic distribution in its algorithms to represent or process the data.

In this sense, for example, we consider that those approaches that model the geographic influence by means of power-law distributions such as~\cite{DBLP:conf/ijcai/WangSOC18} and~\cite{DBLP:journals/ijon/RenSES17}, those that make use of the Kernel Density Estimation (KDE) like~\cite{DBLP:conf/gis/ZhangCL14}, or those using Bayesian algorithms in the inference or in the optimization steps as in~\cite{DBLP:journals/eswa/LiXCZ15} fit into this category.
Another example can be found in \WWO from~\cite{DBLP:conf/kdd/LiuLLQX16}, which is a model that exploits the sequential preferences of the users to recommend POIs within a time duration; for this, it estimates the distribution of the temporal intervals and creates a low-rank graph to deal with the sparse conditions of the data. 

It is important to mention that many proposals can be classified as members of the probabilistic and factorization categories, such as Probabilistic Matrix Factorization (PMF) or some formal topic modeling algorithms, like Latent Dirichlet Allocation (LDA);
some examples can be found in~\cite{DBLP:conf/webi/GuoHT15,DBLP:journals/ijon/RenSES17}.
For instance, 
\ReviewOne{Poisson Geo-PFM, the algorithm proposed in \cite{DBLP:journals/tkde/LiuXPFY15}, is a geographical probabilistic factor method that models the geographical influence by using a parametric power-law distribution to represent the users activity areas over a set of latent regions, and 
\HILDA~\citep{DBLP:journals/ijon/XiongQHXBLYY20} and \MMBE~\citep{DBLP:journals/fgcs/HuangMLS20} are latent probabilistic models based on LDA. \HILDA exploits three different factors: community-behavior (social information), region-POI component (geographical information) and the sentiment-word (textual data), while 
\MMBE is a multi-modal Bayesian embedding model that exploits several influences: social (using user embeddings), sequential (using skip-gram and DeepWalk, a mechanism to learn embeddings of vertices in a graph~\cite{DBLP:conf/kdd/PerozziAS14}), geographical (exploiting different regions), content (topics) and temporal (used for modeling the distribution over topics). A different approach is \GAIMC, a method that first models the geographical influence by using a Gaussian Mixture Model (GMM) and then uses a matrix completion approach to perform the recommendations. 
}

In the same way, we consider proposals based on Markov Chains (MC) as probabilistic since they model the probability of going to the next POI using the immediately previous visited POIs.
In fact, this is one of the most popular approaches because of its simplicity and expressiveness.
For example, the authors of~\cite{DBLP:conf/ijcai/ChengYLK13} propose \FPMCLR, an approach that makes use of Factorized Personalized Markov Chains (FPMC) but adding physical restrictions: instead of building the entire transition tensor, only neighbor POIs are considered after dividing the Earth in different grids; then, a modified version of the BPR optimization technique is used to take into account the sequential components.
\PRMEG is a next-POI metric embedding method proposed in~\cite{DBLP:conf/ijcai/FengLZCCY15} that models the sequential influence by borrowing ideas from Markov Chains: instead of computing the transition probabilities by counting, they are estimated by computing the Euclidean distance of the POIs in a latent space. 

A more complex method is proposed in~\cite{DBLP:journals/www/YingWXLLZX19}, where the approach called \MEAPT considers the sequential component between the POIs (using a first-order MC) and also the temporal influence by modeling the periodicity and the time intervals between the POIs; then, the user preferences, POI transitions, and POI and temporal relationships are transformed into three latent spaces, while exploiting the Euclidean distance and using BPR as optimization criterion.

\subsubsection{Deep Learning}
Deep Learning (DL) encompasses a set of techniques from the Machine Learning area.
While they emerged throughout the 20th century, in the area of recommendation their popularity has begun to grow in the last 10 years.
When processing and learning from the data, these types of techniques make use of layers of artificial neurons in order to obtain different representations of the data by optimizing a differentiable function.
Although there are many types of neural networks, some of the best known are~\cite{DBLP:journals/csur/ZhangYST19}: the Multilayer Perceptron (MLP), that is the most basic neural network composed by one or more hidden layers between the output and the input layer using different activation functions in each neuron; the Autoencoder (AE) and Variational Autoencoder (VAE), that are unsupervised techniques oriented as compressing and then rebuilding the original data (VAE also assumes that the input data follows a probability distribution and tries to learn the parameters of that distribution); Convolutional Neural Networks (CNN), oriented at processing images using pooling operations and convolutional layers; and Recurrent Neural Networks (RNN) that memorize previous computations for processing sequential information.
As we shall see later, these approaches for POI recommendation have become paramount in the last \ReviewOne{$4$} years \ReviewOne{(and in the year 2020 it has been the most extended type of model in the area)}; some paradigmatic examples are the following.

First, \PACE is a deep learning technique proposed in~\cite{DBLP:conf/kdd/YangBZY017}, where an architecture with three main components is presented: an embedding layer that takes as inputs the embeddings of the POI and user, the context layer used for context prediction, and the preference layer composed by multiple feed-forward layers.
Second, \VPOI from~\cite{DBLP:conf/www/WangWTSRL17} is one of the few approaches that use images for POI recommendation, because of this, here the authors use CNNs to extract the visual contents from the images and which are later exploited in the learning process.
Other examples are \CARA, an approach based on RNNs proposed in~\cite{DBLP:conf/sigir/ManotumruksaMO18}, that consists in two gating mechanisms: the first one to control the influence of ordinary contexts and the second one to model the sequential influence by analyzing time intervals and geographical distance between successive \checkins and \ReviewOne{\SAENAD, an autoencoder model in which the encoder uses a self-attentive mechanism in which the most representative POIs contribute more to the hidden representation of the user, while the decoder incorporates geographical influence with a radial basis function kernel using the pairwise distance between the POIs}.
\ReviewOne{Finally, \STGN from~\cite{DBLP:conf/aaai/ZhaoZLXLZSZ19} is a spatio-temporal gated network model that incorporates two temporal and two distance gates to LSTM, to control the influence of short (recent visited POIs) and long (all previous visited POIs) term preferences.}

Other deep learning techniques that have been used more recently in the area of POI recommendation are embeddings, specifically graph and word embeddings.
The latter consists of learning a latent representation of the words so that those that have a similar meaning also have a similar representation~\cite{DBLP:conf/kes/NailiCG17}, while in graph embeddings the objective is to transform a graph into one or more $d$-dimensional vectors which preserve the graph information as much as possible~\cite{DBLP:journals/tkde/CaiZC18}.
Nevertheless, other techniques such as matrix factorization are used to learn these embeddings, so in this review we will include these proposals within the family of deep learning techniques or factorization depending on the specific case.
One example from the POI recommendation domain is \STA from~\cite{DBLP:journals/tois/QianLNY19}, where the authors define a graph embedding approach that incorporates both temporal and geographical information.

\subsubsection{\GraphLink}
%%Link analysis or Social Graph. We will consider graph if they use random Walk, HITS, PageRank or social link between the users. However if they use knowledge graph to exploit content information or similar, They will not be considered as graph.
Link-based or graph-based techniques build one or more graphs using the data stored in the system, which in our case is a LBSN.
They typically consider the users or POIs as nodes, and exploit various influences (e.g., geographical, social, temporal, etc.) to create and weight links between these nodes.
There are a great number of models based on graphs, among which the following are the most commonly used in POI recommendation approaches: Random Walk, Hypertext Induced Topic Selection (HITS), PageRank, etc.
However, as we shall see in the next sections, its popularity in the area of POI recommendation is not very high.

Among the few (representative) examples we have found in the literature, the following is a paradigmatic example of how this type of methods are used.
In~\cite{DBLP:conf/socialcom/NoulasSLM12}, a Random Walk approach is proposed, where a graph is built in which both the venues and users are nodes of the graphs, and where a link exists between a user and an item whenever the user has checked-in in that item; additionally, users are linked to each other based on their social relationships.
In the model proposed in~\cite{DBLP:conf/cikm/YuanCS14}, \GTAG, also two types of links are used, but considering different information: geographical and temporal influence; on the one hand, POIs are connected by distance to the nearest venues and weighted according to a power-law distribution, on the other hand, users and POIs are connected according to sessions defined based on their \checkins and using an exponential function to weight the edges to account for the temporal influence; with all this information, a Breadth-first Preference Propagation algorithm is used.

Thanks to the flexibility of these models, they can exploit almost any type of information source.
For instance, in~\cite{DBLP:conf/gis/0003ZM12} the authors propose an online POI recommendation model where a weighted categorical tree is built for each user, where a HITS-based approach is used to obtain local experts, which are later used to produce recommendations.
A more complex approach is found in \UPOIWALK, where a Dynamic HITS-based Random Walk model is proposed in~\cite{DBLP:journals/tist/YingKTL14} that combines several relationships captured in the data: popularity (between POIs and \checkins), social (between POIs and users' social circles), and categorical (between semantic labels and user preferences).

\subsubsection{Hybrid}
Contrary to the more classical understanding of how hybrid methods are defined~\cite{DBLP:conf/adaptive/Burke07}, in this review we do not classify approaches using and combining several components within the same algorithm as such -- for example, adding a similarity computation in a matrix factorization formulation or using a matrix factorization algorithm in a more complex deep learning model.
We decided to follow this procedure because,
as we have discussed previously and it can be observed in most of the examples shown before,
most algorithms combine several sources (geographical, temporal, sequential, social, etc.) in different ways, in such a way that if we took the more traditional and strict definition of hybrid recommender, almost every recommendation approach could be considered as such. 
\ReviewOne{Hence, most of the analyzed hybrid approaches follow this formulation:}
\begin{equation}
	%\ReviewOne{\hat{s}(u,l,\theta,C) = \sum_{i=1}^{|C|} w_i \cdot c_i(u,l,\theta)}
  %% C is already checkins!
	\ReviewOne{\hat{g}(u,l,\theta,H) = \sum_{h=1}^{|H|} w_h \cdot \hat{g}_h(u,l,\theta)}
\end{equation}
\ReviewOne{where $\theta$ represents again the contextual information%(timestamp, user's current geographical position, etc)
, whereas $H$ denotes the set of components of the hybrid approach and $w_i$ describes the weight associated to the corresponding component. Note that sometimes the aggregation function is not a sum but a product operation.}

In the following, we present some of the approaches that we do consider as hybrids, starting with those that integrate social information, since it was identified in many hybrid methods.
One possible reason for this is that this information source cannot be easily modeled under a unified framework together with other sources due to its different nature, hence, it needs tailored combinations or aggregations as the ones we present next.
For example, the \UPOIMINE approach proposed in~\cite{DBLP:conf/kdd/YingLKT12} besides considering the individual preferences of the users (through the tags of the previously visited POIs), it exploits the social information from the target user friends and uses the popularity of the venues in order to counter the data sparsity; all of this is then combined into a regression tree model, focused on predicting the next restaurant to visit, instead of general POIs.
On the other hand, the \USG model proposed in~\cite{DBLP:conf/sigir/YeYLL11} combines three different scores: user preferences, social information through a combination of the classical user-based CF formulation, and geographical influence with a power-law distribution.
Similarly, various information sources are combined in~\cite{DBLP:journals/isci/GengJGLW19}, although in this case the authors model POI recommendation as a multi-objective optimization problem considering social, geographical, and user similarity influences.

As in the previously described method, we found several approaches where two other sources of information besides social are exploited.
\LORE is a method proposed in~\cite{DBLP:conf/gis/ZhangCL14} that combines social (by computing similarities between friends), geographical (using a two-dimensional Kernel Density Estimation), and sequential (with an additive Markov Chain trained with the transition probabilities between all the users) information.
Categorical information is exploited in \GEOSOCA, a model proposed in~\cite{DBLP:conf/sigir/ZhangC15} where social, geographical, and categorical influences are combined, using a power-law distribution for the first and last models, whereas a similar method to the one described in \LORE is used for the geographical one.

Besides social information, geographical (as already discussed in other parts of this review) is another source that is exploited frequently; indeed, this is evidenced in the following hybrid approaches which exploit this type of signal among others. \ReviewOne{In \cite{DBLP:conf/cikm/LiuLAM13}, the authors propose a hybrid model that combines a matrix factorization approach using a transition matrix to model the transitions between POIs with a power-law distribution.} 
The so-called \LBPR method from~\cite{DBLP:conf/ijcai/HeLL17} adapts the BPR technique to predict the next category and then obtain a ranking of POIs using the predicted category and after incorporating a geographical score for the candidate POIs; the main difference with other approaches based on BPR is that this model uses lists of categories instead of category pairs in the learning step.
The \APRASA model as proposed in~\cite{DBLP:journals/kbs/SiZL19}, on the other hand, takes into account geographical and temporal information by computing the popularity of the POIs in different time periods and using a Kernel Density Estimation component.
Finally, the \GE method proposed in~\cite{DBLP:conf/cikm/XieYWXCW16} consists of a graph-based embedding model where four types of graphs are considered: a POI-POI graph (to capture the \checkin sequences of POIs), a POI-region graph (to exploit the geographical information), a POI-time graph (for temporal and cyclic behaviour), and a POI-word graph (to exploit semantics). \ReviewOne{Finally, \TECF from~\cite{DBLP:journals/tii/WangCWCLG20} is a hybrid approach that combines user-based collaborative filtering (based on DeepWalk), a temporal user-based collaborative filtering, and a power-law distribution for modeling the geographical influence.}

\begin{table}[tb]
\caption{Evaluation protocols characterized by their application level and the type of split; %.
where X+Y denotes the type (X, either \underline{R}andom or \underline{T}emporal) and application level (Y, either \underline{S}ystem or \underline{U}ser) of the split,
and \cmark, \xmark, and \qmark\ indicate if that characteristic is known to occur, never occur, or it is impossible to know in a protocol.
}
\label{tab:evProtocolFixCC}
%\small
\footnotesize
\begin{tabular}{rcccc}
\toprule
Characteristics & R+S & R+U & T+S & T+U \\
\midrule
All users are evaluated & \qmark & \cmark & \qmark & \cmark \\
Leaks future data in test from training & Very likely & Very likely & \xmark & For some users \\
Test contains recent \checkins & \qmark & \qmark & \cmark & \cmark \\
Test is made as a random subset & \cmark & \cmark & \xmark & \xmark \\
\bottomrule
\end{tabular}
\end{table}

\subsection{Characterization of evaluation methodologies}
\label{ss:evaluationProtocols}
The evaluation methodologies used in the POI recommendation domain are not too different from those traditionally used in classical recommendation and presented in Section~\ref{ss:classicEv}.
However, considering the importance some dimensions have in this domain -- i.e., time and geographical information, mostly -- we describe now in more detail those time-aware evaluation methodologies used in the area~\cite{DBLP:journals/umuai/CamposDC14}, together with some variations inherent to the POI recommendation problem.

As explained in Section~\ref{ss:classicEv}, the first step in any offline evaluation is to divide the available data into different sets: at least training and test, although an additional validation set is preferred to tune the parameters of the models and not overfit the test set.
How the original data is divided is critical to imitate the use of the recommendation algorithm in a real scenario, that is why in the POI recommendation \ReviewOne{task} we observe that the \ReviewOne{temporal} dimension is often used when splitting the dataset, even though these methodologies were already used and formalized in the area, nonetheless, random partitions of the data are still very popular~\cite{DBLP:journals/umuai/CamposDC14}.

More specifically, we consider a split is \textit{temporal} whenever the \checkins are ordered according to the temporal dimension (either by the actual timestamps or because there is some sequential information in the data) so that the \checkins in the test set are more recent than those in the training set; otherwise we consider the split is \textit{random}, including the cross-validation setting presented in previous sections.
An additional criteria that may have a great impact on the final results is whether the split is done at the system or user level; this reflects whether the previous criteria (temporal or random) is applied to the whole dataset or in a user basis -- i.e., for each user independently.
This criteria affects the number and type of users that belong to each set, since performing the split at the user level would guarantee that all the users exist in both training and test sets.
Finally, a parameter that defines different variations of these protocols is whether the subset is selected according to a percentage or ratio between training and test -- typical values are 80\% for training and the rest for test -- or based on a fixed number of elements, for example, the last $1$ or $2$ interactions go into the test set; for the sake of a cleaner presentation we will not consider this parameter in the classification we use in the next sections.
We present in Table~\ref{tab:evProtocolFixCC} a summary of the implications for the four possibilities regarding data splitting that will be used in the rest of the paper.
%For example, by making a user partition, we ensure that all users appear in the test set (a subset of interactions for every user are selected for evaluation). On the other hand, it may not be too realistic because not all users are equally active on a system and even if we perform a temporal partition by users, we would be training the models with interactions belonging to different temporal contexts.
Even though we have found some papers where the evaluation was performed in other ways (e.g., temporal windows or splits by distance between locations), most of the analyzed articles fit into the aforementioned evaluation protocol classification.
Based on this, we argue that the most realistic scenario is a temporal partition at the system level, as it takes into account the temporal dimension while avoiding any leaking of the user interactions from the future into the training set.

Because of the paramount importance of the geographical dimension, some authors include in their experimental settings variations tailored for the POI recommendation problem.
In particular, those approaches that exploit geographical information or neighbor venues tend to filter the data by cities or, in general, by geographical regions, such as country or continent.
Another important characteristic of the data produced by LBSNs is that users may visit more than once the POIs, hence, this leads to two ways of producing the splits explained before: at the \checkin level (keeping all the \checkins, even the repeated ones) or at POI level (removing the duplicated user-POI pairs and keeping only one instance before running the splitting strategy).
These repetitions may have a significant effect in training since it allows to capture item frequencies at the user level, but its effect is even more dramatic in the test set, since it may hide the fact that some uninteresting baselines (such as returning those items previously interacted by the user) would perform very well~\cite{DBLP:journals/umuai/SanchezB20}.

Finally, regarding the metrics used when evaluating POI recomender systems, it should be noted that error-based metrics are not very interesting when the interaction to be predicted is a \checkin, since that value is always $1$; when the user interaction is different (such as ratings or those described in Section~\ref{sss:int_types}), then these metrics can be applied, considering the limitations already described in Section~\ref{ss:classicEv}.
Nonetheless, it is important to mention that in recent years, researchers dealing with the problem of POI recommendation and related tasks (see Section~\ref{s:RecommendationTasks}) have adapted ranking metrics to consider distances between the recommended POIs and the actual order followed by the users in the test set; some examples can be found in~\cite{DBLP:conf/cikm/ChenOX16}, where the authors use a metric based on $F_1$ that takes into account the pairwise order between POIs, and~\cite{DBLP:journals/umuai/SanchezB20}, where the Longest Common Subsequence algorithm is introduced in ranking metrics to penalize those recommendations less similar with the sequence followed by the user.

\subsection{Relation to other recommendation tasks}
\label{s:RecommendationTasks}
POI recommendation is not the only task that can be performed using data from LBSNs, due to the richness of the data of this kind of social network, a large number of related tasks/problems have arisen.
Since they are not the focus of this review, we discuss them briefly now:

\begin{itemize}
	\item Trajectory (or route) recommendation: typically, POI recommendation approaches provide each user with a list of POIs that hopefully may be of interest; however, there is generally no intrinsic relationship between these recommended POIs. Instead, in route recommendation, a complete trajectory is generated and provided to the target user. Because of this, additional restrictions must be taken into account, such as the duration or length of the route or the schedule of the venues~\cite{DBLP:conf/cikm/ChenOX16}.

	\item Friend recommendation: this is a well studied problem in the context of traditional social networks, like Twitter or Facebook.
  Considering the importance of the social dimension in LBSNs in general, and of social information for POI recommendation (discussed in previous sections), there is an increasing interest in this problem by the community.

	\item Group recommendation: quite frequently users visit a city in groups, either composed by friends, family, or even as organized tours.
  In this case, instead of recommending POIs to a unique target user, the algorithms should be tailored to groups of users. This problem needs to take into account additional factors, such as the difference between passive and active users in such groups, or the balance between individual preferences.
\end{itemize}

\section{Systematic review of state-of-the-art algorithms}
\label{s:soa_algo}
In this section, we analyze the state-of-the-art algorithms according to the classification presented in Section~\ref{ss:models} for the papers considered in this study (described in Section~\ref{ss:how}).
Since the number of papers included in this review is very large (more than \ReviewOne{$300$}), we selected the \textit{most representative} papers for each year and include their whole characterization in Table~\ref{tab:examplesTableRepresentatives}.
When selecting the most representative papers, we considered the top-$5$ most cited articles per year according to Scopus with at least one citation\footnote{\ReviewOne{The number of cites reported have been obtained on April 14, 2021.}}.
We also include in the table two summary rows that count how many papers (among the sets of most representative or the entire collection) satisfy each condition.
Each of the conditions (columns) correspond to the categories described in Sections~\ref{ss:informationSource}, \ref{ss:models}, and~\ref{ss:evaluationProtocols}, respectively.

%
%%

%Revised May 2021
\pgfplotstableread[col sep=semicolon]{
Year;Reference;Acronym;KNN;Factorization;Probabilistic;DL;GraphLink;Hybrid;Other;Geographical;Social;Content;Textual;Sequential;Temporal;RandomS;nfoldS;TemporalS;OtherS;CC;Fix
2011;\cite{DBLP:conf/sigir/YeYLL11};{\USG};{\cmark};;{\cmark};;;{\cmark};;{\cmark};{\cmark};;;;;{\cmark};;;;;{\cmark}
2012;\cite{DBLP:conf/icde/LevandoskiSEM12};{\LARS};{\cmark};;;;;;;{\cmark};;;;;;{\cmark};;;;{\cmark};
2012;\cite{DBLP:conf/gis/0003ZM12};{\NA};{\cmark};;;;{\cmark};;;{\cmark};;{\cmark};;;;;;;{\cmark};;
2012;\cite{DBLP:conf/kdd/YingLKT12};{\UPOIMINE};{\cmark};;;;;{\cmark};;{\cmark};{\cmark};{\cmark};;;;;;;;;
2012;\cite{DBLP:conf/socialcom/NoulasSLM12};RW, Weighted-RW;;;{\cmark};;{\cmark};;;;{\cmark};;;;;;{\qmark};{\cmark};;{\cmark};
2013;\cite{DBLP:conf/cikm/LiuLAM13};{\NA};;{\cmark};{\cmark};;;{\cmark};;{\cmark};;{\cmark};;{\cmark};{\cmark};;;{\cmark};;;{\cmark}
2013;\cite{DBLP:conf/kdd/LiuFYX13};{\GTBNMF};;{\cmark};{\cmark};;;;;{\cmark};;;{\cmark};;;{\cmark};;;;{\cmark};
2013;\cite{DBLP:conf/ijcai/ChengYLK13};{\FPMCLR};;{\cmark};{\cmark};;;;;{\cmark};;;;{\cmark};;;;{\cmark};;{\cmark};
2013;\cite{DBLP:conf/recsys/GaoTHL13};{\LRT};;{\cmark};;;;;;;;;;;{\cmark};{\cmark};;;;;{\cmark}
2013;\cite{DBLP:conf/sigir/YuanCMSM13};{\UTESE};{\cmark};;{\cmark};;;{\cmark};;{\cmark};;;;;{\cmark};{\cmark};;;;;{\cmark}
2014;\cite{DBLP:journals/tist/YingKTL14};{\UPOIWALK};;;;;{\cmark};;;{\cmark};{\cmark};{\cmark};;;;;;;;;
2014;\cite{DBLP:conf/cikm/YuanCS14};{\GTAG};;;;;{\cmark};;;{\cmark};;;;;{\cmark};{\cmark};;;;;{\cmark}
2014;\cite{DBLP:conf/kdd/LianZXSCR14};{\GEOMF};;{\cmark};;;;;;{\cmark};;;;;;{\cmark};;;;;{\cmark}
2014;\cite{DBLP:conf/cikm/LiuWSM14};{\IRENMF};;{\cmark};;;;;;{\cmark};;;;;;{\cmark};;;;;{\cmark}
2014;\cite{DBLP:conf/gis/ZhangCL14};{\LORE};{\cmark};;{\cmark};;;{\cmark};;{\cmark};{\cmark};;;{\cmark};;;;{\cmark};;{\cmark};
2015;\cite{DBLP:journals/tkde/LiuXPFY15};Poisson Geo-PFM;;{\cmark};{\cmark};;;;;{\cmark};;;;;;{\cmark};;;;{\cmark};
2015;\cite{DBLP:conf/sigir/LiCLPK15};{\RANKGEOFM};;{\cmark};;;;;;{\cmark};;;;;{\cmark};;;{\cmark};;;{\cmark}
2015;\cite{DBLP:conf/sigir/ZhangC15};{\GEOSOCA};;;{\cmark};;;{\cmark};;{\cmark};{\cmark};{\cmark};;;;;;{\cmark};;{\cmark};
2015;\cite{DBLP:conf/ijcai/FengLZCCY15};{\PRMEG};;{\cmark};{\cmark};;;;;{\cmark};;;;{\cmark};{\cmark};;;{\cmark};;{\cmark};
2015;\cite{DBLP:conf/aaai/GaoTHL15};{\CAPRF};;{\cmark};;;;;;;;;{\cmark};;;{\cmark};;;;;{\cmark}
2016;\cite{DBLP:conf/cikm/XieYWXCW16};{\GE};;{\cmark};{\cmark};;;{\cmark};;{\cmark};;{\cmark};;{\cmark};{\cmark};;;{\cmark};;;{\cmark}
2016;\cite{DBLP:conf/kdd/LiGHZ16};{\ASMF};;{\cmark};;;{\cmark};;;{\cmark};{\cmark};{\cmark};;;;;;{\cmark};;;{\cmark}
2016;\cite{DBLP:conf/aaai/ZhaoZYLK16};{\STELLAR};;{\cmark};;;;;;;;;;;{\cmark};;;{\cmark};;{\cmark};
2016;\cite{DBLP:conf/aaai/HeLLSC16};{\NA};;{\cmark};{\cmark};;;;;{\cmark};;;;{\cmark};;;;;;;
2016;\cite{DBLP:conf/kdd/LiuLLQX16};{\WWO};;;{\cmark};;;;;;;;;{\cmark};{\cmark};;;{\cmark};;{\cmark};
2017;\cite{DBLP:conf/www/ZhaoZKL17};{\GEOTEASER};;{\cmark};{\cmark};{\cmark};;;;{\cmark};;;;{\cmark};{\cmark};;;{\cmark};;;{\cmark}
2017;\cite{DBLP:conf/kdd/YangBZY017};{\PACE};;;;{\cmark};;;;{\cmark};{\cmark};;;;;;;{\cmark};;;{\cmark}
2017;\cite{DBLP:journals/ijon/RenSES17};{\TGSCPMF};;{\cmark};{\cmark};;;;;{\cmark};{\cmark};{\cmark};{\cmark};;;{\cmark};;;;{\cmark};
2017;\cite{DBLP:conf/ijcai/HeLL17};{\LBPR};;{\cmark};{\cmark};;;{\cmark};;{\cmark};;{\cmark};;{\cmark};;;;{\cmark};;;{\cmark}
2017;\cite{DBLP:conf/www/WangWTSRL17};{\VPOI};;{\cmark};{\cmark};{\cmark};;;;;;{\cmark};;;;{\cmark};;;;;{\cmark}
2018;\cite{DBLP:conf/cikm/MaZWL18};{\SAENAD};;;;{\cmark};;;;{\cmark};;;;;;{\cmark};;;;;{\cmark}
2018;\cite{DBLP:conf/sigir/ManotumruksaMO18};{\CARA};;;;{\cmark};;;;{\cmark};;;;{\cmark};{\cmark};;;{\cmark};;;{\cmark}
2018;\cite{DBLP:journals/toit/YaoSWZQ17};{\TENMF};;{\cmark};;;;;;{\cmark};{\cmark};;;;{\cmark};{\cmark};;;;;{\cmark}
2018;\cite{DBLP:journals/ijon/GaoLLSZ18};{\GEOEISO};;{\cmark};{\cmark};;;;;{\cmark};{\cmark};;;;;{\cmark};;;;{\cmark};
2018;\cite{DBLP:conf/ijcai/WangSOC18};{\GEOIE};;{\cmark};{\cmark};;;;;{\cmark};;;;;;;;{\cmark};;;{\cmark}
2019;\cite{DBLP:journals/www/YingWXLLZX19};{\MEAPT};;{\cmark};{\cmark};;;;;;;;;{\cmark};{\cmark};;;{\cmark};;;{\cmark}
2019;\cite{DBLP:journals/isci/GengJGLW19};{\MLR};{\cmark};;{\cmark};;;;{\cmark};{\cmark};{\cmark};;;;;{\cmark};;;;{\cmark};
2019;\cite{DBLP:journals/kbs/SiZL19};{\APRASA};;;{\cmark};;;{\cmark};;{\cmark};;;;;{\cmark};{\cmark};;;;;{\cmark}
2019;\cite{DBLP:journals/tois/QianLNY19};{\STA};{\cmark};{\cmark};;;;;;{\cmark};;;;;{\cmark};;;{\cmark};;;{\cmark}
2019;\cite{DBLP:conf/aaai/ZhaoZLXLZSZ19};{\STGN};;;;{\cmark};;;;{\cmark};;;;{\cmark};{\cmark};;;{\cmark};;;{\cmark}
2020;\cite{DBLP:journals/ijon/XiongQHXBLYY20};{\HILDA};;{\cmark};{\cmark};;;;;{\cmark};{\cmark};;{\cmark};;;{\cmark};{\cmark};;;{\cmark};
2020;\cite{DBLP:journals/iotj/WangCWCG20};{\GAIMC};;{\cmark};{\cmark};;;;;{\cmark};;;;;;{\cmark};;;;{\cmark};
2020;\cite{DBLP:journals/kbs/ZhaoLQH20};{\SPR};;{\cmark};;;;;;{\cmark};;;{\cmark};;;{\cmark};{\cmark};;;{\cmark};
2020;\cite{DBLP:journals/fgcs/HuangMLS20};{\MMBE};;{\cmark};{\cmark};{\cmark};;;;{\cmark};{\cmark};{\cmark};;{\cmark};{\cmark};;;{\cmark};;{\cmark};
2020;\cite{DBLP:journals/tii/WangCWCLG20};{\TECF};{\cmark};;{\cmark};{\cmark};{\cmark};{\cmark};;{\cmark};;;;;{\cmark};{\cmark};;;;;{\cmark} 
}\examplesTableRepresentatives

\begin{table*}
\caption{Summary of analyzed POI recommendation approaches sorted by publication year. The \cmark mark denotes that the proposed model has the feature indicated in the column, whereas \NA shows that no acronym was given.
}
\label{tab:examplesTableRepresentatives}
\vspace{-0.25cm}
\ifloadtables
\pgfplotstabletypeset[
		font=\fontsize{7pt}{7.5pt}\selectfont,
		outfile=tab_examplesTableRepresentatives.tex,
	every head row/.style={
            before row={%
                \hline
                \multicolumn{3}{? c ? }{Details} & \multicolumn{6}{c?}{Information used} & \multicolumn{7}{c?}{Model} & \multicolumn{3}{c?}{Split type}\\
                \hline
            },
            after row={
            	\hline
            },
            typeset cell/.code={
            \ifnum\pgfplotstablecol=\pgfplotstablecols
            	\pgfkeyssetvalue{/pgfplots/table/@cell content}{\rotatebox{90}{##1}\\}%
            \else
            	\pgfkeyssetvalue{/pgfplots/table/@cell content}{\rotatebox{90}{##1}&}%
            \fi
            },
        },
        every last row/.style={
		after row={%
                \hline
		\multicolumn{3}{? c ? }{Most Representatives} & 38 & 14 & 11 & 5 & 13 & 18 & 9 & 27 & 26 & 8 & 6 & 10 & 1 & 21 & 20 & 1 \\
		\hline
		\multicolumn{3}{? c ? }{Total} & 218 & 116 & 108 & 42 & 73 & 134 & 90 & 141 & 139 & 66 & 44 & 98 & 19 & 152 &  118 & 14 \\
		\hline
		},
	},
        column type=c,
        %column type/.add={|}{},
        every even row/.style={
        	before row={
        		\hline
        		\rowcolor{lightgray!50}},
        	after row={\hline}
        },
        %%%%%%%%%%%%%%%%%%%%%%%%%%%%%%%%%%%%%%%%%%%%%
        % Year,Reference,Acronym
        columns/Year/.style={
			string type,
			column type/.add={?}{|},
        },
		columns/Reference/.style={
			string type,
			column type/.add={}{|},
        },
        columns/Acronym/.style={
			string type,
			column type/.add={}{?},
        },
        %%%%%%%%%%%%%%%%%
        % Methodology: KNN,Factorization,Probabilistic,DNN,GraphLink,Hybrid,Other
        columns/KNN/.style={
			string type,
			column name={\SIMILARITY},
			column type/.add={}{|},
        },
        columns/Factorization/.style={
			string type,
			column type/.add={}{|},
        },
        columns/Probabilistic/.style={
			string type,
			column type/.add={}{|},
        },
        columns/DL/.style={
			string type,
			column type/.add={}{|},
        },
        columns/GraphLink/.style={
			string type,
			column name={\GraphLink},
			column type/.add={}{|},
        },
        columns/Hybrid/.style={
			string type,
			column type/.add={}{|},
        },
        columns/Other/.style={
			string type,
			column type/.add={}{?},
        },
        %%%%%%%%%%%%%%%%
        %Information Used: Geographical,Social,Content,Textual, Sequential,Temporal
        columns/Geographical/.style={
			string type,
			column type/.add={}{|},
        },
        columns/Social/.style={
			string type,
			column type/.add={}{|},
        },
        columns/Content/.style={
			string type,
			column type/.add={}{|},
        },
        columns/Textual/.style={
			string type,
			column type/.add={}{|},
        },
        columns/Sequential/.style={
			string type,
			column type/.add={}{|},
        },
        columns/Temporal/.style={
			string type,
			column type/.add={}{?},
        },
        %%%%%%%%%%%%%%%%%%
        %% Type of split: TCC,TFix,RCC,RFix,nfold,Other
        columns/RandomS/.style={
			string type,
			column name={Random},
			column type/.add={}{|},
        },
        columns/nfoldS/.style={
			string type,
			column name={\nfold},
			column type/.add={}{|},
        },
        columns/TemporalS/.style={
			string type,
			column name={Temporal},
			column type/.add={}{|},
        },
        columns/OtherS/.style={
			string type,
			column name={Other},
			column type/.add={}{?},
        },
        columns/CC/.style={
			string type,
			column name={\CC},
			column type/.add={}{|},
        },
        columns/Fix/.style={
			string type,
			column name={\Fix},
			column type/.add={}{?},
        },
        fixed,
        fixed zerofill,
        columns={Year,Reference,Acronym,Geographical,Social,Content,Textual,Sequential,Temporal,KNN,Factorization,Probabilistic,DL,GraphLink,Hybrid,Other,RandomS,TemporalS,OtherS},
    ]{\examplesTableRepresentatives}
\else
	\begingroup \fontsize {7pt}{7.5pt}\selectfont %
\begin {tabular}{?c|c|c?c|c|c|c|c|c?c|c|c|c|c|c|c?c|c|c?}%
\hline \multicolumn {3}{? c ? }{Details} & \multicolumn {6}{c?}{Information used} & \multicolumn {7}{c?}{Model} & \multicolumn {3}{c?}{Split type}\\ \hline \rotatebox {90}{Year}&\rotatebox {90}{Reference}&\rotatebox {90}{Acronym}&\rotatebox {90}{Geographical}&\rotatebox {90}{Social}&\rotatebox {90}{Content}&\rotatebox {90}{Textual}&\rotatebox {90}{Sequential}&\rotatebox {90}{Temporal}&\rotatebox {90}{\SIMILARITY }&\rotatebox {90}{Factorization}&\rotatebox {90}{Probabilistic}&\rotatebox {90}{DL}&\rotatebox {90}{\GraphLink }&\rotatebox {90}{Hybrid}&\rotatebox {90}{Other}&\rotatebox {90}{Random}&\rotatebox {90}{Temporal}&\rotatebox {90}{Other}\\\hline %
\hline \rowcolor {lightgray!50}2011&\cite {DBLP:conf/sigir/YeYLL11}&\USG &\cmark &\cmark &&&&&\cmark &&\cmark &&&\cmark &&\cmark &&\\\hline %
2012&\cite {DBLP:conf/icde/LevandoskiSEM12}&\LARS &\cmark &&&&&&\cmark &&&&&&&\cmark &&\\%
\hline \rowcolor {lightgray!50}2012&\cite {DBLP:conf/gis/0003ZM12}&\NA &\cmark &&\cmark &&&&\cmark &&&&\cmark &&&&&\cmark \\\hline %
2012&\cite {DBLP:conf/kdd/YingLKT12}&\UPOIMINE &\cmark &\cmark &\cmark &&&&\cmark &&&&&\cmark &&&&\\%
\hline \rowcolor {lightgray!50}2012&\cite {DBLP:conf/socialcom/NoulasSLM12}&RW, Weighted-RW&&\cmark &&&&&&&\cmark &&\cmark &&&&\cmark &\\\hline %
2013&\cite {DBLP:conf/cikm/LiuLAM13}&\NA &\cmark &&\cmark &&\cmark &\cmark &&\cmark &\cmark &&&\cmark &&&\cmark &\\%
\hline \rowcolor {lightgray!50}2013&\cite {DBLP:conf/kdd/LiuFYX13}&\GTBNMF &\cmark &&&\cmark &&&&\cmark &\cmark &&&&&\cmark &&\\\hline %
2013&\cite {DBLP:conf/ijcai/ChengYLK13}&\FPMCLR &\cmark &&&&\cmark &&&\cmark &\cmark &&&&&&\cmark &\\%
\hline \rowcolor {lightgray!50}2013&\cite {DBLP:conf/recsys/GaoTHL13}&\LRT &&&&&&\cmark &&\cmark &&&&&&\cmark &&\\\hline %
2013&\cite {DBLP:conf/sigir/YuanCMSM13}&\UTESE &\cmark &&&&&\cmark &\cmark &&\cmark &&&\cmark &&\cmark &&\\%
\hline \rowcolor {lightgray!50}2014&\cite {DBLP:journals/tist/YingKTL14}&\UPOIWALK &\cmark &\cmark &\cmark &&&&&&&&\cmark &&&&&\\\hline %
2014&\cite {DBLP:conf/cikm/YuanCS14}&\GTAG &\cmark &&&&&\cmark &&&&&\cmark &&&\cmark &&\\%
\hline \rowcolor {lightgray!50}2014&\cite {DBLP:conf/kdd/LianZXSCR14}&\GEOMF &\cmark &&&&&&&\cmark &&&&&&\cmark &&\\\hline %
2014&\cite {DBLP:conf/cikm/LiuWSM14}&\IRENMF &\cmark &&&&&&&\cmark &&&&&&\cmark &&\\%
\hline \rowcolor {lightgray!50}2014&\cite {DBLP:conf/gis/ZhangCL14}&\LORE &\cmark &\cmark &&&\cmark &&\cmark &&\cmark &&&\cmark &&&\cmark &\\\hline %
2015&\cite {DBLP:journals/tkde/LiuXPFY15}&Poisson Geo-PFM&\cmark &&&&&&&\cmark &\cmark &&&&&\cmark &&\\%
\hline \rowcolor {lightgray!50}2015&\cite {DBLP:conf/sigir/LiCLPK15}&\RANKGEOFM &\cmark &&&&&\cmark &&\cmark &&&&&&&\cmark &\\\hline %
2015&\cite {DBLP:conf/sigir/ZhangC15}&\GEOSOCA &\cmark &\cmark &\cmark &&&&&&\cmark &&&\cmark &&&\cmark &\\%
\hline \rowcolor {lightgray!50}2015&\cite {DBLP:conf/ijcai/FengLZCCY15}&\PRMEG &\cmark &&&&\cmark &\cmark &&\cmark &\cmark &&&&&&\cmark &\\\hline %
2015&\cite {DBLP:conf/aaai/GaoTHL15}&\CAPRF &&&&\cmark &&&&\cmark &&&&&&\cmark &&\\%
\hline \rowcolor {lightgray!50}2016&\cite {DBLP:conf/cikm/XieYWXCW16}&\GE &\cmark &&\cmark &&\cmark &\cmark &&\cmark &\cmark &&&\cmark &&&\cmark &\\\hline %
2016&\cite {DBLP:conf/kdd/LiGHZ16}&\ASMF &\cmark &\cmark &\cmark &&&&&\cmark &&&\cmark &&&&\cmark &\\%
\hline \rowcolor {lightgray!50}2016&\cite {DBLP:conf/aaai/ZhaoZYLK16}&\STELLAR &&&&&&\cmark &&\cmark &&&&&&&\cmark &\\\hline %
2016&\cite {DBLP:conf/aaai/HeLLSC16}&\NA &\cmark &&&&\cmark &&&\cmark &\cmark &&&&&&&\\%
\hline \rowcolor {lightgray!50}2016&\cite {DBLP:conf/kdd/LiuLLQX16}&\WWO &&&&&\cmark &\cmark &&&\cmark &&&&&&\cmark &\\\hline %
2017&\cite {DBLP:conf/www/ZhaoZKL17}&\GEOTEASER &\cmark &&&&\cmark &\cmark &&\cmark &\cmark &\cmark &&&&&\cmark &\\%
\hline \rowcolor {lightgray!50}2017&\cite {DBLP:conf/kdd/YangBZY017}&\PACE &\cmark &\cmark &&&&&&&&\cmark &&&&&\cmark &\\\hline %
2017&\cite {DBLP:journals/ijon/RenSES17}&\TGSCPMF &\cmark &\cmark &\cmark &\cmark &&&&\cmark &\cmark &&&&&\cmark &&\\%
\hline \rowcolor {lightgray!50}2017&\cite {DBLP:conf/ijcai/HeLL17}&\LBPR &\cmark &&\cmark &&\cmark &&&\cmark &\cmark &&&\cmark &&&\cmark &\\\hline %
2017&\cite {DBLP:conf/www/WangWTSRL17}&\VPOI &&&\cmark &&&&&\cmark &\cmark &\cmark &&&&\cmark &&\\%
\hline \rowcolor {lightgray!50}2018&\cite {DBLP:conf/cikm/MaZWL18}&\SAENAD &\cmark &&&&&&&&&\cmark &&&&\cmark &&\\\hline %
2018&\cite {DBLP:conf/sigir/ManotumruksaMO18}&\CARA &\cmark &&&&\cmark &\cmark &&&&\cmark &&&&&\cmark &\\%
\hline \rowcolor {lightgray!50}2018&\cite {DBLP:journals/toit/YaoSWZQ17}&\TENMF &\cmark &\cmark &&&&\cmark &&\cmark &&&&&&\cmark &&\\\hline %
2018&\cite {DBLP:journals/ijon/GaoLLSZ18}&\GEOEISO &\cmark &\cmark &&&&&&\cmark &\cmark &&&&&\cmark &&\\%
\hline \rowcolor {lightgray!50}2018&\cite {DBLP:conf/ijcai/WangSOC18}&\GEOIE &\cmark &&&&&&&\cmark &\cmark &&&&&&\cmark &\\\hline %
2019&\cite {DBLP:journals/www/YingWXLLZX19}&\MEAPT &&&&&\cmark &\cmark &&\cmark &\cmark &&&&&&\cmark &\\%
\hline \rowcolor {lightgray!50}2019&\cite {DBLP:journals/isci/GengJGLW19}&\MLR &\cmark &\cmark &&&&&\cmark &&\cmark &&&&\cmark &\cmark &&\\\hline %
2019&\cite {DBLP:journals/kbs/SiZL19}&\APRASA &\cmark &&&&&\cmark &&&\cmark &&&\cmark &&\cmark &&\\%
\hline \rowcolor {lightgray!50}2019&\cite {DBLP:journals/tois/QianLNY19}&\STA &\cmark &&&&&\cmark &\cmark &\cmark &&&&&&&\cmark &\\\hline %
2019&\cite {DBLP:conf/aaai/ZhaoZLXLZSZ19}&\STGN &\cmark &&&&\cmark &\cmark &&&&\cmark &&&&&\cmark &\\%
\hline \rowcolor {lightgray!50}2020&\cite {DBLP:journals/ijon/XiongQHXBLYY20}&\HILDA &\cmark &\cmark &&\cmark &&&&\cmark &\cmark &&&&&\cmark &&\\\hline %
2020&\cite {DBLP:journals/iotj/WangCWCG20}&\GAIMC &\cmark &&&&&&&\cmark &\cmark &&&&&\cmark &&\\%
\hline \rowcolor {lightgray!50}2020&\cite {DBLP:journals/kbs/ZhaoLQH20}&\SPR &\cmark &&&\cmark &&&&\cmark &&&&&&\cmark &&\\\hline %
2020&\cite {DBLP:journals/fgcs/HuangMLS20}&\MMBE &\cmark &\cmark &\cmark &&\cmark &\cmark &&\cmark &\cmark &\cmark &&&&&\cmark &\\%
\hline \rowcolor {lightgray!50}2020&\cite {DBLP:journals/tii/WangCWCLG20}&\TECF &\cmark &&&&&\cmark &\cmark &&\cmark &\cmark &\cmark &\cmark &&\cmark &&\\\hline \multicolumn {3}{? c ? }{Most Representatives} & 38 & 14 & 11 & 5 & 13 & 18 & 9 & 27 & 26 & 8 & 6 & 10 & 1 & 21 & 20 & 1 \\ \hline \multicolumn {3}{? c ? }{Total} & 218 & 116 & 108 & 42 & 73 & 134 & 90 & 141 & 139 & 66 & 44 & 98 & 19 & 152 & 118 & 14 \\ \hline %
\end {tabular}%
\endgroup %
\fi
\end{table*}

%%%%%%%%%%%%%%
%% 

%Revised May 2021
%\ifloadtables

%% Question: is it really necessary to have it in a file? Why don't you load the information directly into the table as in the other files?
\pgfplotstableread{DataUsedYear.csv}\tableDataUsedYear
\pgfplotstabletranspose[string type,
    colnames from=Years,
    input colnames to=Years
]\tableDataUsedYearT{\tableDataUsedYear}
%\fi

%Revised May 2021
%\ifloadtables

\pgfplotstableread{AlgorithmMethodologyYear.csv}\tableAlgorithmMethodologyYear
\pgfplotstabletranspose[string type,
    colnames from=Years,
    input colnames to=Years
]\tableAlgorithmMethodologyYearT{\tableAlgorithmMethodologyYear}
%\fi

%Revised May 2021
%\ifloadtables

\pgfplotstableread{EvMethodologyYear.csv}\tableEvMethodologyYear
\pgfplotstabletranspose[string type,
    colnames from=Years,
    input colnames to=Years
]\tableEvMethodologyYearT{\tableEvMethodologyYear}
%\fi

\ifloadtables
\begin{figure}[tb]
\begin{minipage}[t]{0.49\textwidth}
\begin{center}
\begin{tikzpicture}
\begin{axis}[
	xlabel=Years,
	ylabel=Number of Papers,
	width  = 1\textwidth,
	%width  = 0.5\columnwidth,
	height  = 0.2\textheight,
	font=\footnotesize,
	title = {Information usage evolution per year},
	%xminorgrids, xmajorgrids, minor x tick num=1, %<- Vertical grid
	%yminorgrids, ymajorgrids, minor y tick num=1, %<- Horizontal grid
	%xmin = 5, xmax = 100,
	%ymin = 0.10, ymax = 0.22,
	%xtick={5,20,40,60,80,100},
	%y tick label style={
	%					/pgf/number format/.cd,
	%							fixed,
	%							fixed zerofill,
	%							precision=2,
	%					/tikz/.cd
	%			},
	x tick label style={
						/pgf/number format/.cd,%
						scaled x ticks = false,
          				set thousands separator={},
	   					fixed,
	},%
	label style={font=\footnotesize},
	ticklabel style = {font=\footnotesize},
	legend style={at={(0.5, 0.1)}, anchor=north, legend columns = 3, font=\footnotesize},
	legend to name=namedInformationYear,
]
\addplot table [x={Years}, y={Geographical}] {\tableDataUsedYearT};
\addlegendentry{Geographical};
\addplot table [x={Years}, y={Social}] {\tableDataUsedYearT};
\addlegendentry{Social};
\addplot[brown, mark=triangle*] table [x={Years}, y={Content}] {\tableDataUsedYearT};
\addlegendentry{Content};
\addplot table [x={Years}, y={Textual}] {\tableDataUsedYearT};
\addlegendentry{Textual};
\addplot table [x={Years}, y={Sequential}] {\tableDataUsedYearT};
\addlegendentry{Sequential};
\addplot table [x={Years}, y={Temporal}] {\tableDataUsedYearT};
\addlegendentry{Temporal};
\end{axis}
\end{tikzpicture}
\end{center}
\centering
\hspace*{0.5cm}
\ref{namedInformationYear}
\caption{Number of papers using different information sources by year.}
\label{fig:evol_info}
\end{minipage}
\hfill
\begin{minipage}[t]{0.49\textwidth}
\begin{center}
\begin{tikzpicture}
\begin{axis}[
	xlabel=Years,
	ylabel=Number of Papers,
	width  = 1\textwidth,
	%width  = 0.5\columnwidth,
	height  = 0.2\textheight,
	font=\footnotesize,
	title = {Evaluation methodology evolution per year},
	%xminorgrids, xmajorgrids, minor x tick num=1, %<- Vertical grid
	%yminorgrids, ymajorgrids, minor y tick num=1, %<- Horizontal grid
	%xmin = 5, xmax = 100,
	%ymin = 0.10, ymax = 0.22,
	%xtick={5,20,40,60,80,100},
	%y tick label style={
	%					/pgf/number format/.cd,
	%							fixed,
	%							fixed zerofill,
	%							precision=2,
	%					/tikz/.cd
	%			},
	x tick label style={
						/pgf/number format/.cd,%
						scaled x ticks = false,
          				set thousands separator={},
	   					fixed,
	},%
	label style={font=\footnotesize},
	ticklabel style = {font=\footnotesize},
	legend style={at={(0.5,-0.1)}, anchor=north, legend columns = 2, font=\footnotesize},
	legend to name=namedEvaluationYear,
]
\addplot table [x={Years}, y={Temporal}] {\tableEvMethodologyYearT};
\addlegendentry{Temporal};
\addplot table [x={Years}, y={Random}] {\tableEvMethodologyYearT};
\addlegendentry{Random};
\addplot[brown, mark=triangle*] table [x={Years}, y={Other}] {\tableEvMethodologyYearT};
\addlegendentry{Other};
\end{axis}
\end{tikzpicture}
\end{center}
\centering
\hspace*{0.5cm}
\ref{namedEvaluationYear}
\caption{Number of papers using different evaluation methodologies by year.}
\label{fig:evol_eval}
\end{minipage}
\end{figure}
\else
\begin{figure}[tb]
\begin{minipage}[t]{0.49\textwidth}
\begin{center}
	\includegraphics[]{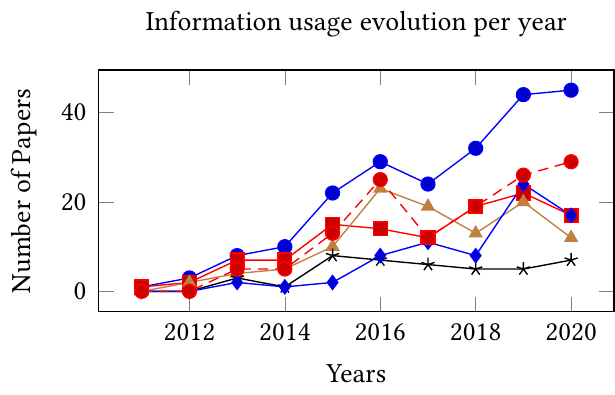}
\end{center}
\centering
\hspace*{0.5cm}
\includegraphics[]{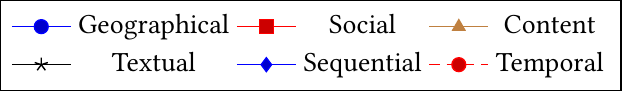}
\caption{Number of papers using different information sources by year.}
\label{fig:evol_info}
\end{minipage}
\hfill
\begin{minipage}[t]{0.49\textwidth}
\begin{center}
	\includegraphics[]{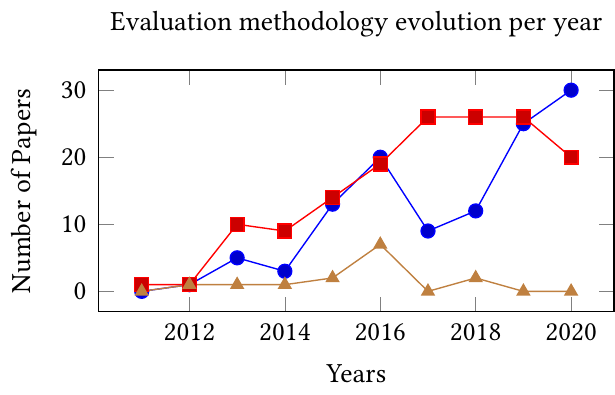}
\end{center}
\centering
\hspace*{0.5cm}
\includegraphics[]{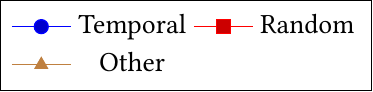}
\caption{Number of papers using different evaluation methodologies by year.}
\label{fig:evol_eval}
\end{minipage}
\end{figure}
\fi

\ifloadtables
\begin{figure}[tb]
%\begin{minipage}[b]{0.8\textwidth}
%\begin{minipage}[t][0.25\textheight][t]{\textwidth}
\begin{minipage}[b]{\textwidth}
%\begin{center}
\begin{tikzpicture}
\begin{axis}[
	xlabel=Years,
	ylabel=Number of Papers,
	width  = 1\textwidth,
	height  = 0.2\textheight,
	%width  = 0.5\columnwidth,
	font=\footnotesize,
	title = {Algorithm methodology evolution per year},
	%xminorgrids, xmajorgrids, minor x tick num=1, %<- Vertical grid
	%yminorgrids, ymajorgrids, minor y tick num=1, %<- Horizontal grid
	%xmin = 5, xmax = 100,
	%ymin = 0.10, ymax = 0.22,
	%xtick={5,20,40,60,80,100},
	%y tick label style={
	%					/pgf/number format/.cd,
	%							fixed,
	%							fixed zerofill,
	%							precision=2,
	%					/tikz/.cd
	%			},
	x tick label style={
						/pgf/number format/.cd,%
						scaled x ticks = false,
          				set thousands separator={},
	   					fixed,
	},%
	label style={font=\footnotesize},
	ticklabel style = {font=\footnotesize},
	legend style={at={(0.5,-0.1)}, anchor=north, legend columns = 4, font=\footnotesize},
	legend to name=namedMethodologyYear,
]
\addplot table [x={Years}, y={SimCF}] {\tableAlgorithmMethodologyYearT};
\addlegendentry{\SIMILARITY};
\addplot table [x={Years}, y={Factorization}] {\tableAlgorithmMethodologyYearT};
\addlegendentry{Factorization};
\addplot[brown, mark=triangle*] table [x={Years}, y={Probabilistic}] {\tableAlgorithmMethodologyYearT};
\addlegendentry{Probabilistic};
\addplot table [x={Years}, y={DeepLearning}] {\tableAlgorithmMethodologyYearT};
\addlegendentry{DL};
\addplot table [x={Years}, y={SocialGraph}] {\tableAlgorithmMethodologyYearT};
\addlegendentry{\GraphLink};
\addplot table [x={Years}, y={Hybrid}] {\tableAlgorithmMethodologyYearT};
\addlegendentry{Hybrid};
\addplot table [x={Years}, y={Other}] {\tableAlgorithmMethodologyYearT};
\addlegendentry{Other};
\end{axis}
\end{tikzpicture}
%\end{center}
\centering
\hspace*{1cm}
\ref{namedMethodologyYear}
\caption{Number of papers using different algorithms by year.}
\label{fig:evol_algo}
\end{minipage}
\end{figure}
\else
\begin{figure}[tb]
%\begin{minipage}[b]{0.8\textwidth}
%\begin{minipage}[t][0.25\textheight][t]{\textwidth}
\begin{minipage}[b]{\textwidth}
\begin{center}
	\includegraphics[]{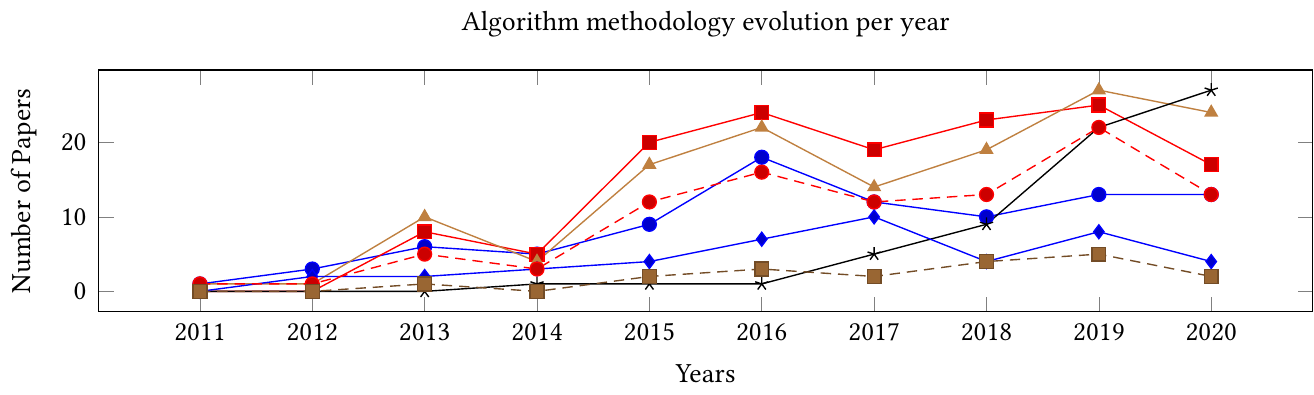}
\end{center}
\centering
\hspace*{1cm}
\includegraphics[]{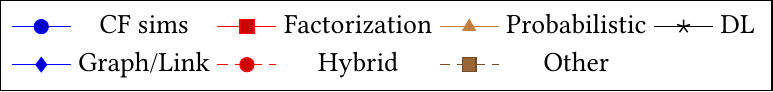}
\caption{Number of papers using different algorithms by year.}
\label{fig:evol_algo}
\end{minipage}
\end{figure}
\fi

%%%%%%%%%%%%%%
%Revised 8 of july 2020

\pgfplotstableread[col sep=semicolon]{
Year;Reference;Acronym;FilteringUsersItems;RegionSplit;Error;Ranking;Metrics;Baselines;ClassicNPBaselines;ClassicPBaselines;GeoBaselines;UseValidation;SplitCheckPOIs;ColdStart;RandomS;nfoldS;TemporalS;OtherS;CC;Fix
2011;\cite{DBLP:conf/sigir/YeYLL11};{\USG};;;;{\cmark};{\PRECISION, \RECALL};{\MODS, \RANDOMWALK};;{\cmark};{\cmark};;{\xmark};{\cmark};{\cmark};;;;;{\cmark}
2012;\cite{DBLP:conf/icde/LevandoskiSEM12};{\LARS};;{\cmark};;;Storage, Locality loss;{\MODS, \IB};;{\cmark};{\cmark};;{\xmark};;{\cmark};;;;{\cmark};
2012;\cite{DBLP:conf/gis/0003ZM12};{\NA};{\cmark};{\cmark};;{\cmark};{\PRECISION, \RECALL};MPC, LCF, PCF;;{\cmark};;;{\xmark};;;;;{\cmark};;
2012;\cite{DBLP:conf/kdd/YingLKT12};{\UPOIMINE};;;{\cmark};{\cmark};{\NDCG, \MAE};{TrustWalker, \USG};;{\cmark};{\cmark};;{\xmark};;;;;;;
2012;\cite{DBLP:conf/socialcom/NoulasSLM12};RW, Weighted-RW;;{\cmark};;{\cmark};{\PRECISION, \RECALL, APR};{\POPULARITY, \RANDOM, \UB, \MF,...};{\cmark};{\cmark};{\cmark};;{\cmark};;;;{\cmark};;{\cmark};
2013;\cite{DBLP:conf/cikm/LiuLAM13};{\NA};;{\cmark};;{\cmark};{\PRECISION, \RECALL};{MF, GeoCF, MGMMF, Markov, MF};;{\cmark};{\cmark};;{\cmark};;;;{\cmark};;;{\cmark}
2013;\cite{DBLP:conf/kdd/LiuFYX13};{\GTBNMF};;{\cmark};;{\cmark};{\PRECISION, \RECALL, r\PRECISION, r\RECALL};SVD, PMF, NMF, BNMF;;{\cmark};;;{\cmark};;{\cmark};;;;{\cmark};
2013;\cite{DBLP:conf/ijcai/ChengYLK13};{\FPMCLR};{\cmark};;;{\cmark};{\PRECISION, \RECALL};PMF, PTF, FPMC;;;;;{\cmark};;;;{\cmark};;{\cmark};
2013;\cite{DBLP:conf/recsys/GaoTHL13};{\LRT};{\cmark};;;{\cmark};{\PRECISION, \RECALL};{\USG, R-LRT, NMF};;{\cmark};;;{\xmark};;{\cmark};;;;;{\cmark}
2013;\cite{DBLP:conf/sigir/YuanCMSM13};{\UTESE};{\cmark};{\cmark};;{\cmark};{\PRECISION, \RECALL};{\UB, UBT, \MODS};;{\cmark};{\cmark};{\cmark};{\xmark};;{\cmark};;;;;{\cmark}
2014;\cite{DBLP:journals/tist/YingKTL14};{\UPOIWALK};;{\cmark};{\cmark};{\cmark};{\NDCG, \MAE};{\USG, HITS, TrustWalker};;{\cmark};{\cmark};;{\qmark};;;;;;;
2014;\cite{DBLP:conf/cikm/YuanCS14};{\GTAG};{\cmark};{\cmark};;{\cmark};{\PRECISION, \RECALL};{\UB, \UTESE, \LRT...};;{\cmark};{\cmark};{\cmark};{\xmark};;{\cmark};;;;;{\cmark}
2014;\cite{DBLP:conf/kdd/LianZXSCR14};{\GEOMF};{\cmark};;;{\cmark};{\PRECISION, \RECALL};{\UB, MF};;{\cmark};;;{\xmark};;{\cmark};;;;;{\cmark}
2014;\cite{DBLP:conf/cikm/LiuWSM14};{\IRENMF};;{\cmark};;{\cmark};{\PRECISION, \RECALL};{\MODS, \UB, \IB, BPRMF, ...};;{\cmark};{\cmark};{\cmark};{\xmark};;{\cmark};;;;;{\cmark}
2014;\cite{DBLP:conf/gis/ZhangCL14};{\LORE};;;;{\cmark};{\PRECISION, \RECALL};{MC, \IGSLR, GS2D};;;{\cmark};;{\cmark};;;;{\cmark};;{\cmark};
2015;\cite{DBLP:journals/tkde/LiuXPFY15};Poisson Geo-PFM;{\cmark};{\cmark};;{\cmark};;;;{\cmark};{\cmark};;{\cmark};;{\cmark};;;;{\cmark};
2015;\cite{DBLP:conf/sigir/LiCLPK15};{\RANKGEOFM};;{\cmark};;{\cmark};{\PRECISION, \RECALL};{\UB, BPRMF, \GEOMF, PMF,...};;{\cmark};{\cmark};{\cmark};{\cmark};;;;{\cmark};;;{\cmark}
2015;\cite{DBLP:conf/sigir/ZhangC15};{\GEOSOCA};;{\cmark};;{\cmark};{\PRECISION, \RECALL};{\USG, \CORE, DRW,...};;;{\cmark};;{\cmark};;;;{\cmark};;{\cmark};
2015;\cite{DBLP:conf/ijcai/FengLZCCY15};{\PRMEG};{\cmark};{\cmark};;{\cmark};{\PRECISION, \RECALL};{\POPULARITY, \UB, MC, MF,...};{\cmark};{\cmark};;{\cmark};{\cmark};;;;{\cmark};;{\cmark};
2015;\cite{DBLP:conf/aaai/GaoTHL15};{\CAPRF};{\cmark};{\cmark};;{\cmark};{\PRECISION, \RECALL};{\UB, MF, STLR, SELR};;{\cmark};;;{\xmark};;{\cmark};;;;;{\cmark}
2016;\cite{DBLP:conf/cikm/XieYWXCW16};{\GE};;{\cmark};;{\cmark};{Accuracy};{SVD, JIM, PRME-G, Geo-SAGE};;;{\cmark};{\cmark};{\cmark};{\cmark};;;{\cmark};;;{\cmark}
2016;\cite{DBLP:conf/kdd/LiGHZ16};{\ASMF};{\cmark};{\cmark};;{\cmark};{\PRECISION, \RECALL, MAP};{\USG, \IRENMF, MF, BPRMF,...};;{\cmark};{\cmark};;{\xmark};{\cmark};;;{\cmark};;;{\cmark}
2016;\cite{DBLP:conf/aaai/ZhaoZYLK16};{\STELLAR};{\cmark};;;{\cmark};{\PRECISION, \RECALL};{BPRMF, \LRT, FPMC, WRMF,...};;{\cmark};;;{\cmark};;;;{\cmark};;{\cmark};
2016;\cite{DBLP:conf/aaai/HeLLSC16};{\NA};{\cmark};{\cmark};;{\cmark};{\PRECISION};{MF, PMF,FPMC-LR};;{\cmark};;;{\qmark};;;;;;;
2016;\cite{DBLP:conf/kdd/LiuLLQX16};{\WWO};{\cmark};;;{\cmark};{\NDCG, F};{PMF, FPMC, PIMF};;;;;{\cmark};;;;{\cmark};;{\cmark};
2017;\cite{DBLP:conf/www/ZhaoZKL17};{\GEOTEASER};{\cmark};;;{\cmark};{\PRECISION, \RECALL};{BPRMF, \LRT, \LORE, \RANKGEOFM,...};;{\cmark};{\cmark};;{\cmark};;;;{\cmark};;;{\cmark}
2017;\cite{DBLP:conf/kdd/YangBZY017};{\PACE};{\cmark};;;{\cmark};{\PRECISION, \RECALL, NDCG, MAP};{\IRENMF, \USG, \LORE, \ASMF,...};;;{\cmark};;{\cmark};;;;{\cmark};;;{\cmark}
2017;\cite{DBLP:journals/ijon/RenSES17};{\TGSCPMF};{\cmark};;{\cmark};{\cmark};{\PRECISION, \RECALL, \MAE, \RMSE};{\USG, \ASMF, LCARS,...};;;{\cmark};;{\cmark};;{\cmark};;;;{\cmark};
2017;\cite{DBLP:conf/ijcai/HeLL17};{\LBPR};;{\cmark};;{\cmark};{\PRECISION};{MF, PMF, FPMC,...};;{\cmark};{\cmark};;{\cmark};;;;{\cmark};;;{\cmark}
2017;\cite{DBLP:conf/www/WangWTSRL17};{\VPOI};{\cmark};{\cmark};;{\cmark};{\PRECISION, \RECALL};{\UB, NMF, VBPR, ...};;{\cmark};;;{\xmark};{\cmark};{\cmark};;;;;{\cmark}
2018;\cite{DBLP:conf/cikm/MaZWL18};{\SAENAD};{\cmark};{\cmark};;{\cmark};{\PRECISION, \RECALL, \MAP};{WRMF, BPRMF, MGMMF, IrenMF, RankGeoFM, PACE, DeepAE};;{\cmark};{\cmark};;{\xmark};;{\cmark};;;;;{\cmark}
2018;\cite{DBLP:conf/sigir/ManotumruksaMO18};{\CARA};{\cmark};;;{\cmark};{\NDCG, HR};{\MF, BPR, RNN, \STELLAR,...};;{\cmark};{\cmark};;{\cmark};{\cmark};;;{\cmark};;;{\cmark}
2018;\cite{DBLP:journals/toit/YaoSWZQ17};{\TENMF};{\cmark};;;{\cmark};{\PRECISION, \RECALL};{NMF, UCF, ICF, FA, GA, LIM};;{\cmark};{\cmark};;{\xmark};;{\cmark};;;;;{\cmark}
2018;\cite{DBLP:journals/ijon/GaoLLSZ18};{\GEOEISO};{\cmark};{\cmark};;{\cmark};{\PRECISION, \RECALL};{\USG, \IRENMF, \IGSLR,...};;{\cmark};{\cmark};;{\cmark};;{\cmark};;;;{\cmark};
2018;\cite{DBLP:conf/ijcai/WangSOC18};{\GEOIE};{\cmark};;;{\cmark};{\PRECISION, \RECALL};{\GEOMF, \RANKGEOFM, \GEOTEASER,...};;;{\cmark};{\cmark};{\cmark};;;;{\cmark};;;{\cmark}
2019;\cite{DBLP:journals/www/YingWXLLZX19};{\MEAPT};{\cmark};{\cmark};;{\cmark};{\PRECISION, \RECALL};{Popular, BPR, FPMC, PRME};{\cmark};{\cmark};;{\cmark};{\cmark};;;;{\cmark};;;{\cmark}
2019;\cite{DBLP:journals/isci/GengJGLW19};{\MLR};;{\cmark};;{\cmark};{\PRECISION, \RECALL, F};{\UB, \IGSLR, KDE,...};;{\cmark};{\cmark};;{\xmark};;{\cmark};;;;{\cmark};
2019;\cite{DBLP:journals/kbs/SiZL19};{\APRASA};;;{\cmark};{\cmark};{\PRECISION, \RECALL, F, \MAE};{\UB, \UTESE, \GTAG, \CTFARA,...};;{\cmark};{\cmark};;{\cmark};;{\cmark};;;;;{\cmark}
2019;\cite{DBLP:journals/tois/QianLNY19};{\STA};;;;{\cmark};{\NDCG, \RECALL};{\USG, \GEOMF, \RANKGEOFM, \GE,...};;;{\cmark};{\cmark};{\cmark};{\cmark};;;{\cmark};;;{\cmark}
2019;\cite{DBLP:conf/aaai/ZhaoZLXLZSZ19};{\STGN};{\cmark};{\cmark};;{\cmark};;;;;{\cmark};;{\cmark};{\cmark};;;{\cmark};;;{\cmark}
2020;\cite{DBLP:journals/ijon/XiongQHXBLYY20};{\HILDA};;{\cmark};;{\cmark};;;;;{\cmark};;{\cmark};;{\cmark};{\cmark};;;{\cmark};
2020;\cite{DBLP:journals/iotj/WangCWCG20};{\GAIMC};{\cmark};{\cmark};;{\cmark};;;;;{\cmark};;{\cmark};;{\cmark};;;;{\cmark};
2020;\cite{DBLP:journals/kbs/ZhaoLQH20};{\SPR};;{\cmark};{\cmark};;;;;{\cmark};{\cmark};;{\cmark};;{\cmark};{\cmark};;;{\cmark};
2020;\cite{DBLP:journals/fgcs/HuangMLS20};{\MMBE};{\cmark};;;{\cmark};;;;;{\cmark};;{\cmark};;;;{\cmark};;{\cmark};
2020;\cite{DBLP:journals/tii/WangCWCLG20};{\TECF};{\cmark};{\cmark};;{\cmark};;;;{\cmark};{\cmark};;{\xmark};;{\cmark};;;;;{\cmark}
}\examplesTableRepresentativesEvaluation

\begin{table*}
\caption{Evaluation details of analyzed POI recommendation approaches sorted by publication year.
}
\label{tab:examplesTableRepresentativesEvaluation}
\vspace{-0.25cm}
\ifloadtables
\pgfplotstabletypeset[
		font=\fontsize{7pt}{7.5pt}\selectfont,
		outfile=tab_examplesTableRepEval.tex,
	every head row/.style={
            before row={%
                \hline
                \multicolumn{3}{? c ? }{Details} & \multicolumn{7}{c?}{Evaluation configuration} & \multicolumn{3}{c?}{Baselines} & \multicolumn{3}{c?}{Split type} & \multicolumn{2}{c?}{Split level}\\
                \hline
            },
            after row={
            	\hline
            },
            typeset cell/.code={
            \ifnum\pgfplotstablecol=\pgfplotstablecols
            	\pgfkeyssetvalue{/pgfplots/table/@cell content}{\rotatebox{90}{##1}\\}%
            \else
            	\pgfkeyssetvalue{/pgfplots/table/@cell content}{\rotatebox{90}{##1}&}%
            \fi
            },
        },
	   every last row/.style={
		after row={%
                \hline
		\multicolumn{3}{? c ? }{Most Representatives} & 27 & 9 & 5 & 43 & 28 & C:27 P:16 & 7 & 3 & 32 & 33 & 21 & 20 & 1 & 17 & 24\\
		\hline
		\multicolumn{3}{? c ? }{Total} & 171 & 53 & 25 & 294 & 144 & C:205 P:71 & 38 & 35 & 173 & 198 & 152 & 118 & 14 & 135 & 135\\
		\hline
		},
	},
        column type=c,
        %column type/.add={|}{},
        every even row/.style={
        	before row={
        		\hline
        		\rowcolor{lightgray!50}},
        	after row={\hline}
        },
        %%%%%%%%%%%%%%%%%%%%%%%%%%%%%%%%%%%%%%%%%%%%%
        % Year,Reference,Acronym
        columns/Year/.style={
			string type,
			column type/.add={?}{|},
        },
		columns/Reference/.style={
			string type,
			column type/.add={}{|},
        },
        columns/Acronym/.style={
			string type,
			column type/.add={}{?},
        },
        %%%%%%%%%%%%%%%%%
        % Evaluation configuration: FilteringUsersItems;RegionSplit;Error;Baselines;ClassicNPBaselines;ClassicPBaselines;GeoBaselines;UseValidation;SplitCheckPOIs;ColdStart
        columns/FilteringUsersItems/.style={
			string type,
			column name={Filter data},
			column type/.add={}{|},
        },
        columns/RegionSplit/.style={
			string type,
			column name={Region Split},
			column type/.add={}{|},
        },
        columns/Error/.style={
			string type,
			column type/.add={}{|},
        },
 	columns/Ranking/.style={
			string type,
			column type/.add={}{|},
        },
 	columns/Metrics/.style={
			string type,
			column name={Metrics},
			column type/.add={}{|},
        },
        columns/Baselines/.style={
			string type,
			column type/.add={}{|},
        },
	columns/ClassicNPBaselines/.style={
			string type,
			column name={Classic Non Personalized}, %Basically Popularity and/or random
			column type/.add={}{|},
        },
	columns/ClassicPBaselines/.style={
			string type,
			column name={Classic Personalized}, %Basically UB, IB, BPR, MF... (MC, LDA is not considered as personalized as it is not standard)
			column type/.add={}{|},
        },
	columns/GeoBaselines/.style={
			string type,
			column name={Geographical}, %Any modification or algorithm that exploit distance or geographical
			column type/.add={}{?},
        },
        columns/UseValidation/.style={
			string type,
			column name={Validation},
			column type/.add={}{|},
        },
        columns/SplitCheckPOIs/.style={
			string type,
			column name={Check-in(\cmark) or POI(\xmark) split},
			column type/.add={}{|},
        },
        columns/ColdStart/.style={
			string type,
			column name={Cold Start Analysis},
			column type/.add={}{?},
        },
 	%%%%%%%%%%%%%%%%%%
        %% Type of split: TCC,TFix,RCC,RFix,nfold,Other
        columns/RandomS/.style={
			string type,
			column name={Random},
			column type/.add={}{|},
        },
        columns/nfoldS/.style={
			string type,
			column name={\nfold},
			column type/.add={}{|},
        },
        columns/TemporalS/.style={
			string type,
			column name={Temporal},
			column type/.add={}{|},
        },
        columns/OtherS/.style={
			string type,
			column name={Other},
			column type/.add={}{?},
        },
        columns/CC/.style={
			string type,
			column name={\CC},
			column type/.add={}{|},
        },
        columns/Fix/.style={
			string type,
			column name={\Fix},
			column type/.add={}{?},
        },
        fixed,
        fixed zerofill,
        columns={Year,Reference,Acronym,FilteringUsersItems,UseValidation,Error,Ranking,RegionSplit,SplitCheckPOIs,ColdStart,ClassicNPBaselines,ClassicPBaselines,GeoBaselines,RandomS,TemporalS,OtherS,CC,Fix},
    ]{\examplesTableRepresentativesEvaluation}
\else
	\begingroup \fontsize {7pt}{7.5pt}\selectfont %
\begin {tabular}{?c|c|c?c|c|c|c|c|c|c?c|c|c?c|c|c?c|c?}%
\hline \multicolumn {3}{? c ? }{Details} & \multicolumn {7}{c?}{Evaluation configuration} & \multicolumn {3}{c?}{Baselines} & \multicolumn {3}{c?}{Split type} & \multicolumn {2}{c?}{Split level}\\ \hline \rotatebox {90}{Year}&\rotatebox {90}{Reference}&\rotatebox {90}{Acronym}&\rotatebox {90}{Filter data}&\rotatebox {90}{Validation}&\rotatebox {90}{Error}&\rotatebox {90}{Ranking}&\rotatebox {90}{Region Split}&\rotatebox {90}{Check-in(\cmark ) or POI(\xmark ) split}&\rotatebox {90}{Cold Start Analysis}&\rotatebox {90}{Classic Non Personalized}&\rotatebox {90}{Classic Personalized}&\rotatebox {90}{Geographical}&\rotatebox {90}{Random}&\rotatebox {90}{Temporal}&\rotatebox {90}{Other}&\rotatebox {90}{\CC }&\rotatebox {90}{\Fix }\\\hline %
\hline \rowcolor {lightgray!50}2011&\cite {DBLP:conf/sigir/YeYLL11}&\USG &&&&\cmark &&\xmark &\cmark &&\cmark &\cmark &\cmark &&&&\cmark \\\hline %
2012&\cite {DBLP:conf/icde/LevandoskiSEM12}&\LARS &&&&&\cmark &\xmark &&&\cmark &\cmark &\cmark &&&\cmark &\\%
\hline \rowcolor {lightgray!50}2012&\cite {DBLP:conf/gis/0003ZM12}&\NA &\cmark &&&\cmark &\cmark &\xmark &&&\cmark &&&&\cmark &&\\\hline %
2012&\cite {DBLP:conf/kdd/YingLKT12}&\UPOIMINE &&&\cmark &\cmark &&\xmark &&&\cmark &\cmark &&&&&\\%
\hline \rowcolor {lightgray!50}2012&\cite {DBLP:conf/socialcom/NoulasSLM12}&RW, Weighted-RW&&&&\cmark &\cmark &\cmark &&\cmark &\cmark &\cmark &&\cmark &&\cmark &\\\hline %
2013&\cite {DBLP:conf/cikm/LiuLAM13}&\NA &&&&\cmark &\cmark &\cmark &&&\cmark &\cmark &&\cmark &&&\cmark \\%
\hline \rowcolor {lightgray!50}2013&\cite {DBLP:conf/kdd/LiuFYX13}&\GTBNMF &&&&\cmark &\cmark &\cmark &&&\cmark &&\cmark &&&\cmark &\\\hline %
2013&\cite {DBLP:conf/ijcai/ChengYLK13}&\FPMCLR &\cmark &&&\cmark &&\cmark &&&&&&\cmark &&\cmark &\\%
\hline \rowcolor {lightgray!50}2013&\cite {DBLP:conf/recsys/GaoTHL13}&\LRT &\cmark &&&\cmark &&\xmark &&&\cmark &&\cmark &&&&\cmark \\\hline %
2013&\cite {DBLP:conf/sigir/YuanCMSM13}&\UTESE &\cmark &\cmark &&\cmark &\cmark &\xmark &&&\cmark &\cmark &\cmark &&&&\cmark \\%
\hline \rowcolor {lightgray!50}2014&\cite {DBLP:journals/tist/YingKTL14}&\UPOIWALK &&&\cmark &\cmark &\cmark &\qmark &&&\cmark &\cmark &&&&&\\\hline %
2014&\cite {DBLP:conf/cikm/YuanCS14}&\GTAG &\cmark &\cmark &&\cmark &\cmark &\xmark &&&\cmark &\cmark &\cmark &&&&\cmark \\%
\hline \rowcolor {lightgray!50}2014&\cite {DBLP:conf/kdd/LianZXSCR14}&\GEOMF &\cmark &&&\cmark &&\xmark &&&\cmark &&\cmark &&&&\cmark \\\hline %
2014&\cite {DBLP:conf/cikm/LiuWSM14}&\IRENMF &&\cmark &&\cmark &\cmark &\xmark &&&\cmark &\cmark &\cmark &&&&\cmark \\%
\hline \rowcolor {lightgray!50}2014&\cite {DBLP:conf/gis/ZhangCL14}&\LORE &&&&\cmark &&\cmark &&&&\cmark &&\cmark &&\cmark &\\\hline %
2015&\cite {DBLP:journals/tkde/LiuXPFY15}&Poisson Geo-PFM&\cmark &&&\cmark &\cmark &\cmark &&&\cmark &\cmark &\cmark &&&\cmark &\\%
\hline \rowcolor {lightgray!50}2015&\cite {DBLP:conf/sigir/LiCLPK15}&\RANKGEOFM &&\cmark &&\cmark &\cmark &\cmark &&&\cmark &\cmark &&\cmark &&&\cmark \\\hline %
2015&\cite {DBLP:conf/sigir/ZhangC15}&\GEOSOCA &&&&\cmark &\cmark &\cmark &&&&\cmark &&\cmark &&\cmark &\\%
\hline \rowcolor {lightgray!50}2015&\cite {DBLP:conf/ijcai/FengLZCCY15}&\PRMEG &\cmark &\cmark &&\cmark &\cmark &\cmark &&\cmark &\cmark &&&\cmark &&\cmark &\\\hline %
2015&\cite {DBLP:conf/aaai/GaoTHL15}&\CAPRF &\cmark &&&\cmark &\cmark &\xmark &&&\cmark &&\cmark &&&&\cmark \\%
\hline \rowcolor {lightgray!50}2016&\cite {DBLP:conf/cikm/XieYWXCW16}&\GE &&\cmark &&\cmark &\cmark &\cmark &\cmark &&&\cmark &&\cmark &&&\cmark \\\hline %
2016&\cite {DBLP:conf/kdd/LiGHZ16}&\ASMF &\cmark &&&\cmark &\cmark &\xmark &\cmark &&\cmark &\cmark &&\cmark &&&\cmark \\%
\hline \rowcolor {lightgray!50}2016&\cite {DBLP:conf/aaai/ZhaoZYLK16}&\STELLAR &\cmark &&&\cmark &&\cmark &&&\cmark &&&\cmark &&\cmark &\\\hline %
2016&\cite {DBLP:conf/aaai/HeLLSC16}&\NA &\cmark &&&\cmark &\cmark &\qmark &&&\cmark &&&&&&\\%
\hline \rowcolor {lightgray!50}2016&\cite {DBLP:conf/kdd/LiuLLQX16}&\WWO &\cmark &&&\cmark &&\cmark &&&&&&\cmark &&\cmark &\\\hline %
2017&\cite {DBLP:conf/www/ZhaoZKL17}&\GEOTEASER &\cmark &&&\cmark &&\cmark &&&\cmark &\cmark &&\cmark &&&\cmark \\%
\hline \rowcolor {lightgray!50}2017&\cite {DBLP:conf/kdd/YangBZY017}&\PACE &\cmark &&&\cmark &&\cmark &&&&\cmark &&\cmark &&&\cmark \\\hline %
2017&\cite {DBLP:journals/ijon/RenSES17}&\TGSCPMF &\cmark &&\cmark &\cmark &&\cmark &&&&\cmark &\cmark &&&\cmark &\\%
\hline \rowcolor {lightgray!50}2017&\cite {DBLP:conf/ijcai/HeLL17}&\LBPR &&&&\cmark &\cmark &\cmark &&&\cmark &\cmark &&\cmark &&&\cmark \\\hline %
2017&\cite {DBLP:conf/www/WangWTSRL17}&\VPOI &\cmark &&&\cmark &\cmark &\xmark &\cmark &&\cmark &&\cmark &&&&\cmark \\%
\hline \rowcolor {lightgray!50}2018&\cite {DBLP:conf/cikm/MaZWL18}&\SAENAD &\cmark &&&\cmark &\cmark &\xmark &&&\cmark &\cmark &\cmark &&&&\cmark \\\hline %
2018&\cite {DBLP:conf/sigir/ManotumruksaMO18}&\CARA &\cmark &&&\cmark &&\cmark &\cmark &&\cmark &\cmark &&\cmark &&&\cmark \\%
\hline \rowcolor {lightgray!50}2018&\cite {DBLP:journals/toit/YaoSWZQ17}&\TENMF &\cmark &&&\cmark &&\xmark &&&\cmark &\cmark &\cmark &&&&\cmark \\\hline %
2018&\cite {DBLP:journals/ijon/GaoLLSZ18}&\GEOEISO &\cmark &&&\cmark &\cmark &\cmark &&&\cmark &\cmark &\cmark &&&\cmark &\\%
\hline \rowcolor {lightgray!50}2018&\cite {DBLP:conf/ijcai/WangSOC18}&\GEOIE &\cmark &\cmark &&\cmark &&\cmark &&&&\cmark &&\cmark &&&\cmark \\\hline %
2019&\cite {DBLP:journals/www/YingWXLLZX19}&\MEAPT &\cmark &\cmark &&\cmark &\cmark &\cmark &&\cmark &\cmark &&&\cmark &&&\cmark \\%
\hline \rowcolor {lightgray!50}2019&\cite {DBLP:journals/isci/GengJGLW19}&\MLR &&&&\cmark &\cmark &\xmark &&&\cmark &\cmark &\cmark &&&\cmark &\\\hline %
2019&\cite {DBLP:journals/kbs/SiZL19}&\APRASA &&&\cmark &\cmark &&\cmark &&&\cmark &\cmark &\cmark &&&&\cmark \\%
\hline \rowcolor {lightgray!50}2019&\cite {DBLP:journals/tois/QianLNY19}&\STA &&\cmark &&\cmark &&\cmark &\cmark &&&\cmark &&\cmark &&&\cmark \\\hline %
2019&\cite {DBLP:conf/aaai/ZhaoZLXLZSZ19}&\STGN &\cmark &&&\cmark &\cmark &\cmark &\cmark &&&\cmark &&\cmark &&&\cmark \\%
\hline \rowcolor {lightgray!50}2020&\cite {DBLP:journals/ijon/XiongQHXBLYY20}&\HILDA &&&&\cmark &\cmark &\cmark &&&&\cmark &\cmark &&&\cmark &\\\hline %
2020&\cite {DBLP:journals/iotj/WangCWCG20}&\GAIMC &\cmark &&&\cmark &\cmark &\cmark &&&&\cmark &\cmark &&&\cmark &\\%
\hline \rowcolor {lightgray!50}2020&\cite {DBLP:journals/kbs/ZhaoLQH20}&\SPR &&&\cmark &&\cmark &\cmark &&&\cmark &\cmark &\cmark &&&\cmark &\\\hline %
2020&\cite {DBLP:journals/fgcs/HuangMLS20}&\MMBE &\cmark &&&\cmark &&\cmark &&&&\cmark &&\cmark &&\cmark &\\%
\hline \rowcolor {lightgray!50}2020&\cite {DBLP:journals/tii/WangCWCLG20}&\TECF &\cmark &&&\cmark &\cmark &\xmark &&&\cmark &\cmark &\cmark &&&&\cmark \\\hline \multicolumn {3}{? c ? }{Most Representatives} & 27 & 9 & 5 & 43 & 28 & C:27 P:16 & 7 & 3 & 32 & 33 & 21 & 20 & 1 & 17 & 24\\ \hline \multicolumn {3}{? c ? }{Total} & 171 & 53 & 25 & 294 & 144 & C:205 P:71 & 38 & 35 & 173 & 198 & 152 & 118 & 14 & 135 & 135\\ \hline %
\end {tabular}%
\endgroup %
\fi
\end{table*}

%% let's analyse it in the new order: first information source, then models, and leave the evaluation protocols for the next section
Based on this table, let us first analyze the trends on information sources used throughout the years.
We observe that most of the algorithms make use of geographic information in some way (either by calculating distances between POIs, grouping users and POIs in clusters according to regions, modeling movement distributions, etc.).
Most researchers argue that this type of information is critical since users tend to visit POIs close to where they are and this conclusion can be obtained by performing a preliminary analysis of the LBSNs data.
On the other hand, social information is also widely modeled, partly because many of the datasets used, such as Gowalla, also provide the links of friendship between users.
However, while there is more or less consensus on the importance of geographical information, this is not so clear for social information, where some researchers claim it is not so important~\cite{DBLP:conf/cikm/GaoTL12,DBLP:journals/tist/ChengYKL16} while others state it plays an important role~\cite{DBLP:conf/icdm/ChengC13,DBLP:journals/ijon/GaoLLSZ18}.
One possible explanation for this effect is that even though users may share their tastes with friends (from the same or different cities), they may not visit the same POIs, in part because it is common for users to visit the locations closest to their centers of activity (home and work, basically) and most likely they will be different from their friends', even if they are ``similar'' in terms of tastes.

Textual or content information is also exploited by many approaches, especially those using some kind of probabilistic model such as topic modeling or the POI categories.
%It is important to emphasize that each user has particular tastes and although there may be many POIs candidates to be recommended, POI categories (if it is a restaurant, a museum, etc.) can be decisive.
This is because the features of the items (categories) in this domain are very distinctive and may even discriminate between different types of users in a LBSN; for example, a tourist may prefer to visit museums and restaurants, whereas a local may prefer a bar or a shopping center.
Finally, regarding temporal and sequential information, we observe that the latter is not so exploited (although some deep learning techniques make use of sequential information implicitly),
but temporal information is taken into account in many approaches regardless of the technique used by the model under analysis, probably because of its flexibility to be introduced in almost any recommendation technique (usually at the cost of increased sparsity).
The same trend can be found in Figure~\ref{fig:evol_info}, where all the selected papers in the review, not only the representative ones, are shown in a year basis.

Now, when we analyze the type of model, a change in trend can be seen in the years between 2011 and 2015 and subsequent years, since for the former, proposals that used some type of collaborative system based on neighbors were reasonably popular, but not anymore.
However, in subsequent years there has been a greater dominance of probabilistic and factorization proposals.
This is something that already happened in traditional recommendation, where since the Netflix Prize in 2009~\cite{DBLP:journals/sigkdd/BellK07} in which matrix factorization models outperformed other traditional approaches, they have received more attention from the RS research community.
In the same way, deep learning techniques have experienced a significant growth since 2017 in the POI recommendation area.
This becomes evident in Figure~\ref{fig:evol_algo} where, again, all the papers are included.
Here, we observe that until 2017 there are less than $5$ DL techniques included in the selected papers, but this type of model increases steadily year after year \ReviewOne{and in 2020 is the most extended approach}.
Finally, with respect to graph/link-based and hybrid models we observe that, in general, graph models are not widely used, whereas hybrid techniques, even though they are not widely used, they have been used throughout all the years collected in our analysis.
One reasonable assumption for this is that hybrid proposals allow several elements to be combined into one, thus alleviating the possible drawbacks that each of them may show separately.

%% refer to the next section to analyse in detail the evaluation aspects
In the next section, we analyze in detail the evaluation aspects, including the  split types that appear in the already discussed Table~\ref{tab:examplesTableRepresentatives}.

\section{Systematic review of state-of-the-art evaluation methodologies}
\label{s:soa_eval}
In this section, we continue the analysis on the state-of-the-art in POI recommendation presented previously but focusing on the evaluation aspects.
Thus, we analyze the last columns of Table~\ref{tab:examplesTableRepresentatives}, together with Figure~\ref{fig:evol_eval}, which shows the characterization of some evaluation protocols, as shown in Section~\ref{ss:evaluationProtocols}.
We observe they are well distributed between random and temporal splits, although it is interesting to note that until 2014 the random partition predominated over the temporal split. However, from that year onwards the use of temporal splitting has increased steadily.

Nonetheless, there is still no common evaluation protocol to evaluate the performance of POI recommenders; this is an interesting but also a concerning conclusion, since this means that we might be comparing models that try to solve the same problem (POI recommendation) but, at the same time, we are evaluating them in very different ways, which in turn affects the performance of the algorithms.
In particular, we have found surprising combinations regarding this, such as works with models using temporal information in their formulation that were running a random evaluation protocol in their evaluation, like~\cite{DBLP:conf/sigir/YuanCMSM13,DBLP:conf/recsys/GaoTHL13,DBLP:conf/aaai/ZhaoZYLK16,DBLP:conf/cikm/YuanCS14}.

In any case, it should be noted that even if the entire community moves to a common splitting method, there are other aspects of the experimental settings that could affect the final performance of the algorithms and, hence, the validity of the published results.
For example, how the candidate items to be ranked are selected is a well-known source for bias in RS evaluation~\cite{DBLP:journals/ir/BelloginCC17}, and it is not obvious how to compare results when all the items in the system are scored and ranked against strategies where only one item from the test set is ranked together with a random set of POIs, as it has become recently popular among deep learning techniques such as~\cite{DBLP:conf/sigir/ManotumruksaMO18}.

Because of this, in Table~\ref{tab:examplesTableRepresentativesEvaluation} we extend the evaluation aspects to be considered for the same works presented before.
We now include whether some kind of data filtering is performed (to avoid both users and POIs with very few interactions), if a validation subset is used, the type of metric (error or ranking) reported, if the split was used based on geographical information, and if repetitions were considered or not (i.e., if the split was done by \checkins or POIs), as discussed in Section~\ref{ss:evaluationProtocols}.
We also decided to include if the authors performed some kind of cold-start analysis and the types of baselines considered in the experiments: whether they use classic non-customized recommendation baselines (popularity and/or random), classic and personalized baselines (user/item-based, BPR, or MFs), and geographic baselines (any algorithm that uses a geographical component).

Based on this information, we first observe that a relatively large number of articles apply some kind of filter in the data, the most typical one being to remove users or POIs with less than $n$ interactions.
It is important to note that we only put the mark \cmark if the authors specifically state in their paper that they filter the data, so there might be other proposals that use a pre-filtered dataset that do not count in the table; hence, these numbers are probably underestimating this aspect.
Nonetheless, it is true that in some situations it might be necessary to make some pre-filtering of the data, but we must be careful since, if the filtering is too strict, we may end up evaluating the system with very little data, making the obtained results not generalizable.
On the other hand, sometimes, instead of performing a simple training and test splits, researchers obtain a third subset of data to tune the parameters (called validation but different from a k-fold cross-validation). However, as we observe in the table, it is not very common in POI recommendation (as occurs in traditional recommendation).
With respect to the type of metrics reported, there seems to be more consensus, since the vast majority of papers use some kind of ranking-based metric. Besides, most approaches that evaluate using rating prediction also use ranking evaluation, although there are very few approaches that only use rating prediction, like~\cite{DBLP:conf/ht/YangZYW13}.

Regarding the region split column, we believe it is quite important when comparing research works in this domain, since algorithms executed in a worldwide dataset are not comparable against those executed in independent cities, mostly because the geographical influence is indeed affecting the recommenders in a very different way \ReviewOne{(and the obvious correlation between past and future user actions based on this dimension)}, depending on the type of split we are using.
Similarly, depending on whether the split is done by \checkins or by POIs, it might affect the obtained results.
Although this distinction might be subtle, if we analyze this aspect we observe there is a lot of disparity in the works, not leading to any clear conclusion.
Let us consider for example that we select for each user 80\% of their \checkins to train and the rest 20\% to test, as we mentioned before, on many LBSNs there may be repetitions so the test set may be composed by \checkins that appear in the training set; however, if the split is made by POIs, we make sure to remove such repetitions and therefore we would not be recommending POIs that the user has already visited.
At the same time, even though datasets in this domain are very sparse, few researchers perform a specific analysis on cold start, as evidenced from the values shown in the table (we denote as cold start those works that explicitly consider users or POIs with very few interactions, e.g., less than 5).

Now, in terms of the baselines used, although most of the approaches compare against baselines that can be categorized as classic algorithms such as MFs or \KNN, and many others use geographical influence, it is surprising that there are a limited number of works that test their approaches against very basic baselines like popularity, when it has been shown to be quite effective in domains with a high sparsity~\cite{DBLP:journals/ir/BelloginCC17}.

\ReviewOne{Finally, in Table~\ref{tab:numberPapersEvMethdology} we show the articles processed according to the type of evaluation methodology used to evaluate their models (offline, online, and user studies). As we observe in this table, the vast majority of the researchers use an offline evaluation methodology. 
We argue this might be due to how expensive running such experiments become, in particular for those authors who are already exploiting \checkins from LBSNs, so in most cases they do not need (even though it would provide complementary evidence) to check the user feedback directly.} 
\ReviewOne{We would like to highlight, however, that even though we have not found any work in which the proposed models are evaluated in an online environment, we have found some proposals in which the authors claim that their algorithm is an online POI recommender, even though it is only evaluated in an offline scenario. Some examples of this type of works include: \cite{DBLP:conf/icde/LevandoskiSEM12, DBLP:conf/gis/0003ZM12, DBLP:journals/tist/YingKTL14, DBLP:conf/apweb/WangLF14, DBLP:journals/tkdd/YinCCHZ15, DBLP:journals/tois/QianLNY19}.
}
\ReviewTwo{Although this may be surprising, we are not the only researchers to notice that there are almost no online models in POI recommendation, at least using data from LBSNs~\citep{DBLP:series/sbcs/ZhaoLK18}.} %% 

\section{Systematic review of datasets used in state-of-the-art}
\label{s:soa_data}
There are several LBSNs that researchers use to explore the problem of POI recommendation and related tasks, but among all of them, there are four that stand out: Foursquare\footnote{\url{https://foursquare.com}}, Gowalla\footnote{It does not exist anymore since 2012.}, Brightkite\footnote{It was acquired by another social network in 2009 and does not exist anymore since 2012.}, and Yelp\footnote{\url{https://www.yelp.com}}, as it is evidenced by  Table~\ref{tab:numberPapersUsingDataset}, that shows the number of papers reporting data from each LBSN.
Hence, researchers obtained data from these systems and used them for their experiments, even though the same data could be used for different purposes -- i.e., not only the pure POI recommendation task we address here, but also for social or review recommendation.

Besides the differences in the actual recommendation task, which might be more or less obvious when comparing two research works, we noticed remarkable differences in the statistics reported for datasets that (in principle) belong to the same LBSN.
The reason might be obvious: the datasets are obtained and pre-processed differently, however, since there is no \textit{canonical name} for the datasets (as it occurs in other domains, e.g., with the MovieLens or Lastfm datasets), they are indistinctively referred as the name of the corresponding LBSN, which confuse the reader and other researchers into thinking that the same data is used in two works.
\ReviewOne{We would also like to mention that in some works we have found strange statistics that we believe to be inaccurate. For example, in~\cite{DBLP:journals/access/CaoGMLLL20}, the authors claim to use a Foursquare dataset but the same statistics can be found in~\cite{DBLP:journals/complexity/GuoJLX19, Song2019} for a Gowalla dataset. The statistics reported in~\cite{DBLP:conf/apweb/WangLF14, DBLP:journals/wpc/RaviS17} are also strange as they report more users than check-ins (in this case, for a Foursquare dataset).}
To shed some light on this aspect, we now present some details about the most important datasets based on these LBSNs and later analyze some of these differences.

\begin{itemize}
\item Foursquare: it is possibly the most famous LBSN, which agrees with our statistics (see Table~\ref{tab:numberPapersUsingDataset}) that show it is the LBSN most frequently used by researchers, among the works included in our analysis.
Users in Foursquare can visit a place, mark it as visited in the system (by \textit{checking-in} in the venue) so their followers or friends could track it, like a venue, comment on it (by writing \textit{tips}), and obtain recommendations from the system
(since 2014 most of this functionality was derived to Swarm).
In general, these \checkins cannot be obtained directly neither from its website nor its API; because of that, most researchers rely on other social networks where users share their interactions with Foursquare, mostly Twitter.
Even though we will show different datasets from this LBSN in Tables~\ref{tab:EvMethodologyMostCited} and~\ref{tab:versionsFoursquare}, we consider important to emphasize that many papers that report using Foursquare include a url\footnote{\url{http://www.public.asu.edu/~hgao16/dataset.html}} that does not work anymore; these include the original work~\cite{DBLP:conf/cikm/GaoTL12} and many more, such as~\cite{DBLP:conf/recsys/GaoTHL13, DBLP:conf/gis/ZhangCL14, DBLP:journals/tsc/ZhangCL15, DBLP:journals/tcss/StepanMDM16, DBLP:conf/dasfaa/HosseiniL16}.

\item Gowalla: a LBSN that was acquired by Facebook in 2011.
Most papers use the Gowalla dataset that can be found in the SNAP repository\footnote{\url{http://snap.stanford.edu/data/loc-gowalla.html}}, such as~\cite{DBLP:conf/gis/WangTM13, DBLP:conf/webi/GuoHT15, DBLP:conf/dasfaa/ChenLLLX16}.
As it also happens with other LBSNs, some researchers claim they use Gowalla, but they fail to provide any source to obtain such dataset~\cite{DBLP:conf/kdd/YingLKT12, DBLP:conf/socialcom/NoulasSLM12, DBLP:conf/dsaa/ZhouW14, DBLP:conf/ijcai/LiGLL17}.

\item Brightkite: a less popular LBSN, but used in a large number of research works because of its availability.
In the same way as Gowalla, a dataset from this LBSN is included in the SNAP repository\footnote{\url{http://snap.stanford.edu/data/loc-brightkite.html}}, which makes it easy to be used by researchers, since it is not available since 2012; some examples include~\cite{DBLP:conf/dasfaa/HosseiniL16, DBLP:conf/dasfaa/ChenLLLX16, DBLP:conf/sigir/YaoSQWSH15}.

\item Yelp: this is a LBSN that focuses on businesses, rather than generic POIs like other LBSNs. It also differs from the other LBSNs in that users provide a rating based on 5 stars to the different businesses they visit; besides, users can also write a review about them.
The Yelp dataset is available on its website\footnote{\url{https://www.yelp.com/dataset}} and can be obtained after agreeing on the dataset license;
however, many papers refer to a different url\footnote{\url{https://www.yelp.com/dataset_challenge}} that does not work anymore, like~\cite{DBLP:journals/eswa/LiXCZ15, DBLP:conf/pakdd/GuptaPM15, DBLP:conf/kdd/YangBZY017, DBLP:conf/um/BaralZIL18}; this is because this dataset was first released in the context of a challenge ran by Yelp, which has gone at least through $12$ rounds where the data has been increased each time; this makes the comparisons even more difficult since it is not possible to get the dataset corresponding to a specific round, and this information is usually omitted in the papers.

\item Others: in addition to the aforementioned LBSNs, some proposals work with datasets extracted from other systems, such as Jiepang (a Chinese LBSN similar to Foursquare) used in~\cite{DBLP:conf/kdd/LianZXSCR14, DBLP:conf/icdm/LianGZYXZR15, DBLP:conf/icdm/LianZGZYX16, DBLP:journals/wpc/RaviS17}, Weeplaces used in~\cite{DBLP:conf/recsys/BaralL16, DBLP:conf/iri/BaralWLC16}, GeoLife used in~\cite{DBLP:conf/adc/Abdel-FataoLL15, DBLP:journals/jnca/ZhuXGZ17}, and others less popular in our context, like Twitter and TripAdvisor.
\end{itemize}

\pgfplotstableread[col sep=semicolon]{
NumberPapers;KNN;Factorization;Probabilistic;DNN;GraphLink;Hybrid;Other;Geographical;Social;Content;Sequential;Temporal;RandomS;nfoldS;TemporalS;OtherS;CC;Fix
Most Representatives;7;19;14;3;5;10;1;31;12;13;9;10;15;2;17;4;14;20
Total;57;81;83;35;36;75;14;174;95;111;43;100;94;12;83;30;91;98
}\numberPapersHaving

%Revised May 2021
\pgfplotstableread[col sep=semicolon]{
NumberPapers;Gowalla;Foursquare;Yelp;Brightkite;Other
Most Representatives;30;34;4;6;8
Total;156;199;54;40;43
}\numberPapersUsingDataset

%Revised May 2021
\pgfplotstableread[col sep=semicolon]{
NumberPapers;Offline;Online;UserStudy
Most Representatives;45;0;0
Total;306;0;5
}\numberPaperEvMethodology

\begin{table*}
\caption{\ReviewOne{Evaluation methodologies used by papers included in our review.}}
\label{tab:numberPapersEvMethdology}
\vspace{-0.25cm}
\ifloadtables
\pgfplotstabletypeset[
		font=\fontsize{7pt}{7.5pt}\selectfont,
		outfile=tab_numberPapersEvMethodology.tex,
		every head row/.style={
            before row={%
                \hline
                \multirow{2}{*}{ Number of Papers}  & \multicolumn{3}{c?}{Evaluation methodology}\\
                \cline{2-4}
            },
            after row={
            	\hline
            },
        },
        every even row/.style={
        	before row={
        		\hline},
        	after row={\hline}
        },
        every last row/.style={after row=\hline},
        column type=c,
        %column type/.add={|}{},
        %%%%%%%%%%%%%%%%%%%%%%%%%%%%%%%%%%%%%%%%%%%%%
        % NumberPapers
        columns/NumberPapers/.style={
			string type,
			column name={},
			column type/.add={?}{|},
        },
        %%%%%%%%%%%%%%%%%
        % Ev Methodology: Offline, Online, User study
        columns/Offline/.style={
			string type,
			column type/.add={}{|},
        },
        columns/Online/.style={
			string type,
			column type/.add={}{|},
        },
        columns/UserStudy/.style={
			string type,
			column type/.add={}{?},
			column name={User study},
        },
        fixed,
        fixed zerofill,
        columns={NumberPapers,Offline,Online,UserStudy},
    ]{\numberPaperEvMethodology}
\else
	\begingroup \fontsize {7pt}{7.5pt}\selectfont %
\begin {tabular}{?c|c|c|c?}%
\hline \multirow {2}{*}{ Number of Papers} & \multicolumn {3}{c?}{Evaluation methodology}\\ \cline {2-4}&Offline&Online&User study\\\hline %
\hline Most Representatives&45&0&0\\\hline %
Total&306&0&5\\\hline %
\end {tabular}%
\endgroup %
\fi
\end{table*}

\begin{table*}
\caption{Papers included in our review that use a dataset from each LBSN.}
\label{tab:numberPapersUsingDataset}
\vspace{-0.25cm}
\ifloadtables
\pgfplotstabletypeset[
		font=\fontsize{7pt}{7.5pt}\selectfont,
		outfile=tab_numberPapersUsingDataset.tex,
		every head row/.style={
            before row={%
                \hline
                \multirow{2}{*}{ Number of Papers}  & \multicolumn{5}{c?}{LBSN}\\
                \cline{2-6}
            },
            after row={
            	\hline
            },
        },
        every even row/.style={
        	before row={
        		\hline},
        	after row={\hline}
        },
        every last row/.style={after row=\hline},
        column type=c,
        %column type/.add={|}{},
        %%%%%%%%%%%%%%%%%%%%%%%%%%%%%%%%%%%%%%%%%%%%%
        % NumberPapers
        columns/NumberPapers/.style={
			string type,
			column name={},
			column type/.add={?}{|},
        },
        %%%%%%%%%%%%%%%%%
        % Methodology: Gowalla,Foursquare,Yelp,Brightkite,Other
        columns/Gowalla/.style={
			string type,
			column type/.add={}{|},
        },
        columns/Foursquare/.style={
			string type,
			column type/.add={}{|},
        },
        columns/Yelp/.style={
			string type,
			column type/.add={}{|},
        },
        columns/Brightkite/.style={
			string type,
			column type/.add={}{|},
        },
        columns/Other/.style={
			string type,
			column type/.add={}{?},
        },
        fixed,
        fixed zerofill,
        columns={NumberPapers,Gowalla,Foursquare,Yelp,Brightkite,Other},
    ]{\numberPapersUsingDataset}
\else
	\begingroup \fontsize {7pt}{7.5pt}\selectfont %
\begin {tabular}{?c|c|c|c|c|c?}%
\hline \multirow {2}{*}{ Number of Papers} & \multicolumn {5}{c?}{LBSN}\\ \cline {2-6}&Gowalla&Foursquare&Yelp&Brightkite&Other\\\hline %
\hline Most Representatives&30&34&4&6&8\\\hline %
Total&156&199&54&40&43\\\hline %
\end {tabular}%
\endgroup %
\fi
\end{table*}

\pgfplotstableread[col sep=semicolon]{
Year;Reference;Acronym;NCites;Dataset;NUsers;NPOIs;NCheckins;MetricsUsed;TypeSplit
2011;\cite{DBLP:conf/sigir/YeYLL11};\USG;784;Foursquare;153,577;96,229;\unk;P, R;Random \Fix
2011;\cite{DBLP:conf/sigir/YeYLL11};\USG;784;Whrrl;5,892;53,432;\unk;P, R;Random \Fix
2012;\cite{DBLP:conf/gis/0003ZM12};\NA;477;Foursquare (NY);2,886;\unk;10,687;P, R;Other
2012;\cite{DBLP:conf/gis/0003ZM12};\NA;477;Foursquare (LA);228;\unk;9,836;P, R;Other
2013;\cite{DBLP:conf/sigir/YuanCMSM13};\UTE, \SE, \UTESE;535;Foursquare;2,321;5,596;194,108;P, R;Random \Fix
2013;\cite{DBLP:conf/sigir/YuanCMSM13};\UTE, \SE, \UTESE;535;Gowalla;10,162;24,250;456,988;P, R;Random \Fix
2014;\cite{DBLP:conf/kdd/LianZXSCR14};\GEOMF;377;Jiepang;276,450;574,095;\unk;P, R;Random \Fix
2015;\cite{DBLP:conf/sigir/LiCLPK15};\RANKGEOFM;220;Foursquare;2,321;5,596;194,108;P, R;Temporal \Fix
2015;\cite{DBLP:conf/sigir/LiCLPK15};\RANKGEOFM;220;Gowalla;10,162;24,250;456,988;P, R;Temporal \Fix
2016;\cite{DBLP:conf/cikm/XieYWXCW16};\GE;184;Foursquare;114,508;62,462;1,434,668;Accuracy;Temporal \Fix
2016;\cite{DBLP:conf/cikm/XieYWXCW16};\GE;184;Gowalla;107,092;1,280,969;6,442,892;Accuracy;Temporal \Fix
2017;\cite{DBLP:conf/kdd/YangBZY017};\PACE;153;Gowalla;18,737;32,510;1,278,274;P, R, NDCG, MAP;Temporal \Fix
2017;\cite{DBLP:conf/kdd/YangBZY017};\PACE;153;Yelp;30,887;18,995;860,888;P, R, NDCG, MAP;Temporal \Fix
2018;\cite{DBLP:conf/ijcai/WangSOC18};\GEOIE;49;Foursquare;6,118;88,193;172,961;P, R;Temporal \Fix
2018;\cite{DBLP:conf/ijcai/WangSOC18};\GEOIE;49;Gowalla;1,624;3,585;115,890;P, R;Temporal \Fix
2019;\cite{DBLP:conf/aaai/ZhaoZLXLZSZ19};\STGN;45;Foursquare (CA);49,005;206,097;425,691;Accuracy, MAP;Temporal \Fix
2019;\cite{DBLP:conf/aaai/ZhaoZLXLZSZ19};\STGN;45;Foursquare (SIN);30,887;18,995;860,888;Accuracy, MAP;Temporal \Fix
2019;\cite{DBLP:conf/aaai/ZhaoZLXLZSZ19};\STGN;45;Gowalla;18,737;32,510;1,278,274;Accuracy, MAP;Temporal \Fix
2019;\cite{DBLP:conf/aaai/ZhaoZLXLZSZ19};\STGN;45;Brightkite;51,406;772,967;4,747,288;Accuracy, MAP;Temporal \Fix
2020;\cite{DBLP:journals/tii/WangCWCLG20};\TECF;14;Foursquare;2,321;5,596;194,108;P, R; Random \Fix
2020;\cite{DBLP:journals/tii/WangCWCLG20};\TECF;14;Foursquare;10,162;24,250;456,988;P, R; Random \Fix
}\exampleTableMostCited

\begin{table*}
\caption{
Details of the experimental settings for the work with most citations each year (note each work appears once for each reported dataset).
\unk denotes that value is not provided in the paper.
The columns Ref. and Cit. denote the original reference for that work and the number of citations, as of \ReviewOne{April 2021}.
}
%Citations were obtained 14 April 2021.
\label{tab:EvMethodologyMostCited}
\vspace{-0.25cm}
\ifloadtables
\pgfplotstabletypeset[
		font=\fontsize{7pt}{7.5pt}\selectfont,
		outfile=tab_evMethodologyMostCited.tex,
        every last row/.style={after row=\hline},
        column type=c,
        every head row/.style={
            before row={%
                \hline
                \multicolumn{4}{? c ? }{Details} & \multicolumn{4}{c?}{Statistics} & \multicolumn{2}{ c ? }{Evaluation Details}\\
                \hline
            },
            after row={
            	\hline
            },
            typeset cell/.code={
            \ifnum\pgfplotstablecol=\pgfplotstablecols
						%\pgfkeyssetvalue{/pgfplots/table/@cell content}{\multicolumn{1}{|c|}{\rotatebox{90}{##1}}\\}%
						\pgfkeyssetvalue{/pgfplots/table/@cell content}{\multicolumn{1}{|c|}{##1}\\}%
            \else
							%\pgfkeyssetvalue{/pgfplots/table/@cell content}{\rotatebox{90}{##1}&}%
							%\pgfkeyssetvalue{/pgfplots/table/@cell content}{\multicolumn{1}{|c}{\rotatebox{90}{##1}}&}%
							\pgfkeyssetvalue{/pgfplots/table/@cell content}{\multicolumn{1}{|c}{##1}&}%
            \fi
            },
						%% Missing: put this in header
						% \multicolumn{1}{c}{}
        },
        %column type/.add={|}{},
        every even row/.style={
        	before row={
        		\hline},
        	after row={\hline}
        },
        every row no 0/.style={
        	before row={\hline \rowcolor{lightgray!50}},
        },
        every row no 1/.style={
        	before row={\rowcolor{lightgray!50}},
        },
        every row no 4/.style={
        	before row={\hline \rowcolor{lightgray!50}},
        },
        every row no 5/.style={
        	before row={\rowcolor{lightgray!50}},
        },
		every row no 7/.style={
        	before row={\rowcolor{lightgray!50}},
        },
        every row no 8/.style={
        	before row={\hline \rowcolor{lightgray!50}},
        },
        every row no 11/.style={
        	before row={\rowcolor{lightgray!50}},
        },
        every row no 12/.style={
        	before row={\hline \rowcolor{lightgray!50}},
        },
        every row no 15/.style={
        	before row={\rowcolor{lightgray!50}},
        },
        every row no 16/.style={
        	before row={\hline\rowcolor{lightgray!50}},
        },
        every row no 17/.style={
        	before row={\rowcolor{lightgray!50}},
        },
        every row no 18/.style={
        	before row={\hline\rowcolor{lightgray!50}},
        },
        %%%%%%%%%%%%%%%%%%%%%%%%%%%%%%%%%%%%%%%%%%%%%
        % First part (details): Year, Reference, Acronym, NCites
        columns/Year/.style={
			string type,
			column type/.add={?}{|},
        },
		columns/Reference/.style={
			string type,
			column name={Ref.},
			column type/.add={}{|},
        },
        columns/Acronym/.style={
			string type,
			column type/.add={}{|},
        },
        columns/NCites/.style={
				column type=r,
			string type,
			column name={Cit.},
			%column name={\multicolumn{1}{c}{N\textordmasculine\xspace Cites}},
			column type/.add={}{?},
        },
        %%%%%%%%%%%%%%%%%
        % Second part: Dataset;NUsers;NPOIs;NCheckins
        columns/Dataset/.style={
			string type,
			column type/.add={}{|},
        },
        columns/NUsers/.style={
				column type=r,
			string type,
			column name={Users},
			column type/.add={}{|},
        },
        columns/NPOIs/.style={
				column type=r,
			string type,
			column name={POIs},
			column type/.add={}{|},
        },
        columns/NCheckins/.style={
				column type=r,
			string type,
			column name={Check-ins},
			column type/.add={}{?},
        },
        %%%%%%%%%%%%%%%%%
        % Third part (Evaluation details): MetricsUsed, TypeSplit
          columns/MetricsUsed/.style={
			string type,
			column name={Metrics used},
			column type/.add={}{|},
        },
        columns/TypeSplit/.style={
			string type,
			column name={Type of Split},
			column type/.add={}{?},
        },
        fixed,
        fixed zerofill,
        columns={Year,Reference,Acronym,NCites,Dataset,NUsers,NPOIs,NCheckins,MetricsUsed,TypeSplit},
    ]{\exampleTableMostCited}
\else
	\begingroup \fontsize {7pt}{7.5pt}\selectfont %
\begin {tabular}{?c|c|c|r?c|r|r|r?c|c?}%
\hline \multicolumn {4}{? c ? }{Details} & \multicolumn {4}{c?}{Statistics} & \multicolumn {2}{ c ? }{Evaluation Details}\\ 
\hline Year&Ref.& Acronym & Cit.& Dataset & Users & POIs & Check-ins & Metrics used & Type of Split \\\hline %
\hline \rowcolor {lightgray!50}2011&\cite {DBLP:conf/sigir/YeYLL11}&\USG &784&Foursquare&153,577&96,229&\unk &P, R&Random \Fix \\\hline %
\rowcolor {lightgray!50}2011&\cite {DBLP:conf/sigir/YeYLL11}&\USG &784&Whrrl&5,892&53,432&\unk &P, R&Random \Fix \\%
\hline 2012&\cite {DBLP:conf/gis/0003ZM12}&\NA &477&Foursquare (NY)&2,886&\unk &10,687&P, R&Other\\\hline %
2012&\cite {DBLP:conf/gis/0003ZM12}&\NA &477&Foursquare (LA)&228&\unk &9,836&P, R&Other\\%
\hline \rowcolor {lightgray!50}2013&\cite {DBLP:conf/sigir/YuanCMSM13}&\UTE , \SE , \UTESE &535&Foursquare&2,321&5,596&194,108&P, R&Random \Fix \\\hline %
\rowcolor {lightgray!50}2013&\cite {DBLP:conf/sigir/YuanCMSM13}&\UTE , \SE , \UTESE &535&Gowalla&10,162&24,250&456,988&P, R&Random \Fix \\%
\hline 2014&\cite {DBLP:conf/kdd/LianZXSCR14}&\GEOMF &377&Jiepang&276,450&574,095&\unk &P, R&Random \Fix \\\hline %
\rowcolor {lightgray!50}2015&\cite {DBLP:conf/sigir/LiCLPK15}&\RANKGEOFM &220&Foursquare&2,321&5,596&194,108&P, R&Temporal \Fix \\%
\hline \rowcolor {lightgray!50}2015&\cite {DBLP:conf/sigir/LiCLPK15}&\RANKGEOFM &220&Gowalla&10,162&24,250&456,988&P, R&Temporal \Fix \\\hline %
2016&\cite {DBLP:conf/cikm/XieYWXCW16}&\GE &184&Foursquare&114,508&62,462&1,434,668&Accuracy&Temporal \Fix \\%
\hline 2016&\cite {DBLP:conf/cikm/XieYWXCW16}&\GE &184&Gowalla&107,092&1,280,969&6,442,892&Accuracy&Temporal \Fix \\\hline %
\rowcolor {lightgray!50}2017&\cite {DBLP:conf/kdd/YangBZY017}&\PACE &153&Gowalla&18,737&32,510&1,278,274&P, R, NDCG, MAP&Temporal \Fix \\%
\hline \rowcolor {lightgray!50}2017&\cite {DBLP:conf/kdd/YangBZY017}&\PACE &153&Yelp&30,887&18,995&860,888&P, R, NDCG, MAP&Temporal \Fix \\\hline %
2018&\cite {DBLP:conf/ijcai/WangSOC18}&\GEOIE &49&Foursquare&6,118&88,193&172,961&P, R&Temporal \Fix \\%
\hline 2018&\cite {DBLP:conf/ijcai/WangSOC18}&\GEOIE &49&Gowalla&1,624&3,585&115,890&P, R&Temporal \Fix \\\hline %
\rowcolor {lightgray!50}2019&\cite {DBLP:conf/aaai/ZhaoZLXLZSZ19}&\STGN &45&Foursquare (CA)&49,005&206,097&425,691&Accuracy, MAP&Temporal \Fix \\%
\hline \rowcolor {lightgray!50}2019&\cite {DBLP:conf/aaai/ZhaoZLXLZSZ19}&\STGN &45&Foursquare (SIN)&30,887&18,995&860,888&Accuracy, MAP&Temporal \Fix \\\hline %
\rowcolor {lightgray!50}2019&\cite {DBLP:conf/aaai/ZhaoZLXLZSZ19}&\STGN &45&Gowalla&18,737&32,510&1,278,274&Accuracy, MAP&Temporal \Fix \\%
\hline \rowcolor {lightgray!50}2019&\cite {DBLP:conf/aaai/ZhaoZLXLZSZ19}&\STGN &45&Brightkite&51,406&772,967&4,747,288&Accuracy, MAP&Temporal \Fix \\\hline %
2020&\cite {DBLP:journals/tii/WangCWCLG20}&\TECF &14&Foursquare&2,321&5,596&194,108&P, R&Random \Fix \\%
\hline 2020&\cite {DBLP:journals/tii/WangCWCLG20}&\TECF &14&Foursquare&10,162&24,250&456,988&P, R&Random \Fix \\\hline %
\end {tabular}%
\endgroup %
\fi
\end{table*}

%%UPDATE 2 JUNE 2021. Only consider the datasets at least used in three different publications. All statistics of the publications must match (number of users, number of items and number of check-ins)

\pgfplotstableread[col sep=semicolon]{
url;OriginalPaper;Users;POIs;Checkins;PapersUsingDataset
https:/sites.google.com/site/dbhongzhi;\cite{DBLP:conf/icwsm/ChengCLS11};114,508;62,462;1,434,668;\cite{DBLP:conf/cikm/XieYWXCW16, DBLP:journals/jcst/QianLHY18, DBLP:journals/tois/QianLNY19, DBLP:conf/dsaa/ChristoforidisK18,DBLP:journals/jcloudc/KuangTZY20}
http://www.public.asu.edu/~hgao16/dataset.html;\cite{DBLP:conf/cikm/GaoTL12};11,326;182,968;1,385,223;\cite{DBLP:conf/gis/ZhangC13, DBLP:conf/gis/ZhangCL14, DBLP:journals/tsc/ZhangCL15, DBLP:journals/isci/ZhangC15, DBLP:journals/tcss/StepanMDM16, DBLP:journals/tsc/ZhangC16, DBLP:journals/ijon/GaoLLSCLW18, Gao2018, DBLP:journals/access/PanCWZLC19}
https://archive.org/details/201309\_foursquare\_dataset\_umn;\cite{DBLP:conf/cikm/ManotumruksaMO17};10,766;10,695;1,336,278;\cite{DBLP:conf/cikm/ManotumruksaMO17, DBLP:conf/sigir/ManotumruksaMO18, Manotumruksa2019, DBLP:conf/ecir/ManotumruksaRMO19}
https://sites.google.com/site/yangdingqi/home/foursquare-dataset;\cite{DBLP:journals/pvldb/LiuPCY17};24,941;28,593;1,196,248;\cite{DBLP:conf/cikm/MaZWL18, DBLP:journals/soco/0008WL19, DBLP:journals/access/XuLLWGY19, DBLP:conf/mir/LiuLWWW19, DBLP:conf/icycsee/YueZZM20, DBLP:conf/cikm/ChangJKK20, DBLP:journals/www/ZhongZZZTW20, DBLP:conf/pakdd/WangSC20a}
???;???;2,293;61,858;573,703;\cite{DBLP:conf/bigdataconf/MaroulisBK16, DBLP:conf/iccS/ZhongMZW20,DBLP:conf/wasa/ZhongMZW20}
https://sites.google.com/site/yangdingqi/home/foursquare-dataset;\cite{DBLP:journals/pvldb/LiuPCY17};7,642;28,484;512,523;\cite{DBLP:conf/kdd/LianZXSCR14,DBLP:conf/airs/RahmaniAABAC19,DBLP:conf/icc/Su0LZXTG20,DBLP:conf/ecir/RahmaniABC20}
???;\cite{DBLP:conf/aaai/GaoTHL15};4,163;121,142;483,813;\cite{DBLP:conf/sigir/ZhangC15, DBLP:conf/dasfaa/HosseiniL16, DBLP:conf/wise/XieYXWZ16, DBLP:conf/dasfaa/HosseiniYZZS17, DBLP:journals/www/HosseiniYZSKC19,Mighan2019, DBLP:journals/ipm/QiaoLLTM20}
???;???;1,083;38,333;227,428;\cite{DBLP:conf/bigdataconf/MaroulisBK16, DBLP:conf/socinfo/DebnathTE16, Elangovan2020, DBLP:conf/iccS/ZhongMZW20, DBLP:conf/wasa/ZhongMZW20}
http://www.ntu.edu.sg/home/gaocong/datacode.htm;\cite{DBLP:conf/sigir/YuanCMSM13};2,321;5,596;194,108;\cite{DBLP:conf/sigir/YuanCMSM13, DBLP:conf/cikm/YuanCS14, DBLP:conf/webi/KojimaT15, DBLP:conf/sigir/LiCLPK15, DBLP:conf/icdm/ZhaoXLZ0SX17, DBLP:journals/kbs/SiZL17, DBLP:conf/pakdd/YuWSG17, DBLP:conf/icaci/XuLYZ18, DBLP:journals/complexity/ChenZZYNZ18, Kala2019, DBLP:conf/ijcnn/GaoLLZGH19, DBLP:conf/seke/ZengTLH19,DBLP:journals/kbs/SiZL19, Liu2020,DBLP:journals/tkde/AliannejadiRC20,DBLP:journals/tii/WangCWCLG20,DBLP:journals/iotj/WangCWCG20, DBLP:journals/ijgi/LiuZLQZ20}
???;\cite{DBLP:conf/kdd/LiGHZ16};2,551;13,474;124,933;\cite{DBLP:conf/kdd/LiGHZ16, DBLP:conf/ijcai/LiGLL17, DBLP:conf/ijcnn/SuLTZXG19, DBLP:conf/ijcnn/SuZ0ZXTG20}
}\tablesDatasetsFoursquare

\iffalse %% to make paper shorter
\begin{table*}
\caption{Statistics of reported versions for the Foursquare dataset in works included in our review, sorted by number of \checkins.}
\label{tab:versionsFoursquare}
\vspace{-0.25cm}
\ifloadtables
\pgfplotstabletypeset[
		font=\fontsize{7pt}{7.5pt}\selectfont,
		outfile=tab_versionsFoursquare.tex,
        every last row/.style={after row=\hline},
		column type=r,
        every head row/.style={
            before row={%
                \hline
            },
            after row={
            	\hline
            },
            %typeset cell/.code={
            %\ifnum\pgfplotstablecol=\pgfplotstablecols
            % 	\pgfkeyssetvalue{/pgfplots/table/@cell content}{\rotatebox{90}{##1}\\}%
            %\else
            %	\pgfkeyssetvalue{/pgfplots/table/@cell content}{\rotatebox{90}{##1}&}%
            %\fi
            %},
        },
	every even row/.style={
        	before row={
        		\hline
        		\rowcolor{lightgray!50}},
        	after row={\hline}
        },
        %%%%%%%%%%%%%%%%%%%%%%%%%%%%%%%%%%%%%%%%%%%%%
        % First part url, users and checkimns
        columns/url/.style={
			string type,
			column type/.add={?}{|},
        },
	columns/Users/.style={
			string type,
			column type/.add={?}{|},
        },
	columns/OriginalPaper/.style={
			string type,
			column name={Original Paper},
			column type/.add={?}{|},
        },
	columns/POIs/.style={
			string type,
			column type/.add={}{|},
        },
        columns/Checkins/.style={
			string type,
			column name={Check-ins},
			column type/.add={}{|},
        },
	columns/PapersUsingDataset/.style={
			string type,
			column name={References using this dataset},
			column type/.add={c?}{},
        },
        fixed,
        fixed zerofill,
        columns={Users,POIs,Checkins,PapersUsingDataset},
    ]{\tablesDatasetsFoursquare}
\else
	\begingroup \fontsize {7pt}{7.5pt}\selectfont %
\begin {tabular}{?r|r|r|c?r}%
\hline Users&POIs&Check-ins&References using this dataset\\\hline %
\hline \rowcolor {lightgray!50}114,508&62,462&1,434,668&\cite {DBLP:conf/cikm/XieYWXCW16, DBLP:journals/jcst/QianLHY18, DBLP:journals/tois/QianLNY19, DBLP:conf/dsaa/ChristoforidisK18,DBLP:journals/jcloudc/KuangTZY20}\\\hline %
11,326&182,968&1,385,223&\cite {DBLP:conf/gis/ZhangC13, DBLP:conf/gis/ZhangCL14, DBLP:journals/tsc/ZhangCL15, DBLP:journals/isci/ZhangC15, DBLP:journals/tcss/StepanMDM16, DBLP:journals/tsc/ZhangC16, DBLP:journals/ijon/GaoLLSCLW18, Gao2018, DBLP:journals/access/PanCWZLC19}\\%
\hline \rowcolor {lightgray!50}10,766&10,695&1,336,278&\cite {DBLP:conf/cikm/ManotumruksaMO17, DBLP:conf/sigir/ManotumruksaMO18, Manotumruksa2019, DBLP:conf/ecir/ManotumruksaRMO19}\\\hline %
24,941&28,593&1,196,248&\cite {DBLP:conf/cikm/MaZWL18, DBLP:journals/soco/0008WL19, DBLP:journals/access/XuLLWGY19, DBLP:conf/mir/LiuLWWW19, DBLP:conf/icycsee/YueZZM20, DBLP:conf/cikm/ChangJKK20, DBLP:journals/www/ZhongZZZTW20, DBLP:conf/pakdd/WangSC20a}\\%
\hline \rowcolor {lightgray!50}2,293&61,858&573,703&\cite {DBLP:conf/bigdataconf/MaroulisBK16, DBLP:conf/iccS/ZhongMZW20,DBLP:conf/wasa/ZhongMZW20}\\\hline %
7,642&28,484&512,523&\cite {DBLP:conf/kdd/LianZXSCR14,DBLP:conf/airs/RahmaniAABAC19,DBLP:conf/icc/Su0LZXTG20,DBLP:conf/ecir/RahmaniABC20}\\%
\hline \rowcolor {lightgray!50}4,163&121,142&483,813&\cite {DBLP:conf/sigir/ZhangC15, DBLP:conf/dasfaa/HosseiniL16, DBLP:conf/wise/XieYXWZ16, DBLP:conf/dasfaa/HosseiniYZZS17, DBLP:journals/www/HosseiniYZSKC19,Mighan2019, DBLP:journals/ipm/QiaoLLTM20}\\\hline %
1,083&38,333&227,428&\cite {DBLP:conf/bigdataconf/MaroulisBK16, DBLP:conf/socinfo/DebnathTE16, Elangovan2020, DBLP:conf/iccS/ZhongMZW20, DBLP:conf/wasa/ZhongMZW20}\\%
\hline \rowcolor {lightgray!50}2,321&5,596&194,108&\cite {DBLP:conf/sigir/YuanCMSM13, DBLP:conf/cikm/YuanCS14, DBLP:conf/webi/KojimaT15, DBLP:conf/sigir/LiCLPK15, DBLP:conf/icdm/ZhaoXLZ0SX17, DBLP:journals/kbs/SiZL17, DBLP:conf/pakdd/YuWSG17, DBLP:conf/icaci/XuLYZ18, DBLP:journals/complexity/ChenZZYNZ18, Kala2019, DBLP:conf/ijcnn/GaoLLZGH19, DBLP:conf/seke/ZengTLH19,DBLP:journals/kbs/SiZL19, Liu2020,DBLP:journals/tkde/AliannejadiRC20,DBLP:journals/tii/WangCWCLG20,DBLP:journals/iotj/WangCWCG20, DBLP:journals/ijgi/LiuZLQZ20}\\\hline %
2,551&13,474&124,933&\cite {DBLP:conf/kdd/LiGHZ16, DBLP:conf/ijcai/LiGLL17, DBLP:conf/ijcnn/SuLTZXG19, DBLP:conf/ijcnn/SuZ0ZXTG20}\\\hline %
\end {tabular}%
\endgroup %
\fi
\end{table*}
\else
\begin{table*}
\caption{Statistics of reported versions for the Foursquare dataset in works included in our review, sorted by number of \checkins.}
\label{tab:versionsFoursquare}
\vspace{-0.25cm}
\ifloadtables
\pgfplotstabletypeset[
		font=\fontsize{7pt}{7.5pt}\selectfont,
		outfile=tab_versionsFoursquare.tex,
        every last row/.style={after row=\hline},
		column type=r,
        every head row/.style={
            before row={%
                \hline
            },
            after row={
            	\hline
            },
            %typeset cell/.code={
            %\ifnum\pgfplotstablecol=\pgfplotstablecols
            % 	\pgfkeyssetvalue{/pgfplots/table/@cell content}{\rotatebox{90}{##1}\\}%
            %\else
            %	\pgfkeyssetvalue{/pgfplots/table/@cell content}{\rotatebox{90}{##1}&}%
            %\fi
            %},
        },
	every even row/.style={
        	before row={
        		\hline
        		\rowcolor{lightgray!50}},
        	after row={\hline}
        },
        %%%%%%%%%%%%%%%%%%%%%%%%%%%%%%%%%%%%%%%%%%%%%
        % First part url, users and checkimns
        columns/url/.style={
			string type,
			column type/.add={?}{|},
        },
	columns/Users/.style={
			string type,
			column type/.add={?}{|},
        },
	columns/OriginalPaper/.style={
			string type,
			column name={Original Paper},
			column type/.add={?}{|},
        },
	columns/POIs/.style={
			string type,
			column type/.add={}{|},
        },
        columns/Checkins/.style={
			string type,
			column name={Check-ins},
			column type/.add={}{|},
        },
	columns/PapersUsingDataset/.style={
			string type,
			column name={Papers using this dataset},
			column type/.add={c?}{},
        },
        fixed,
        fixed zerofill,
        columns={Users,POIs,Checkins,PapersUsingDataset},
    ]{\tablesDatasetsFoursquare}
\else
	\begingroup \fontsize {7pt}{7.5pt}\selectfont %
\begin {tabular}{?r|r|r|c?r}%
\hline Users&POIs&Check-ins& Papers using this dataset\\\hline %
\hline \rowcolor {lightgray!50}114,508&62,462&1,434,668& 5\\\hline %
11,326&182,968&1,385,223& 9 \\ 
\hline \rowcolor {lightgray!50}10,766&10,695&1,336,278& 4 \\\hline 
24,941&28,593&1,196,248& 8 \\ 
\hline \rowcolor {lightgray!50}2,293&61,858&573,703& 3 \\\hline
7,642&28,484&512,523& 4 \\ 
\hline \rowcolor {lightgray!50}4,163&121,142&483,813& 7 \\\hline 
1,083&38,333&227,428& 5 \\ 
\hline \rowcolor {lightgray!50}2,321&5,596&194,108& 18 \\\hline 
2,551&13,474&124,933& 4 \\\hline 
\end {tabular}%
\endgroup %
\fi
\end{table*}
\fi

While doing our systematic review, we found several versions of datasets coming from the same LBSN.
For the sake of space and clarity, we show in Table~\ref{tab:EvMethodologyMostCited} the LBSNs used by the research work with more citations (according to Scopus) for each year, together with some statistics of the dataset and other evaluation details reported in the experiments, such as the type of split and the evaluation metrics.
Based on this information, we observe that all of them evaluate based on some notion of ranking quality; while it is true that this evidences researchers are taking into account the guidelines provided in the recommendation field as a whole~\cite{DBLP:conf/chi/McNeeRK06}, a possible reason for this is that the data to be predicted is not ratings anymore, but binary feedback: whether the user visited the POI or not.
In any case, this table emphasizes an even more important problem: most researchers are only focused on accuracy, disregarding additional dimensions such as novelty, diversity, or serendipity that are becoming prevalent in recent years in the evaluation of RSs~\cite{DBLP:reference/sp/CastellsHV15}.

Table~\ref{tab:EvMethodologyMostCited} also shows an interesting paradigm shift: those works prior to 2015 used a random split, and those more recent used a temporal split \ReviewOne{(except in 2020)}.
We consider this a \ReviewOne{decisive} signal, since it indicates that (at least for the works that are later more cited by colleagues) a more realistic type of split is being used, which would indeed make the proposed approaches easier to put in context in a real scenario.
It is also positive that most of these works (it is by no means the same in general) contrast their approaches against two data sources, which makes the results easier to generalize.
On the other hand, what can be considered as a worrying sign is that there are not two articles sharing the number of \checkins or users, except~\cite{DBLP:conf/sigir/YuanCMSM13, DBLP:conf/sigir/LiCLPK15} and~\cite{DBLP:journals/tii/WangCWCLG20}, but even in this case, each work performs a different data splitting; moreover, there are even cases where some of the statistics are not included (like the number of items or \checkins).
This makes it almost impossible to compare two research works without implementing everything from scratch, hence hindering reproducibility and the advancement of the field~\cite{DBLP:conf/recsys/SaidB14}.

We were also surprised that in most cases the source code of the proposed method is not provided. In particular, among the papers with more citations, only \cite{DBLP:conf/www/ZhaoZKL17, DBLP:conf/kdd/YangBZY017, DBLP:conf/cikm/MaZWL18, DBLP:conf/sigir/ManotumruksaMO18, DBLP:journals/kbs/ZhaoLQH20} redirect to a repository with source code.
With respect to the rest of the analyzed papers (that is, out of the \ReviewOne{$310$} works), only~\cite{DBLP:conf/icdm/HuE14, DBLP:journals/eswa/LiXCZ15, DBLP:conf/www/ZhaoZKL17, DBLP:conf/kdd/YangBZY017, DBLP:conf/cikm/MaZWL18, DBLP:conf/sigir/ManotumruksaMO18, DBLP:conf/icdm/LiSZ18, DBLP:conf/dsaa/ChristoforidisK18, DBLP:conf/seke/YuXW19, DBLP:conf/ecir/RahmaniABC20,https://doi.org/10.1002/ett.3889, DBLP:journals/kbs/ZhaoLQH20, DBLP:conf/kdd/LianWG0C20} provide a url to download the source code of their algorithm. 

As a final analysis, we present in Table~\ref{tab:versionsFoursquare} different versions of datasets extracted from Foursquare, considering this is the most widely used LBSN in the articles included in our review.
In this selection we show the datasets used by more than \ReviewOne{two} articles, since there are works using other variations not reported here but, for the sake of space, we focused only on those reported 
a minimum number of times
among the papers considered in our analysis\footnote{%
For instance, a widely used version called Global-scale dataset from \citeauthor{DBLP:journals/tist/YangZQ16} presented in~\cite{DBLP:journals/tist/YangZQ16} is not included in this table because only one paper reported the exact statistics as the original paper (which is actually not considered because it does not perform POI recommendation), whereas other works take subsets of it.}.
Nevertheless, it is remarkable to observe the large difference in the number of \checkins, ranging from 45k to 2M interactions; as a consequence, the experiments presented in the different works are probably not comparable at all -- even if they belong to the same LBSN -- since the inherent properties of the system are not preserved: for instance, in some cases we have more users than items whereas in other cases it is the other way around; it is also possible that the levels of sparsity change dramatically, together with the number of cities/regions included in each dataset.
It is interesting to observe that most datasets are seldom used, and the few cases where the same dataset is used by many works, it is because they belong to the same authors; see the supporting materials for an explicit list of the references using these datasets.
%, 

\section{Future research directions and open issues}
\label{s:future}
In this section, we present some open issues we have identified after performing an analysis on the state-of-the-art on POI recommendation \ReviewTwo{based on LBSNs}, after that we list some potential future research lines we believe are in line with parallel developments in the field of recommender systems.

\subsection{Open issues and research challenges}
Although several research efforts have been devoted to the problem of POI recommendation, it is still possible to find unresolved issues in the field, which opens up opportunities to improve the area as a whole, for instance, because they are more aligned with the necessities of the final users and, probably, with industry practitioners.
By analyzing the current proposals in POI recommendation \ReviewTwo{based on LBSNs}, we have observed some important open issues that need to be addressed.
In the following, we group them according to the three main systematic reviews we performed: models or algorithms, evaluation methodologies, and datasets.

\subsubsection{Open issues}
Regarding algorithms,
matrix factorization and, more recently, Deep Learning
are very popular approaches in POI recommendation \ReviewTwo{when using data from LBSNs}; however, it is often difficult to explain why the recommendations from these methods are made since they behave like a \textit{black box} and this can be problematic in some domains, in particular in tourism.
%
% baselines
In addition, we have also observed that most researchers do not test their approaches against other classic recommendation algorithms like simple CF methods or non-personalized item popularity, comparing only with other POI recommendation approaches.
Similarly, the sequential information, despite its relevance in this domain, is not usually exploited, which sometimes prompts incorrect or not realistic evaluation methodologies.

In fact, about evaluation methodologies we consider the comparisons between different algorithms must always be as transparent and as fair as possible in order to determine which proposals are superior to others.
Therefore, although in the papers analyzed in this survey there seems to be consensus in evaluating the approaches using IR metrics like 
Precision or Recall, 
this is not the case about how to perform the splits, as there are both random and temporal partitions (each of them with different variations), even though the latter ones are the only strategies that could simulate real scenarios.
At the same time, the sparsity of the datasets used, whether or not they have been pre-filtered, etc., also affects the performance of the models, which in particular may prevent from having research works that are comparable between each other.

Regarding the datasets, although most proposals extract data from well-known LBSNs such as Foursquare, Gowalla, or Yelp, these datasets are often not comparable among them due to different decisions considered when filtering users or items, or even how the data were captured, which produces a large number of versions for each LBSN.
Comparing datasets is even more difficult when some researchers do not provide complete statistics about the actual dataset used in the experiments, leading to data with completely different characteristics and inherent properties (sparsity, granularity of temporal, geographical, and social information, POI attributes, and so on) even when they belong to the same LBSN.

\subsubsection{Discussion}
As we have seen along the survey, the problem of POI recommendation is attractive to a growing number of researchers in the area of recommender systems.
However, it may seem as if most of these issues have something in common: they make both the reproducibility and the generality of the proposed algorithms very difficult.
Thus, in order to advance towards better systems and foster high-quality research, we recommend to:

\begin{itemize}
  %-
  \item Explain in detail how the algorithms have been evaluated indicating the metrics used, the type of split and the rest of the models that have been used as baselines. In this regard, we suggest to test the proposed algorithm against specific POI recommendation models while also analyzing its performance against other baselines used in classical recommendation, such as neighbor-based algorithms, matrix factorization approaches, and the most-popular method.
  The evaluation methodology must be the same for all the algorithms and if it is necessary to make different experiments for choosing the parameters, this needs to be done for all the algorithms involved in the experiments and, if possible, with a validation subset independent of the test set.
	\item Clearly indicate the statistics of the used datasets, stating if any pre-processing step has been performed and showing the final details of the used data, including the number of users, POIs, and \checkins. This would help to detect the percentage of data that was removed to critically analyze if the filtered dataset is actually representative of the original dataset. We also strongly recommend researchers to use more than one dataset or, at least, to use different types of splits or more than one split from the same data if enough information is available.
  %-
  %-
  \item Finally, the easiest way to replicate a research work is by providing the code with a detailed description to achieve the same results mentioned in the paper; if this is not possible, the next best option is to, at least, provide the final datasets with which the algorithms were evaluated, so anyone interested in replicating it should not worry about that step of the evaluation pipeline.
\end{itemize}

In general, these recommendations aim to fix a lack of reproducible experimental settings that could hinder whether there is a significant improvement in the field, as already discussed in the RS and IR communities~\cite{DBLP:conf/recsys/SaidB14,DBLP:conf/cikm/ArmstrongMWZ09}.

\subsection{Future directions}
When preparing this survey, we have identified a number of future research directions, among them are the following.

\subsubsection{Towards realistic methodologies}
% realistic
How realistic the evaluation methodologies or recommendation algorithms proposed really are?
From our perspective, it may seem that sometimes the community is trying to solve a problem that will never arise in the real world, or at least, not in the terms it is being evaluated: using a global or worldwide test set (i.e., not divided by cities or countries) is not realistic, since a user at each point in time is only in one place and only interested in its surroundings.
Because of this, any recommendation algorithm that exploits the geographical information might be artificially benefited from this, but this can also be achieved simply by filtering the venues to be ranked at the evaluation step (as done in some works like~\cite{DBLP:conf/sdm/LiuX13}) instead of when modeling the problem.

Hence, we consider that researchers should formalize and critically think which task they want to solve and whether the evaluation methodology matches the task or if they are making it trivially easy or too much difficult \textit{by design}.
Related examples on this line include presenting a cold-start analysis while filtering users or items with too many interactions (see \cite{DBLP:journals/tist/ChengYKL16}) or not controling for new venues in test in a user basis (since users tend to visit the same venues they did in the past, so a simple baseline like returning the past profile of the user should be considered in the experiments). %

\subsubsection{Consider user types or roles}
As recent studies show, \checkin data can be used to characterize at least four types of travellers~\cite{DBLP:conf/enter/DietzRW19}: vacationers, explorers, voyagers, and globetrotters, thus going beyond the classical tourist roles that are usually considered (either leisure or business).
In the future we expect specialized algorithms would be developed towards each of these user types, since they show different inherent needs and interests, as evidenced by their behavior when visiting POIs in a city, but also because of their personality traits.
Moreover, other roles could be distinguished, even depending on the actual LBSN, since some systems might implicitly appeal to the social side of users, whereas others might be more attractive to POI owners, for instance; because of this, great care must be taken to transfer research results to a different LBSN if their underneath assumptions or interaction philosophies are not compatible.
%~\cite{DBLP:conf/wsdm/OlteanuK018}.

\subsubsection{Adversarial analysis and data quality}
We have found very few papers where the quality of data used in POI recommendation is discussed, one example is~\cite{DBLP:conf/www/PapalexakisPF14}\footnote{This paper is not included in our analysis because it does not satisfy the requirements described in Section~\ref{ss:how}.}.
Whereas in the classical recommendation problem the issue of robust recommendation -- in the sense that the recommendation algorithm should not be too sensitive to attacks from malicious users --  has been researched in the past and revisited recently with a different name~\cite{DBLP:reference/sp/BurkeOH15,DBLP:conf/wsdm/DeldjooNM20}, there are several open issues about this topic regarding POI recommenders and LBSN data, such as: how can these types of systems be attacked? Is it possible to assess if the data already collected has suffered from such attacks? How can we detect and mitigate this malicious content?

It is interesting, however, that in some papers the authors remove some bogus interactions~\cite{DBLP:conf/recsys/Palumbo0TB17}, as an indication that there is information that is better filtered out than left in the model, since these data points might influence the results. Nonetheless, a careful, detailed analysis of the impact of these points and how to detect them is still missing.

\subsubsection{Novel information sources, biases, and privacy}
As surveyed in Section~\ref{ss:informationSource}, the POI recommendation problem typically considers several information sources, however we believe even more information sources will be available in the future, and some of them are ubiquitous at the moment but remain unexplored in this domain.
For example, it is surprising the lack of works dealing with venue schedules; we attribute this to the fact that they tend to be used by optimization approaches which are more common when solving different problems, such as tour recommendation. In any case, we find it strange that they are not exploited for this (more simple) scenario, probably because of the difficulty to obtain trusted and consistent data.
Images, for example, about the POIs have been recently used to infer the preferences of users. %% 
Another paradigmatic example is the Internet-of-Things and all the sensors (such as beacons) that are increasingly common in cities and touristic venues. While specific approaches tailored for a small set of sensor-ready POIs are starting to emerge, general architectures or frameworks aimed at solving the problem at a more global scale, or even interacting with POIs with and without sensors is, to the best of our knowledge, not investigated at the moment, despite its obvious interest and potential to attract users.

However, whenever more user information is exploited, concerns about biases, privacy, and ethical issues should be considered. While parts of these problems have been addressed in the past for LBSN data, we believe it should be revisited according to the novel information sources that might be available, and also because new recommendation approaches may entail or generate different statistical or cultural biases; moreover, recent approaches such as differential privacy aiming to provide valuable personalized recommendations while withholding sensitive information from the users could be an interesting solution for this type of systems~\cite{DBLP:conf/wsdm/YangZ18}.

\subsubsection{Counterfactuals and translation into the real world}
How does all this research translate to the real world with respect to the data?
As we presented in Section~\ref{s:soa_data}, most datasets used in the experimental works come from a limited number of LBSNs (mostly, Foursquare), which may indicate a bias towards the requirements and needs from those systems.
For instance, most works do not gather this information directly from Foursquare but through Twitter~\cite{DBLP:journals/tist/YangZQ16}, evidencing the limitations of the collected data, which are probably incomplete and not uniform across the population of the LBSN under study.

Moreover, as it is common in other problems in the RS domain, the information available only refers to what the user actually did, not all the options presented by the system nor the discarded alternatives.
In particular, this means that no negative information can be inferred, since only positive information (whenever there is an interaction between users and items) is recorded.
Once this type of information would be available, the computation of counterfactuals and definition of intervention policies would allow to better align offline experimentation with online results~\cite{DBLP:conf/wsdm/KicimanS19}.

\subsubsection{Adding constraints}
Constraint-based recommender systems are a family of recommendation approaches that are not among the most popular ones, because they have been applied in very few cases, since a deep knowledge of the domain is typically required~\cite{DBLP:reference/sp/FelfernigFJZ15}.
However, we consider they may fit the POI recommendation problem since, often, the users face the RS with several restrictions or constraints: desired price of attractions, must-see venues, maximum length of the trip, and so on.
Moreover, under special circumstances -- such as an emergency situation or an unexpected crisis --, these constraints may be dictated by the venue owners or even the regional or national authorities; thus, it may become mandatory to satisfy such requirements.
In this context, we foresee novel approaches that allow POI recommendation algorithms to incorporate constraints and adapt their suggestions to these varying conditions, perhaps by exploiting optimization techniques used for the tourist trip design problem~\cite{DBLP:journals/heuristics/GavalasKMP14}.

\subsubsection{Scalability and efficiency}
Last but not least, one major drawback of most of the approaches surveyed in this review is that they exhibit expensive computational costs.
This is because considering additional information dimensions, beyond the user-item interaction matrix, needs memory resources but also complex algorithms that are difficult to scale and execute efficiently.
Therefore, a promising research direction would consist on defining approximated versions of well-known algorithms that could manage large amounts of multi-dimensional data, such as geographical, social, content, and user-item interactions -- the most common information sources.

\begin{acks}
This work has been funded by the Ministerio de Ciencia e Innovaci\'on reference (PID2019-108965GB-I00) and by the European Social Fund (ESF), within the 2017 call for predoctoral contracts.
The authors thank the reviewers for their thoughtful comments and suggestions.
\end{acks}

\bibliographystyle{ACM-Reference-Format}

\bibliography{complete_bibliography_condensed}

%%% -*-BibTeX-*-
%%% Do NOT edit. File created by BibTeX with style
%%% ACM-Reference-Format-Journals [18-Jan-2012].

\begin{thebibliography}{149}

%%% ====================================================================
%%% NOTE TO THE USER: you can override these defaults by providing
%%% customized versions of any of these macros before the \bibliography
%%% command.  Each of them MUST provide its own final punctuation,
%%% except for \shownote{}, \showDOI{}, and \showURL{}.  The latter two
%%% do not use final punctuation, in order to avoid confusing it with
%%% the Web address.
%%%
%%% To suppress output of a particular field, define its macro to expand
%%% to an empty string, or better, \unskip, like this:
%%%
%%% \newcommand{\showDOI}[1]{\unskip}   % LaTeX syntax
%%%
%%% \def \showDOI #1{\unskip}           % plain TeX syntax
%%%
%%% ====================================================================

\ifx \showCODEN    \undefined \def \showCODEN     #1{\unskip}     \fi
\ifx \showDOI      \undefined \def \showDOI       #1{#1}\fi
\ifx \showISBNx    \undefined \def \showISBNx     #1{\unskip}     \fi
\ifx \showISBNxiii \undefined \def \showISBNxiii  #1{\unskip}     \fi
\ifx \showISSN     \undefined \def \showISSN      #1{\unskip}     \fi
\ifx \showLCCN     \undefined \def \showLCCN      #1{\unskip}     \fi
\ifx \shownote     \undefined \def \shownote      #1{#1}          \fi
\ifx \showarticletitle \undefined \def \showarticletitle #1{#1}   \fi
\ifx \showURL      \undefined \def \showURL       {\relax}        \fi
% The following commands are used for tagged output and should be
% invisible to TeX
\providecommand\bibfield[2]{#2}
\providecommand\bibinfo[2]{#2}
\providecommand\natexlab[1]{#1}
\providecommand\showeprint[2][]{arXiv:#2}

\bibitem[\protect\citeauthoryear{Abdel{-}Fatao, Li, and Liu}{Abdel{-}Fatao
  et~al\mbox{.}}{2015}]%
        {DBLP:conf/adc/Abdel-FataoLL15}
\bibfield{author}{\bibinfo{person}{Hamidu Abdel{-}Fatao},
  \bibinfo{person}{Jiuyong Li}, {and} \bibinfo{person}{Jixue Liu}.}
  \bibinfo{year}{2015}\natexlab{}.
\newblock \showarticletitle{Unifying Spatial, Temporal and Semantic Features
  for an Effective {GPS} Trajectory-Based Location Recommendation}. In
  \bibinfo{booktitle}{\emph{{ADC}}} \emph{(\bibinfo{series}{Lecture Notes in
  Computer Science}, Vol.~\bibinfo{volume}{9093})}.
  \bibinfo{publisher}{Springer}, \bibinfo{pages}{41--53}.
\newblock


\bibitem[\protect\citeauthoryear{Adomavicius and Tuzhilin}{Adomavicius and
  Tuzhilin}{2005}]%
        {DBLP:journals/tkde/AdomaviciusT05}
\bibfield{author}{\bibinfo{person}{Gediminas Adomavicius} {and}
  \bibinfo{person}{Alexander Tuzhilin}.} \bibinfo{year}{2005}\natexlab{}.
\newblock \showarticletitle{Toward the Next Generation of Recommender Systems:
  {A} Survey of the State-of-the-Art and Possible Extensions}.
\newblock \bibinfo{journal}{\emph{{IEEE} Trans. Knowl. Data Eng.}}
  \bibinfo{volume}{17}, \bibinfo{number}{6} (\bibinfo{year}{2005}),
  \bibinfo{pages}{734--749}.
\newblock


\bibitem[\protect\citeauthoryear{Armstrong, Moffat, Webber, and
  Zobel}{Armstrong et~al\mbox{.}}{2009}]%
        {DBLP:conf/cikm/ArmstrongMWZ09}
\bibfield{author}{\bibinfo{person}{Timothy~G. Armstrong},
  \bibinfo{person}{Alistair Moffat}, \bibinfo{person}{William Webber}, {and}
  \bibinfo{person}{Justin Zobel}.} \bibinfo{year}{2009}\natexlab{}.
\newblock \showarticletitle{Improvements that don't add up: ad-hoc retrieval
  results since 1998}. In \bibinfo{booktitle}{\emph{{CIKM}}}.
  \bibinfo{publisher}{{ACM}}, \bibinfo{pages}{601--610}.
\newblock


\bibitem[\protect\citeauthoryear{Aujla, Jindal, Chaudhary, Kumar, Vashist,
  Sharma, and Obaidat}{Aujla et~al\mbox{.}}{2019}]%
        {DBLP:conf/icc/AujlaJC0VSO19}
\bibfield{author}{\bibinfo{person}{Gagangeet~Singh Aujla},
  \bibinfo{person}{Anish Jindal}, \bibinfo{person}{Rajat Chaudhary},
  \bibinfo{person}{Neeraj Kumar}, \bibinfo{person}{Sahil Vashist},
  \bibinfo{person}{Neeraj Sharma}, {and} \bibinfo{person}{Mohammad~S.
  Obaidat}.} \bibinfo{year}{2019}\natexlab{}.
\newblock \showarticletitle{{DLRS:} Deep Learning-Based Recommender System for
  Smart Healthcare Ecosystem}. In \bibinfo{booktitle}{\emph{{ICC}}}.
  \bibinfo{publisher}{{IEEE}}, \bibinfo{pages}{1--6}.
\newblock


\bibitem[\protect\citeauthoryear{Baeza{-}Yates and
  Ribeiro{-}Neto}{Baeza{-}Yates and Ribeiro{-}Neto}{2011}]%
        {DBLP:books/aw/Baeza-YatesR2011}
\bibfield{author}{\bibinfo{person}{Ricardo~A. Baeza{-}Yates} {and}
  \bibinfo{person}{Berthier~A. Ribeiro{-}Neto}.}
  \bibinfo{year}{2011}\natexlab{}.
\newblock \bibinfo{booktitle}{\emph{Modern Information Retrieval - the concepts
  and technology behind search, Second edition}}.
\newblock \bibinfo{publisher}{Pearson Education Ltd., Harlow, England}.
\newblock


\bibitem[\protect\citeauthoryear{Balabanovic and Shoham}{Balabanovic and
  Shoham}{1997}]%
        {DBLP:journals/cacm/BalabanovicS97}
\bibfield{author}{\bibinfo{person}{Marko Balabanovic} {and}
  \bibinfo{person}{Yoav Shoham}.} \bibinfo{year}{1997}\natexlab{}.
\newblock \showarticletitle{Content-Based, Collaborative Recommendation}.
\newblock \bibinfo{journal}{\emph{Commun. {ACM}}} \bibinfo{volume}{40},
  \bibinfo{number}{3} (\bibinfo{year}{1997}), \bibinfo{pages}{66--72}.
\newblock


\bibitem[\protect\citeauthoryear{Bao, Zheng, and Mokbel}{Bao
  et~al\mbox{.}}{2012}]%
        {DBLP:conf/gis/0003ZM12}
\bibfield{author}{\bibinfo{person}{Jie Bao}, \bibinfo{person}{Yu Zheng}, {and}
  \bibinfo{person}{Mohamed~F. Mokbel}.} \bibinfo{year}{2012}\natexlab{}.
\newblock \showarticletitle{Location-based and preference-aware recommendation
  using sparse geo-social networking data}. In
  \bibinfo{booktitle}{\emph{{SIGSPATIAL/GIS}}}. \bibinfo{publisher}{{ACM}},
  \bibinfo{pages}{199--208}.
\newblock


\bibitem[\protect\citeauthoryear{Bao, Zheng, Wilkie, and Mokbel}{Bao
  et~al\mbox{.}}{2015}]%
        {DBLP:journals/geoinformatica/0003ZWM15}
\bibfield{author}{\bibinfo{person}{Jie Bao}, \bibinfo{person}{Yu Zheng},
  \bibinfo{person}{David Wilkie}, {and} \bibinfo{person}{Mohamed~F. Mokbel}.}
  \bibinfo{year}{2015}\natexlab{}.
\newblock \showarticletitle{Recommendations in location-based social networks:
  a survey}.
\newblock \bibinfo{journal}{\emph{GeoInformatica}} \bibinfo{volume}{19},
  \bibinfo{number}{3} (\bibinfo{year}{2015}), \bibinfo{pages}{525--565}.
\newblock


\bibitem[\protect\citeauthoryear{Baral and Li}{Baral and Li}{2016}]%
        {DBLP:conf/recsys/BaralL16}
\bibfield{author}{\bibinfo{person}{Ramesh Baral} {and} \bibinfo{person}{Tao
  Li}.} \bibinfo{year}{2016}\natexlab{}.
\newblock \showarticletitle{{MAPS:} {A} Multi Aspect Personalized {POI}
  Recommender System}. In \bibinfo{booktitle}{\emph{RecSys}}.
  \bibinfo{publisher}{{ACM}}, \bibinfo{pages}{281--284}.
\newblock


\bibitem[\protect\citeauthoryear{Baral, Wang, Li, and Chen}{Baral
  et~al\mbox{.}}{2016}]%
        {DBLP:conf/iri/BaralWLC16}
\bibfield{author}{\bibinfo{person}{Ramesh Baral}, \bibinfo{person}{Dingding
  Wang}, \bibinfo{person}{Tao Li}, {and} \bibinfo{person}{Shu{-}Ching Chen}.}
  \bibinfo{year}{2016}\natexlab{}.
\newblock \showarticletitle{GeoTeCS: Exploiting Geographical, Temporal,
  Categorical and Social Aspects for Personalized {POI} Recommendation (Invited
  Paper)}. In \bibinfo{booktitle}{\emph{{IRI}}}. \bibinfo{publisher}{{IEEE}
  Computer Society}, \bibinfo{pages}{94--101}.
\newblock


\bibitem[\protect\citeauthoryear{Baral, Zhu, Iyengar, and Li}{Baral
  et~al\mbox{.}}{2018}]%
        {DBLP:conf/um/BaralZIL18}
\bibfield{author}{\bibinfo{person}{Ramesh Baral}, \bibinfo{person}{Xiaolong
  Zhu}, \bibinfo{person}{S.~S. Iyengar}, {and} \bibinfo{person}{Tao Li}.}
  \bibinfo{year}{2018}\natexlab{}.
\newblock \showarticletitle{ReEL: {R} eview Aware Explanation of Location
  Recommendation}. In \bibinfo{booktitle}{\emph{{UMAP}}}.
  \bibinfo{publisher}{{ACM}}, \bibinfo{pages}{23--32}.
\newblock


\bibitem[\protect\citeauthoryear{Bell and Koren}{Bell and Koren}{2007}]%
        {DBLP:journals/sigkdd/BellK07}
\bibfield{author}{\bibinfo{person}{Robert~M. Bell} {and}
  \bibinfo{person}{Yehuda Koren}.} \bibinfo{year}{2007}\natexlab{}.
\newblock \showarticletitle{Lessons from the Netflix prize challenge}.
\newblock \bibinfo{journal}{\emph{{SIGKDD} Explorations}} \bibinfo{volume}{9},
  \bibinfo{number}{2} (\bibinfo{year}{2007}), \bibinfo{pages}{75--79}.
\newblock


\bibitem[\protect\citeauthoryear{Bellog{\'{\i}}n, Castells, and
  Cantador}{Bellog{\'{\i}}n et~al\mbox{.}}{2017}]%
        {DBLP:journals/ir/BelloginCC17}
\bibfield{author}{\bibinfo{person}{Alejandro Bellog{\'{\i}}n},
  \bibinfo{person}{Pablo Castells}, {and} \bibinfo{person}{Iv{\'{a}}n
  Cantador}.} \bibinfo{year}{2017}\natexlab{}.
\newblock \showarticletitle{Statistical biases in Information Retrieval metrics
  for recommender systems}.
\newblock \bibinfo{journal}{\emph{Inf. Retr. Journal}} \bibinfo{volume}{20},
  \bibinfo{number}{6} (\bibinfo{year}{2017}), \bibinfo{pages}{606--634}.
\newblock


\bibitem[\protect\citeauthoryear{Bernardi, Estevez, Eidis, and Osama}{Bernardi
  et~al\mbox{.}}{2020}]%
        {DBLP:conf/recsys/BernardiEEO20}
\bibfield{author}{\bibinfo{person}{Lucas Bernardi}, \bibinfo{person}{Pablo
  Estevez}, \bibinfo{person}{Matias Eidis}, {and} \bibinfo{person}{Eqbal
  Osama}.} \bibinfo{year}{2020}\natexlab{}.
\newblock \showarticletitle{Recommending Accommodation Filters with Online
  Learning}. In \bibinfo{booktitle}{\emph{ORSUM@RecSys}}
  \emph{(\bibinfo{series}{{CEUR} Workshop Proceedings},
  Vol.~\bibinfo{volume}{2715})}. \bibinfo{publisher}{CEUR-WS.org}.
\newblock


\bibitem[\protect\citeauthoryear{Bothorel, Lathia, Picot{-}Cl{\'{e}}mente, and
  Noulas}{Bothorel et~al\mbox{.}}{2018}]%
        {DBLP:series/lncs/BothorelLPN18}
\bibfield{author}{\bibinfo{person}{C{\'{e}}cile Bothorel},
  \bibinfo{person}{Neal Lathia}, \bibinfo{person}{Romain
  Picot{-}Cl{\'{e}}mente}, {and} \bibinfo{person}{Anastasios Noulas}.}
  \bibinfo{year}{2018}\natexlab{}.
\newblock \showarticletitle{Location Recommendation with Social Media Data}.
\newblock In \bibinfo{booktitle}{\emph{Social Information Access}}.
  \bibinfo{series}{Lecture Notes in Computer Science},
  Vol.~\bibinfo{volume}{10100}. \bibinfo{publisher}{Springer},
  \bibinfo{pages}{624--653}.
\newblock


\bibitem[\protect\citeauthoryear{Burke, O'Mahony, and Hurley}{Burke
  et~al\mbox{.}}{2015}]%
        {DBLP:reference/sp/BurkeOH15}
\bibfield{author}{\bibinfo{person}{Robin Burke}, \bibinfo{person}{Michael~P.
  O'Mahony}, {and} \bibinfo{person}{Neil~J. Hurley}.}
  \bibinfo{year}{2015}\natexlab{}.
\newblock \showarticletitle{Robust Collaborative Recommendation}.
\newblock In \bibinfo{booktitle}{\emph{Recommender Systems Handbook}}.
  \bibinfo{publisher}{Springer}, \bibinfo{pages}{961--995}.
\newblock


\bibitem[\protect\citeauthoryear{Burke}{Burke}{2007}]%
        {DBLP:conf/adaptive/Burke07}
\bibfield{author}{\bibinfo{person}{Robin~D. Burke}.}
  \bibinfo{year}{2007}\natexlab{}.
\newblock \showarticletitle{Hybrid Web Recommender Systems}. In
  \bibinfo{booktitle}{\emph{The Adaptive Web}} \emph{(\bibinfo{series}{Lecture
  Notes in Computer Science}, Vol.~\bibinfo{volume}{4321})}.
  \bibinfo{publisher}{Springer}, \bibinfo{pages}{377--408}.
\newblock


\bibitem[\protect\citeauthoryear{Cai, Zheng, and Chang}{Cai
  et~al\mbox{.}}{2018}]%
        {DBLP:journals/tkde/CaiZC18}
\bibfield{author}{\bibinfo{person}{HongYun Cai}, \bibinfo{person}{Vincent~W.
  Zheng}, {and} \bibinfo{person}{Kevin~Chen{-}Chuan Chang}.}
  \bibinfo{year}{2018}\natexlab{}.
\newblock \showarticletitle{A Comprehensive Survey of Graph Embedding:
  Problems, Techniques, and Applications}.
\newblock \bibinfo{journal}{\emph{{IEEE} Trans. Knowl. Data Eng.}}
  \bibinfo{volume}{30}, \bibinfo{number}{9} (\bibinfo{year}{2018}),
  \bibinfo{pages}{1616--1637}.
\newblock


\bibitem[\protect\citeauthoryear{Campos, D{\'{\i}}ez, and Cantador}{Campos
  et~al\mbox{.}}{2014}]%
        {DBLP:journals/umuai/CamposDC14}
\bibfield{author}{\bibinfo{person}{Pedro~G. Campos}, \bibinfo{person}{Fernando
  D{\'{\i}}ez}, {and} \bibinfo{person}{Iv{\'{a}}n Cantador}.}
  \bibinfo{year}{2014}\natexlab{}.
\newblock \showarticletitle{Time-aware recommender systems: a comprehensive
  survey and analysis of existing evaluation protocols}.
\newblock \bibinfo{journal}{\emph{User Model. User-Adapt. Interact.}}
  \bibinfo{volume}{24}, \bibinfo{number}{1-2} (\bibinfo{year}{2014}),
  \bibinfo{pages}{67--119}.
\newblock


\bibitem[\protect\citeauthoryear{Cantador, Bellog{\'{\i}}n, and
  Vallet}{Cantador et~al\mbox{.}}{2010}]%
        {DBLP:conf/recsys/CantadorBV10}
\bibfield{author}{\bibinfo{person}{Iv{\'{a}}n Cantador},
  \bibinfo{person}{Alejandro Bellog{\'{\i}}n}, {and} \bibinfo{person}{David
  Vallet}.} \bibinfo{year}{2010}\natexlab{}.
\newblock \showarticletitle{Content-based recommendation in social tagging
  systems}. In \bibinfo{booktitle}{\emph{RecSys}}. \bibinfo{publisher}{{ACM}},
  \bibinfo{pages}{237--240}.
\newblock


\bibitem[\protect\citeauthoryear{Cao, Guo, Meng, Liu, Liu, and Li}{Cao
  et~al\mbox{.}}{2020}]%
        {DBLP:journals/access/CaoGMLLL20}
\bibfield{author}{\bibinfo{person}{Keyan Cao}, \bibinfo{person}{Jingjing Guo},
  \bibinfo{person}{Gongjie Meng}, \bibinfo{person}{Haoli Liu},
  \bibinfo{person}{Yefan Liu}, {and} \bibinfo{person}{Gui Li}.}
  \bibinfo{year}{2020}\natexlab{}.
\newblock \showarticletitle{Points-of-Interest Recommendation Algorithm Based
  on {LBSN} in Edge Computing Environment}.
\newblock \bibinfo{journal}{\emph{{IEEE} Access}}  \bibinfo{volume}{8}
  (\bibinfo{year}{2020}), \bibinfo{pages}{47973--47983}.
\newblock


\bibitem[\protect\citeauthoryear{Castells, Hurley, and Vargas}{Castells
  et~al\mbox{.}}{2015}]%
        {DBLP:reference/sp/CastellsHV15}
\bibfield{author}{\bibinfo{person}{Pablo Castells}, \bibinfo{person}{Neil~J.
  Hurley}, {and} \bibinfo{person}{Saul Vargas}.}
  \bibinfo{year}{2015}\natexlab{}.
\newblock \showarticletitle{Novelty and Diversity in Recommender Systems}.
\newblock In \bibinfo{booktitle}{\emph{Recommender Systems Handbook}}.
  \bibinfo{publisher}{Springer}, \bibinfo{pages}{881--918}.
\newblock


\bibitem[\protect\citeauthoryear{Chen, Ong, and Xie}{Chen
  et~al\mbox{.}}{2016b}]%
        {DBLP:conf/cikm/ChenOX16}
\bibfield{author}{\bibinfo{person}{Dawei Chen}, \bibinfo{person}{Cheng~Soon
  Ong}, {and} \bibinfo{person}{Lexing Xie}.} \bibinfo{year}{2016}\natexlab{b}.
\newblock \showarticletitle{Learning Points and Routes to Recommend
  Trajectories}. In \bibinfo{booktitle}{\emph{{CIKM}}}.
  \bibinfo{publisher}{{ACM}}, \bibinfo{pages}{2227--2232}.
\newblock


\bibitem[\protect\citeauthoryear{Chen, Zeng, Cong, Qin, Xiang, and Dai}{Chen
  et~al\mbox{.}}{2015}]%
        {DBLP:conf/aaai/ChenZCQXD15}
\bibfield{author}{\bibinfo{person}{Xuefeng Chen}, \bibinfo{person}{Yifeng
  Zeng}, \bibinfo{person}{Gao Cong}, \bibinfo{person}{Shengchao Qin},
  \bibinfo{person}{Yanping Xiang}, {and} \bibinfo{person}{Yuanshun Dai}.}
  \bibinfo{year}{2015}\natexlab{}.
\newblock \showarticletitle{On Information Coverage for Location Category Based
  Point-of-Interest Recommendation}. In \bibinfo{booktitle}{\emph{{AAAI}}}.
  \bibinfo{publisher}{{AAAI} Press}, \bibinfo{pages}{37--43}.
\newblock


\bibitem[\protect\citeauthoryear{Chen, Li, Li, Liu, and Xu}{Chen
  et~al\mbox{.}}{2016a}]%
        {DBLP:conf/dasfaa/ChenLLLX16}
\bibfield{author}{\bibinfo{person}{Yan Chen}, \bibinfo{person}{Xin Li},
  \bibinfo{person}{Lin Li}, \bibinfo{person}{Guiquan Liu}, {and}
  \bibinfo{person}{Guangdong Xu}.} \bibinfo{year}{2016}\natexlab{a}.
\newblock \showarticletitle{Modeling User Mobility via User Psychological and
  Geographical Behaviors Towards Point of-Interest Recommendation}. In
  \bibinfo{booktitle}{\emph{{DASFAA} {(1)}}} \emph{(\bibinfo{series}{Lecture
  Notes in Computer Science}, Vol.~\bibinfo{volume}{9642})}.
  \bibinfo{publisher}{Springer}, \bibinfo{pages}{364--380}.
\newblock


\bibitem[\protect\citeauthoryear{Cheng, Yang, King, and Lyu}{Cheng
  et~al\mbox{.}}{2016}]%
        {DBLP:journals/tist/ChengYKL16}
\bibfield{author}{\bibinfo{person}{Chen Cheng}, \bibinfo{person}{Haiqin Yang},
  \bibinfo{person}{Irwin King}, {and} \bibinfo{person}{Michael~R. Lyu}.}
  \bibinfo{year}{2016}\natexlab{}.
\newblock \showarticletitle{A Unified Point-of-Interest Recommendation
  Framework in Location-Based Social Networks}.
\newblock \bibinfo{journal}{\emph{{ACM} {TIST}}} \bibinfo{volume}{8},
  \bibinfo{number}{1} (\bibinfo{year}{2016}), \bibinfo{pages}{10:1--10:21}.
\newblock


\bibitem[\protect\citeauthoryear{Cheng, Yang, Lyu, and King}{Cheng
  et~al\mbox{.}}{2013}]%
        {DBLP:conf/ijcai/ChengYLK13}
\bibfield{author}{\bibinfo{person}{Chen Cheng}, \bibinfo{person}{Haiqin Yang},
  \bibinfo{person}{Michael~R. Lyu}, {and} \bibinfo{person}{Irwin King}.}
  \bibinfo{year}{2013}\natexlab{}.
\newblock \showarticletitle{Where You Like to Go Next: Successive
  Point-of-Interest Recommendation}. In \bibinfo{booktitle}{\emph{{IJCAI}}}.
  \bibinfo{publisher}{{IJCAI/AAAI}}, \bibinfo{pages}{2605--2611}.
\newblock


\bibitem[\protect\citeauthoryear{Cheng and Chang}{Cheng and Chang}{2013}]%
        {DBLP:conf/icdm/ChengC13}
\bibfield{author}{\bibinfo{person}{Nai{-}Hung Cheng} {and}
  \bibinfo{person}{Chia{-}Hui Chang}.} \bibinfo{year}{2013}\natexlab{}.
\newblock \showarticletitle{Evaluation of Social, Geography, Location Effects
  for Point-of-Interest Recommendation}. In \bibinfo{booktitle}{\emph{{ICDM}
  Workshops}}. \bibinfo{publisher}{{IEEE} Computer Society},
  \bibinfo{pages}{766--772}.
\newblock


\bibitem[\protect\citeauthoryear{Cho, Myers, and Leskovec}{Cho
  et~al\mbox{.}}{2011}]%
        {DBLP:conf/kdd/ChoML11}
\bibfield{author}{\bibinfo{person}{Eunjoon Cho}, \bibinfo{person}{Seth~A.
  Myers}, {and} \bibinfo{person}{Jure Leskovec}.}
  \bibinfo{year}{2011}\natexlab{}.
\newblock \showarticletitle{Friendship and mobility: user movement in
  location-based social networks}. In \bibinfo{booktitle}{\emph{{KDD}}}.
  \bibinfo{publisher}{{ACM}}, \bibinfo{pages}{1082--1090}.
\newblock


\bibitem[\protect\citeauthoryear{Christoforidis, Kefalas, Papadopoulos, and
  Manolopoulos}{Christoforidis et~al\mbox{.}}{2018}]%
        {DBLP:conf/dsaa/ChristoforidisK18}
\bibfield{author}{\bibinfo{person}{Giannis Christoforidis},
  \bibinfo{person}{Pavlos Kefalas}, \bibinfo{person}{Apostolos Papadopoulos},
  {and} \bibinfo{person}{Yannis Manolopoulos}.}
  \bibinfo{year}{2018}\natexlab{}.
\newblock \showarticletitle{Recommendation of Points-of-Interest Using Graph
  Embeddings}. In \bibinfo{booktitle}{\emph{{DSAA}}}.
  \bibinfo{publisher}{{IEEE}}, \bibinfo{pages}{31--40}.
\newblock


\bibitem[\protect\citeauthoryear{Christoforidis, Kefalas, Papadopoulos, and
  Manolopoulos}{Christoforidis et~al\mbox{.}}{2019}]%
        {DBLP:conf/inista/ChristoforidisK19}
\bibfield{author}{\bibinfo{person}{Giannis Christoforidis},
  \bibinfo{person}{Pavlos Kefalas}, \bibinfo{person}{Apostolos~N.
  Papadopoulos}, {and} \bibinfo{person}{Yannis Manolopoulos}.}
  \bibinfo{year}{2019}\natexlab{}.
\newblock \showarticletitle{Recommending Points of Interest in LBSNs Using Deep
  Learning Techniques}. In \bibinfo{booktitle}{\emph{{INISTA}}}.
  \bibinfo{publisher}{{IEEE}}, \bibinfo{pages}{1--6}.
\newblock


\bibitem[\protect\citeauthoryear{Chuang, Junhao, and Shun}{Chuang
  et~al\mbox{.}}{2019}]%
        {Song2019}
\bibfield{author}{\bibinfo{person}{Song Chuang}, \bibinfo{person}{Wen Junhao},
  {and} \bibinfo{person}{Li Shun}.} \bibinfo{year}{2019}\natexlab{}.
\newblock \showarticletitle{Personalized POI Recommendation Based on Check-in
  Data and Geographical-Regional Influence}. In
  \bibinfo{booktitle}{\emph{Proceedings of the 3rd International Conference on
  Machine Learning and Soft Computing}}. \bibinfo{publisher}{Association for
  Computing Machinery}, \bibinfo{pages}{128--133}.
\newblock
\showISBNx{9781450366120}


\bibitem[\protect\citeauthoryear{de~Gemmis, Lops, Musto, Narducci, and
  Semeraro}{de~Gemmis et~al\mbox{.}}{2015}]%
        {DBLP:reference/sp/GemmisLMNS15}
\bibfield{author}{\bibinfo{person}{Marco de Gemmis}, \bibinfo{person}{Pasquale
  Lops}, \bibinfo{person}{Cataldo Musto}, \bibinfo{person}{Fedelucio Narducci},
  {and} \bibinfo{person}{Giovanni Semeraro}.} \bibinfo{year}{2015}\natexlab{}.
\newblock \showarticletitle{Semantics-Aware Content-Based Recommender Systems}.
\newblock In \bibinfo{booktitle}{\emph{Recommender Systems Handbook}}.
  \bibinfo{publisher}{Springer}, \bibinfo{pages}{119--159}.
\newblock


\bibitem[\protect\citeauthoryear{Deldjoo, Noia, and Merra}{Deldjoo
  et~al\mbox{.}}{2020}]%
        {DBLP:conf/wsdm/DeldjooNM20}
\bibfield{author}{\bibinfo{person}{Yashar Deldjoo}, \bibinfo{person}{Tommaso~Di
  Noia}, {and} \bibinfo{person}{Felice~Antonio Merra}.}
  \bibinfo{year}{2020}\natexlab{}.
\newblock \showarticletitle{Adversarial Machine Learning in Recommender Systems
  (AML-RecSys)}. In \bibinfo{booktitle}{\emph{{WSDM}}}.
  \bibinfo{publisher}{{ACM}}, \bibinfo{pages}{869--872}.
\newblock


\bibitem[\protect\citeauthoryear{Dietz, Roy, and W{\"{o}}rndl}{Dietz
  et~al\mbox{.}}{2019}]%
        {DBLP:conf/enter/DietzRW19}
\bibfield{author}{\bibinfo{person}{Linus~W. Dietz}, \bibinfo{person}{Rinita
  Roy}, {and} \bibinfo{person}{Wolfgang W{\"{o}}rndl}.}
  \bibinfo{year}{2019}\natexlab{}.
\newblock \showarticletitle{Characterisation of Traveller Types Using Check-In
  Data from Location-Based Social Networks}. In
  \bibinfo{booktitle}{\emph{{ENTER}}}. \bibinfo{publisher}{Springer},
  \bibinfo{pages}{15--26}.
\newblock


\bibitem[\protect\citeauthoryear{Ding, Li, Jiang, and Zhou}{Ding
  et~al\mbox{.}}{2018}]%
        {DBLP:journals/csur/DingLJZ18}
\bibfield{author}{\bibinfo{person}{Zhijun Ding}, \bibinfo{person}{Xiaolun Li},
  \bibinfo{person}{Changjun Jiang}, {and} \bibinfo{person}{Mengchu Zhou}.}
  \bibinfo{year}{2018}\natexlab{}.
\newblock \showarticletitle{Objectives and State-of-the-Art of Location-Based
  Social Network Recommender Systems}.
\newblock \bibinfo{journal}{\emph{{ACM} Comput. Surv.}} \bibinfo{volume}{51},
  \bibinfo{number}{1} (\bibinfo{year}{2018}), \bibinfo{pages}{18:1--18:28}.
\newblock


\bibitem[\protect\citeauthoryear{Felfernig, Friedrich, Jannach, and
  Zanker}{Felfernig et~al\mbox{.}}{2015}]%
        {DBLP:reference/sp/FelfernigFJZ15}
\bibfield{author}{\bibinfo{person}{Alexander Felfernig},
  \bibinfo{person}{Gerhard Friedrich}, \bibinfo{person}{Dietmar Jannach}, {and}
  \bibinfo{person}{Markus Zanker}.} \bibinfo{year}{2015}\natexlab{}.
\newblock \showarticletitle{Constraint-Based Recommender Systems}.
\newblock In \bibinfo{booktitle}{\emph{Recommender Systems Handbook}}.
  \bibinfo{publisher}{Springer}, \bibinfo{pages}{161--190}.
\newblock


\bibitem[\protect\citeauthoryear{Feng, Li, Zeng, Cong, Chee, and Yuan}{Feng
  et~al\mbox{.}}{2015}]%
        {DBLP:conf/ijcai/FengLZCCY15}
\bibfield{author}{\bibinfo{person}{Shanshan Feng}, \bibinfo{person}{Xutao Li},
  \bibinfo{person}{Yifeng Zeng}, \bibinfo{person}{Gao Cong},
  \bibinfo{person}{Yeow~Meng Chee}, {and} \bibinfo{person}{Quan Yuan}.}
  \bibinfo{year}{2015}\natexlab{}.
\newblock \showarticletitle{Personalized Ranking Metric Embedding for Next New
  {POI} Recommendation}. In \bibinfo{booktitle}{\emph{{IJCAI}}}.
  \bibinfo{publisher}{{AAAI} Press}, \bibinfo{pages}{2069--2075}.
\newblock


\bibitem[\protect\citeauthoryear{Forhad, Arefin, Kayes, Ahmed, Chowdhury, and
  Kumara}{Forhad et~al\mbox{.}}{2021}]%
        {electronics10161920}
\bibfield{author}{\bibinfo{person}{Md. Shafiul~Alam Forhad},
  \bibinfo{person}{Mohammad~Shamsul Arefin}, \bibinfo{person}{A.~S.~M. Kayes},
  \bibinfo{person}{Khandakar Ahmed}, \bibinfo{person}{Mohammad Jabed~Morshed
  Chowdhury}, {and} \bibinfo{person}{Indika Kumara}.}
  \bibinfo{year}{2021}\natexlab{}.
\newblock \showarticletitle{An Effective Hotel Recommendation System through
  Processing Heterogeneous Data}.
\newblock \bibinfo{journal}{\emph{Electronics}} \bibinfo{volume}{10},
  \bibinfo{number}{16} (\bibinfo{year}{2021}).
\newblock
\showISSN{2079-9292}


\bibitem[\protect\citeauthoryear{Gao, Tang, Hu, and Liu}{Gao
  et~al\mbox{.}}{2013}]%
        {DBLP:conf/recsys/GaoTHL13}
\bibfield{author}{\bibinfo{person}{Huiji Gao}, \bibinfo{person}{Jiliang Tang},
  \bibinfo{person}{Xia Hu}, {and} \bibinfo{person}{Huan Liu}.}
  \bibinfo{year}{2013}\natexlab{}.
\newblock \showarticletitle{Exploring temporal effects for location
  recommendation on location-based social networks}. In
  \bibinfo{booktitle}{\emph{RecSys}}. \bibinfo{publisher}{{ACM}},
  \bibinfo{pages}{93--100}.
\newblock


\bibitem[\protect\citeauthoryear{Gao, Tang, Hu, and Liu}{Gao
  et~al\mbox{.}}{2015}]%
        {DBLP:conf/aaai/GaoTHL15}
\bibfield{author}{\bibinfo{person}{Huiji Gao}, \bibinfo{person}{Jiliang Tang},
  \bibinfo{person}{Xia Hu}, {and} \bibinfo{person}{Huan Liu}.}
  \bibinfo{year}{2015}\natexlab{}.
\newblock \showarticletitle{Content-Aware Point of Interest Recommendation on
  Location-Based Social Networks}. In \bibinfo{booktitle}{\emph{{AAAI}}}.
  \bibinfo{publisher}{{AAAI} Press}, \bibinfo{pages}{1721--1727}.
\newblock


\bibitem[\protect\citeauthoryear{Gao, Tang, and Liu}{Gao et~al\mbox{.}}{2012}]%
        {DBLP:conf/cikm/GaoTL12}
\bibfield{author}{\bibinfo{person}{Huiji Gao}, \bibinfo{person}{Jiliang Tang},
  {and} \bibinfo{person}{Huan Liu}.} \bibinfo{year}{2012}\natexlab{}.
\newblock \showarticletitle{gSCorr: modeling geo-social correlations for new
  check-ins on location-based social networks}. In
  \bibinfo{booktitle}{\emph{{CIKM}}}. \bibinfo{publisher}{{ACM}},
  \bibinfo{pages}{1582--1586}.
\newblock


\bibitem[\protect\citeauthoryear{Gao, Li, Li, Song, and Zhou}{Gao
  et~al\mbox{.}}{2018}]%
        {DBLP:journals/ijon/GaoLLSZ18}
\bibfield{author}{\bibinfo{person}{Rong Gao}, \bibinfo{person}{Jing Li},
  \bibinfo{person}{Xuefei Li}, \bibinfo{person}{Chengfang Song}, {and}
  \bibinfo{person}{Yifei Zhou}.} \bibinfo{year}{2018}\natexlab{}.
\newblock \showarticletitle{A personalized point-of-interest recommendation
  model via fusion of geo-social information}.
\newblock \bibinfo{journal}{\emph{Neurocomputing}}  \bibinfo{volume}{273}
  (\bibinfo{year}{2018}), \bibinfo{pages}{159--170}.
\newblock


\bibitem[\protect\citeauthoryear{Gavalas, Konstantopoulos, Mastakas, and
  Pantziou}{Gavalas et~al\mbox{.}}{2014}]%
        {DBLP:journals/heuristics/GavalasKMP14}
\bibfield{author}{\bibinfo{person}{Damianos Gavalas},
  \bibinfo{person}{Charalampos Konstantopoulos}, \bibinfo{person}{Konstantinos
  Mastakas}, {and} \bibinfo{person}{Grammati~E. Pantziou}.}
  \bibinfo{year}{2014}\natexlab{}.
\newblock \showarticletitle{A survey on algorithmic approaches for solving
  tourist trip design problems}.
\newblock \bibinfo{journal}{\emph{J. Heuristics}} \bibinfo{volume}{20},
  \bibinfo{number}{3} (\bibinfo{year}{2014}), \bibinfo{pages}{291--328}.
\newblock


\bibitem[\protect\citeauthoryear{Geng, Jiao, Gong, Li, and Wu}{Geng
  et~al\mbox{.}}{2019}]%
        {DBLP:journals/isci/GengJGLW19}
\bibfield{author}{\bibinfo{person}{Bingrui Geng}, \bibinfo{person}{Licheng
  Jiao}, \bibinfo{person}{Maoguo Gong}, \bibinfo{person}{Lingling Li}, {and}
  \bibinfo{person}{Yue Wu}.} \bibinfo{year}{2019}\natexlab{}.
\newblock \showarticletitle{A two-step personalized location recommendation
  based on multi-objective immune algorithm}.
\newblock \bibinfo{journal}{\emph{Inf. Sci.}}  \bibinfo{volume}{475}
  (\bibinfo{year}{2019}), \bibinfo{pages}{161--181}.
\newblock


\bibitem[\protect\citeauthoryear{Gunawardana and Shani}{Gunawardana and
  Shani}{2015}]%
        {DBLP:reference/sp/GunawardanaS15}
\bibfield{author}{\bibinfo{person}{Asela Gunawardana} {and}
  \bibinfo{person}{Guy Shani}.} \bibinfo{year}{2015}\natexlab{}.
\newblock \showarticletitle{Evaluating Recommender Systems}.
\newblock In \bibinfo{booktitle}{\emph{Recommender Systems Handbook}}.
  \bibinfo{publisher}{Springer}, \bibinfo{pages}{265--308}.
\newblock


\bibitem[\protect\citeauthoryear{Guo, Jiang, Liu, and Xing}{Guo
  et~al\mbox{.}}{2019a}]%
        {DBLP:journals/complexity/GuoJLX19}
\bibfield{author}{\bibinfo{person}{Lei Guo}, \bibinfo{person}{Haoran Jiang},
  \bibinfo{person}{Xiyu Liu}, {and} \bibinfo{person}{Changming Xing}.}
  \bibinfo{year}{2019}\natexlab{a}.
\newblock \showarticletitle{Network Embedding-Aware Point-of-Interest
  Recommendation in Location-Based Social Networks}.
\newblock \bibinfo{journal}{\emph{Complexity}}  \bibinfo{volume}{2019}
  (\bibinfo{year}{2019}), \bibinfo{pages}{3574194:1--3574194:18}.
\newblock


\bibitem[\protect\citeauthoryear{Guo, Huang, and Theng}{Guo
  et~al\mbox{.}}{2015}]%
        {DBLP:conf/webi/GuoHT15}
\bibfield{author}{\bibinfo{person}{Qing Guo}, \bibinfo{person}{Yi Huang}, {and}
  \bibinfo{person}{Yin{-}Leng Theng}.} \bibinfo{year}{2015}\natexlab{}.
\newblock \showarticletitle{Topic-Sensitive Location Recommendation with
  Spatial Awareness}. In \bibinfo{booktitle}{\emph{{WI-IAT} {(1)}}}.
  \bibinfo{publisher}{{IEEE} Computer Society}, \bibinfo{pages}{237--243}.
\newblock


\bibitem[\protect\citeauthoryear{Guo, Sun, Zhang, and Theng}{Guo
  et~al\mbox{.}}{2019b}]%
        {DBLP:conf/icwe/GuoSZT19}
\bibfield{author}{\bibinfo{person}{Qing Guo}, \bibinfo{person}{Zhu Sun},
  \bibinfo{person}{Jie Zhang}, {and} \bibinfo{person}{Yin{-}Leng Theng}.}
  \bibinfo{year}{2019}\natexlab{b}.
\newblock \showarticletitle{Modeling Heterogeneous Influences for
  Point-of-Interest Recommendation in Location-Based Social Networks}. In
  \bibinfo{booktitle}{\emph{{ICWE}}} \emph{(\bibinfo{series}{Lecture Notes in
  Computer Science}, Vol.~\bibinfo{volume}{11496})}.
  \bibinfo{publisher}{Springer}, \bibinfo{pages}{72--80}.
\newblock


\bibitem[\protect\citeauthoryear{Gupta, Pathak, and Mitra}{Gupta
  et~al\mbox{.}}{2015}]%
        {DBLP:conf/pakdd/GuptaPM15}
\bibfield{author}{\bibinfo{person}{Saurabh Gupta}, \bibinfo{person}{Sayan
  Pathak}, {and} \bibinfo{person}{Bivas Mitra}.}
  \bibinfo{year}{2015}\natexlab{}.
\newblock \showarticletitle{Complementary Usage of Tips and Reviews for
  Location Recommendation in Yelp}. In \bibinfo{booktitle}{\emph{{PAKDD}
  {(2)}}} \emph{(\bibinfo{series}{Lecture Notes in Computer Science},
  Vol.~\bibinfo{volume}{9078})}. \bibinfo{publisher}{Springer},
  \bibinfo{pages}{720--731}.
\newblock


\bibitem[\protect\citeauthoryear{He, Li, and Liao}{He et~al\mbox{.}}{2017}]%
        {DBLP:conf/ijcai/HeLL17}
\bibfield{author}{\bibinfo{person}{Jing He}, \bibinfo{person}{Xin Li}, {and}
  \bibinfo{person}{Lejian Liao}.} \bibinfo{year}{2017}\natexlab{}.
\newblock \showarticletitle{Category-aware Next Point-of-Interest
  Recommendation via Listwise Bayesian Personalized Ranking}. In
  \bibinfo{booktitle}{\emph{{IJCAI}}}. \bibinfo{publisher}{ijcai.org},
  \bibinfo{pages}{1837--1843}.
\newblock


\bibitem[\protect\citeauthoryear{He, Li, Liao, Song, and Cheung}{He
  et~al\mbox{.}}{2016}]%
        {DBLP:conf/aaai/HeLLSC16}
\bibfield{author}{\bibinfo{person}{Jing He}, \bibinfo{person}{Xin Li},
  \bibinfo{person}{Lejian Liao}, \bibinfo{person}{Dandan Song}, {and}
  \bibinfo{person}{William~K. Cheung}.} \bibinfo{year}{2016}\natexlab{}.
\newblock \showarticletitle{Inferring a Personalized Next Point-of-Interest
  Recommendation Model with Latent Behavior Patterns}. In
  \bibinfo{booktitle}{\emph{{AAAI}}}. \bibinfo{publisher}{{AAAI} Press},
  \bibinfo{pages}{137--143}.
\newblock


\bibitem[\protect\citeauthoryear{He and McAuley}{He and McAuley}{2016}]%
        {DBLP:conf/icdm/HeM16}
\bibfield{author}{\bibinfo{person}{Ruining He} {and} \bibinfo{person}{Julian
  McAuley}.} \bibinfo{year}{2016}\natexlab{}.
\newblock \showarticletitle{Fusing Similarity Models with Markov Chains for
  Sparse Sequential Recommendation}. In \bibinfo{booktitle}{\emph{{ICDM}}}.
  \bibinfo{publisher}{{IEEE}}, \bibinfo{pages}{191--200}.
\newblock


\bibitem[\protect\citeauthoryear{Hosseini and Li}{Hosseini and Li}{2016}]%
        {DBLP:conf/dasfaa/HosseiniL16}
\bibfield{author}{\bibinfo{person}{Saeid Hosseini} {and}
  \bibinfo{person}{Lei~Thor Li}.} \bibinfo{year}{2016}\natexlab{}.
\newblock \showarticletitle{Point-Of-Interest Recommendation Using Temporal
  Orientations of Users and Locations}. In \bibinfo{booktitle}{\emph{{DASFAA}
  {(1)}}} \emph{(\bibinfo{series}{Lecture Notes in Computer Science},
  Vol.~\bibinfo{volume}{9642})}. \bibinfo{publisher}{Springer},
  \bibinfo{pages}{330--347}.
\newblock


\bibitem[\protect\citeauthoryear{Hu and Ester}{Hu and Ester}{2014}]%
        {DBLP:conf/icdm/HuE14}
\bibfield{author}{\bibinfo{person}{Bo Hu} {and} \bibinfo{person}{Martin
  Ester}.} \bibinfo{year}{2014}\natexlab{}.
\newblock \showarticletitle{Social Topic Modeling for Point-of-Interest
  Recommendation in Location-Based Social Networks}. In
  \bibinfo{booktitle}{\emph{{ICDM}}}. \bibinfo{publisher}{{IEEE} Computer
  Society}, \bibinfo{pages}{845--850}.
\newblock


\bibitem[\protect\citeauthoryear{Huang, Gartner, Krisp, Raubal, and
  de~Weghe}{Huang et~al\mbox{.}}{2018}]%
        {DBLP:journals/jlbs/HuangGKRW18}
\bibfield{author}{\bibinfo{person}{Haosheng Huang}, \bibinfo{person}{Georg
  Gartner}, \bibinfo{person}{Jukka~Matthias Krisp}, \bibinfo{person}{Martin
  Raubal}, {and} \bibinfo{person}{Nico~Van de Weghe}.}
  \bibinfo{year}{2018}\natexlab{}.
\newblock \showarticletitle{Location based services: ongoing evolution and
  research agenda}.
\newblock \bibinfo{journal}{\emph{J. Locat. Based Serv.}} \bibinfo{volume}{12},
  \bibinfo{number}{2} (\bibinfo{year}{2018}), \bibinfo{pages}{63--93}.
\newblock


\bibitem[\protect\citeauthoryear{Huang, Ma, Liu, and Sangaiah}{Huang
  et~al\mbox{.}}{2020}]%
        {DBLP:journals/fgcs/HuangMLS20}
\bibfield{author}{\bibinfo{person}{Liwei Huang}, \bibinfo{person}{Yutao Ma},
  \bibinfo{person}{Yanbo Liu}, {and} \bibinfo{person}{Arun~Kumar Sangaiah}.}
  \bibinfo{year}{2020}\natexlab{}.
\newblock \showarticletitle{Multi-modal Bayesian embedding for
  point-of-interest recommendation on location-based cyber-physical-social
  networks}.
\newblock \bibinfo{journal}{\emph{Future Gener. Comput. Syst.}}
  \bibinfo{volume}{108} (\bibinfo{year}{2020}), \bibinfo{pages}{1119--1128}.
\newblock


\bibitem[\protect\citeauthoryear{Karimi, Jannach, and Jugovac}{Karimi
  et~al\mbox{.}}{2018}]%
        {DBLP:journals/ipm/KarimiJJ18}
\bibfield{author}{\bibinfo{person}{Mozhgan Karimi}, \bibinfo{person}{Dietmar
  Jannach}, {and} \bibinfo{person}{Michael Jugovac}.}
  \bibinfo{year}{2018}\natexlab{}.
\newblock \showarticletitle{News recommender systems - Survey and roads ahead}.
\newblock \bibinfo{journal}{\emph{Inf. Process. Manage.}} \bibinfo{volume}{54},
  \bibinfo{number}{6} (\bibinfo{year}{2018}), \bibinfo{pages}{1203--1227}.
\newblock


\bibitem[\protect\citeauthoryear{Kaur, Kumar, and Batra}{Kaur
  et~al\mbox{.}}{2018}]%
        {DBLP:journals/fgcs/KaurKB18}
\bibfield{author}{\bibinfo{person}{Harmanjeet Kaur}, \bibinfo{person}{Neeraj
  Kumar}, {and} \bibinfo{person}{Shalini Batra}.}
  \bibinfo{year}{2018}\natexlab{}.
\newblock \showarticletitle{An efficient multi-party scheme for privacy
  preserving collaborative filtering for healthcare recommender system}.
\newblock \bibinfo{journal}{\emph{Future Gener. Comput. Syst.}}
  \bibinfo{volume}{86} (\bibinfo{year}{2018}), \bibinfo{pages}{297--307}.
\newblock


\bibitem[\protect\citeauthoryear{Kiciman and Sharma}{Kiciman and
  Sharma}{2019}]%
        {DBLP:conf/wsdm/KicimanS19}
\bibfield{author}{\bibinfo{person}{Emre Kiciman} {and} \bibinfo{person}{Amit
  Sharma}.} \bibinfo{year}{2019}\natexlab{}.
\newblock \showarticletitle{Causal Inference and Counterfactual Reasoning (3hr
  Tutorial)}. In \bibinfo{booktitle}{\emph{{WSDM}}}.
  \bibinfo{publisher}{{ACM}}, \bibinfo{pages}{828--829}.
\newblock


\bibitem[\protect\citeauthoryear{Koprinska and Yacef}{Koprinska and
  Yacef}{2015}]%
        {DBLP:reference/sp/KoprinskaY15}
\bibfield{author}{\bibinfo{person}{Irena Koprinska} {and}
  \bibinfo{person}{Kalina Yacef}.} \bibinfo{year}{2015}\natexlab{}.
\newblock \showarticletitle{People-to-People Reciprocal Recommenders}.
\newblock In \bibinfo{booktitle}{\emph{Recommender Systems Handbook}}.
  \bibinfo{publisher}{Springer}, \bibinfo{pages}{545--567}.
\newblock


\bibitem[\protect\citeauthoryear{Koren and Bell}{Koren and Bell}{2015}]%
        {DBLP:reference/sp/KorenB15}
\bibfield{author}{\bibinfo{person}{Yehuda Koren} {and}
  \bibinfo{person}{Robert~M. Bell}.} \bibinfo{year}{2015}\natexlab{}.
\newblock \showarticletitle{Advances in Collaborative Filtering}.
\newblock In \bibinfo{booktitle}{\emph{Recommender Systems Handbook}}.
  \bibinfo{publisher}{Springer}, \bibinfo{pages}{77--118}.
\newblock


\bibitem[\protect\citeauthoryear{Kurashima, Iwata, Irie, and
  Fujimura}{Kurashima et~al\mbox{.}}{2010}]%
        {DBLP:conf/cikm/KurashimaIIF10}
\bibfield{author}{\bibinfo{person}{Takeshi Kurashima},
  \bibinfo{person}{Tomoharu Iwata}, \bibinfo{person}{Go Irie}, {and}
  \bibinfo{person}{Ko Fujimura}.} \bibinfo{year}{2010}\natexlab{}.
\newblock \showarticletitle{Travel route recommendation using geotags in photo
  sharing sites}. In \bibinfo{booktitle}{\emph{{CIKM}}}.
  \bibinfo{publisher}{{ACM}}, \bibinfo{pages}{579--588}.
\newblock


\bibitem[\protect\citeauthoryear{Levandoski, Sarwat, Eldawy, and
  Mokbel}{Levandoski et~al\mbox{.}}{2012}]%
        {DBLP:conf/icde/LevandoskiSEM12}
\bibfield{author}{\bibinfo{person}{Justin~J. Levandoski},
  \bibinfo{person}{Mohamed Sarwat}, \bibinfo{person}{Ahmed Eldawy}, {and}
  \bibinfo{person}{Mohamed~F. Mokbel}.} \bibinfo{year}{2012}\natexlab{}.
\newblock \showarticletitle{{LARS:} {A} Location-Aware Recommender System}. In
  \bibinfo{booktitle}{\emph{{ICDE}}}. \bibinfo{publisher}{{IEEE} Computer
  Society}, \bibinfo{pages}{450--461}.
\newblock


\bibitem[\protect\citeauthoryear{Li, Ge, Hong, and Zhu}{Li
  et~al\mbox{.}}{2016}]%
        {DBLP:conf/kdd/LiGHZ16}
\bibfield{author}{\bibinfo{person}{Huayu Li}, \bibinfo{person}{Yong Ge},
  \bibinfo{person}{Richang Hong}, {and} \bibinfo{person}{Hengshu Zhu}.}
  \bibinfo{year}{2016}\natexlab{}.
\newblock \showarticletitle{Point-of-Interest Recommendations: Learning
  Potential Check-ins from Friends}. In \bibinfo{booktitle}{\emph{{KDD}}}.
  \bibinfo{publisher}{{ACM}}, \bibinfo{pages}{975--984}.
\newblock


\bibitem[\protect\citeauthoryear{Li, Ge, Lian, and Liu}{Li
  et~al\mbox{.}}{2017}]%
        {DBLP:conf/ijcai/LiGLL17}
\bibfield{author}{\bibinfo{person}{Huayu Li}, \bibinfo{person}{Yong Ge},
  \bibinfo{person}{Defu Lian}, {and} \bibinfo{person}{Hao Liu}.}
  \bibinfo{year}{2017}\natexlab{}.
\newblock \showarticletitle{Learning User's Intrinsic and Extrinsic Interests
  for Point-of-Interest Recommendation: {A} Unified Approach}. In
  \bibinfo{booktitle}{\emph{{IJCAI}}}. \bibinfo{publisher}{ijcai.org},
  \bibinfo{pages}{2117--2123}.
\newblock


\bibitem[\protect\citeauthoryear{Li, Shen, and Zhu}{Li et~al\mbox{.}}{2018}]%
        {DBLP:conf/icdm/LiSZ18}
\bibfield{author}{\bibinfo{person}{Ranzhen Li}, \bibinfo{person}{Yanyan Shen},
  {and} \bibinfo{person}{Yanmin Zhu}.} \bibinfo{year}{2018}\natexlab{}.
\newblock \showarticletitle{Next Point-of-Interest Recommendation with Temporal
  and Multi-level Context Attention}. In \bibinfo{booktitle}{\emph{{ICDM}}}.
  \bibinfo{publisher}{{IEEE} Computer Society}, \bibinfo{pages}{1110--1115}.
\newblock


\bibitem[\protect\citeauthoryear{Li, Cong, Li, Pham, and Krishnaswamy}{Li
  et~al\mbox{.}}{2015a}]%
        {DBLP:conf/sigir/LiCLPK15}
\bibfield{author}{\bibinfo{person}{Xutao Li}, \bibinfo{person}{Gao Cong},
  \bibinfo{person}{Xiaoli Li}, \bibinfo{person}{Tuan{-}Anh~Nguyen Pham}, {and}
  \bibinfo{person}{Shonali Krishnaswamy}.} \bibinfo{year}{2015}\natexlab{a}.
\newblock \showarticletitle{Rank-GeoFM: {A} Ranking based Geographical
  Factorization Method for Point of Interest Recommendation}. In
  \bibinfo{booktitle}{\emph{{SIGIR}}}. \bibinfo{publisher}{{ACM}},
  \bibinfo{pages}{433--442}.
\newblock


\bibitem[\protect\citeauthoryear{Li, Xu, Chen, and Zong}{Li
  et~al\mbox{.}}{2015b}]%
        {DBLP:journals/eswa/LiXCZ15}
\bibfield{author}{\bibinfo{person}{Xin Li}, \bibinfo{person}{Guandong Xu},
  \bibinfo{person}{Enhong Chen}, {and} \bibinfo{person}{Yu Zong}.}
  \bibinfo{year}{2015}\natexlab{b}.
\newblock \showarticletitle{Learning recency based comparative choice towards
  point-of-interest recommendation}.
\newblock \bibinfo{journal}{\emph{Expert Syst. Appl.}} \bibinfo{volume}{42},
  \bibinfo{number}{9} (\bibinfo{year}{2015}), \bibinfo{pages}{4274--4283}.
\newblock


\bibitem[\protect\citeauthoryear{Lian, Ge, Zhang, Yuan, Xie, Zhou, and
  Rui}{Lian et~al\mbox{.}}{2015}]%
        {DBLP:conf/icdm/LianGZYXZR15}
\bibfield{author}{\bibinfo{person}{Defu Lian}, \bibinfo{person}{Yong Ge},
  \bibinfo{person}{Fuzheng Zhang}, \bibinfo{person}{Nicholas~Jing Yuan},
  \bibinfo{person}{Xing Xie}, \bibinfo{person}{Tao Zhou}, {and}
  \bibinfo{person}{Yong Rui}.} \bibinfo{year}{2015}\natexlab{}.
\newblock \showarticletitle{Content-Aware Collaborative Filtering for Location
  Recommendation Based on Human Mobility Data}. In
  \bibinfo{booktitle}{\emph{{ICDM}}}. \bibinfo{publisher}{{IEEE} Computer
  Society}, \bibinfo{pages}{261--270}.
\newblock


\bibitem[\protect\citeauthoryear{Lian, Wu, Ge, Xie, and Chen}{Lian
  et~al\mbox{.}}{2020}]%
        {DBLP:conf/kdd/LianWG0C20}
\bibfield{author}{\bibinfo{person}{Defu Lian}, \bibinfo{person}{Yongji Wu},
  \bibinfo{person}{Yong Ge}, \bibinfo{person}{Xing Xie}, {and}
  \bibinfo{person}{Enhong Chen}.} \bibinfo{year}{2020}\natexlab{}.
\newblock \showarticletitle{Geography-Aware Sequential Location
  Recommendation}. In \bibinfo{booktitle}{\emph{{KDD}}}.
  \bibinfo{publisher}{{ACM}}, \bibinfo{pages}{2009--2019}.
\newblock


\bibitem[\protect\citeauthoryear{Lian, Zhang, Ge, Zhang, Yuan, and Xie}{Lian
  et~al\mbox{.}}{2016}]%
        {DBLP:conf/icdm/LianZGZYX16}
\bibfield{author}{\bibinfo{person}{Defu Lian}, \bibinfo{person}{Zhenyu Zhang},
  \bibinfo{person}{Yong Ge}, \bibinfo{person}{Fuzheng Zhang},
  \bibinfo{person}{Nicholas~Jing Yuan}, {and} \bibinfo{person}{Xing Xie}.}
  \bibinfo{year}{2016}\natexlab{}.
\newblock \showarticletitle{Regularized Content-Aware Tensor Factorization
  Meets Temporal-Aware Location Recommendation}. In
  \bibinfo{booktitle}{\emph{{ICDM}}}. \bibinfo{publisher}{{IEEE} Computer
  Society}, \bibinfo{pages}{1029--1034}.
\newblock


\bibitem[\protect\citeauthoryear{Lian, Zhao, Xie, Sun, Chen, and Rui}{Lian
  et~al\mbox{.}}{2014}]%
        {DBLP:conf/kdd/LianZXSCR14}
\bibfield{author}{\bibinfo{person}{Defu Lian}, \bibinfo{person}{Cong Zhao},
  \bibinfo{person}{Xing Xie}, \bibinfo{person}{Guangzhong Sun},
  \bibinfo{person}{Enhong Chen}, {and} \bibinfo{person}{Yong Rui}.}
  \bibinfo{year}{2014}\natexlab{}.
\newblock \showarticletitle{GeoMF: joint geographical modeling and matrix
  factorization for point-of-interest recommendation}. In
  \bibinfo{booktitle}{\emph{{KDD}}}. \bibinfo{publisher}{{ACM}},
  \bibinfo{pages}{831--840}.
\newblock


\bibitem[\protect\citeauthoryear{Liu, Fu, Yao, and Xiong}{Liu
  et~al\mbox{.}}{2013a}]%
        {DBLP:conf/kdd/LiuFYX13}
\bibfield{author}{\bibinfo{person}{Bin Liu}, \bibinfo{person}{Yanjie Fu},
  \bibinfo{person}{Zijun Yao}, {and} \bibinfo{person}{Hui Xiong}.}
  \bibinfo{year}{2013}\natexlab{a}.
\newblock \showarticletitle{Learning geographical preferences for
  point-of-interest recommendation}. In \bibinfo{booktitle}{\emph{{KDD}}}.
  \bibinfo{publisher}{{ACM}}, \bibinfo{pages}{1043--1051}.
\newblock


\bibitem[\protect\citeauthoryear{Liu, Meng, Zhang, Xu, and Cao}{Liu
  et~al\mbox{.}}{2020}]%
        {https://doi.org/10.1002/ett.3889}
\bibfield{author}{\bibinfo{person}{Bo Liu}, \bibinfo{person}{Qing Meng},
  \bibinfo{person}{Hengyuan Zhang}, \bibinfo{person}{Kun Xu}, {and}
  \bibinfo{person}{Jiuxin Cao}.} \bibinfo{year}{2020}\natexlab{}.
\newblock \showarticletitle{VGMF: Visual contents and geographical influence
  enhanced point-of-interest recommendation in location-based social network}.
\newblock \bibinfo{journal}{\emph{Transactions on Emerging Telecommunications
  Technologies}} \bibinfo{volume}{n/a}, \bibinfo{number}{n/a}
  (\bibinfo{year}{2020}), \bibinfo{pages}{e3889}.
\newblock
\urldef\tempurl%
\url{https://doi.org/10.1002/ett.3889}
\showDOI{\tempurl}
\showeprint{https://onlinelibrary.wiley.com/doi/pdf/10.1002/ett.3889}


\bibitem[\protect\citeauthoryear{Liu and Xiong}{Liu and Xiong}{2013}]%
        {DBLP:conf/sdm/LiuX13}
\bibfield{author}{\bibinfo{person}{Bin Liu} {and} \bibinfo{person}{Hui Xiong}.}
  \bibinfo{year}{2013}\natexlab{}.
\newblock \showarticletitle{Point-of-Interest Recommendation in Location Based
  Social Networks with Topic and Location Awareness}. In
  \bibinfo{booktitle}{\emph{{SDM}}}. \bibinfo{publisher}{{SIAM}},
  \bibinfo{pages}{396--404}.
\newblock


\bibitem[\protect\citeauthoryear{Liu, Xiong, Papadimitriou, Fu, and Yao}{Liu
  et~al\mbox{.}}{2015}]%
        {DBLP:journals/tkde/LiuXPFY15}
\bibfield{author}{\bibinfo{person}{Bin Liu}, \bibinfo{person}{Hui Xiong},
  \bibinfo{person}{Spiros Papadimitriou}, \bibinfo{person}{Yanjie Fu}, {and}
  \bibinfo{person}{Zijun Yao}.} \bibinfo{year}{2015}\natexlab{}.
\newblock \showarticletitle{A General Geographical Probabilistic Factor Model
  for Point of Interest Recommendation}.
\newblock \bibinfo{journal}{\emph{{IEEE} Trans. Knowl. Data Eng.}}
  \bibinfo{volume}{27}, \bibinfo{number}{5} (\bibinfo{year}{2015}),
  \bibinfo{pages}{1167--1179}.
\newblock


\bibitem[\protect\citeauthoryear{Liu, Nguyen, Zhao, Zha, Yang, Cao, Wu, Zhao,
  and Chen}{Liu et~al\mbox{.}}{2016b}]%
        {DBLP:conf/kdd/LiuNZZYCWZC16}
\bibfield{author}{\bibinfo{person}{Guimei Liu}, \bibinfo{person}{Tam~T.
  Nguyen}, \bibinfo{person}{Gang Zhao}, \bibinfo{person}{Wei Zha},
  \bibinfo{person}{Jianbo Yang}, \bibinfo{person}{Jianneng Cao},
  \bibinfo{person}{Min Wu}, \bibinfo{person}{Peilin Zhao}, {and}
  \bibinfo{person}{Wei Chen}.} \bibinfo{year}{2016}\natexlab{b}.
\newblock \showarticletitle{Repeat Buyer Prediction for E-Commerce}. In
  \bibinfo{booktitle}{\emph{{KDD}}}. \bibinfo{publisher}{{ACM}},
  \bibinfo{pages}{155--164}.
\newblock


\bibitem[\protect\citeauthoryear{Liu, Liu, Aberer, and Miao}{Liu
  et~al\mbox{.}}{2013b}]%
        {DBLP:conf/cikm/LiuLAM13}
\bibfield{author}{\bibinfo{person}{Xin Liu}, \bibinfo{person}{Yong Liu},
  \bibinfo{person}{Karl Aberer}, {and} \bibinfo{person}{Chunyan Miao}.}
  \bibinfo{year}{2013}\natexlab{b}.
\newblock \showarticletitle{Personalized point-of-interest recommendation by
  mining users' preference transition}. In \bibinfo{booktitle}{\emph{{CIKM}}}.
  \bibinfo{publisher}{{ACM}}, \bibinfo{pages}{733--738}.
\newblock


\bibitem[\protect\citeauthoryear{Liu, Liu, Liu, Qu, and Xiong}{Liu
  et~al\mbox{.}}{2016a}]%
        {DBLP:conf/kdd/LiuLLQX16}
\bibfield{author}{\bibinfo{person}{Yanchi Liu}, \bibinfo{person}{Chuanren Liu},
  \bibinfo{person}{Bin Liu}, \bibinfo{person}{Meng Qu}, {and}
  \bibinfo{person}{Hui Xiong}.} \bibinfo{year}{2016}\natexlab{a}.
\newblock \showarticletitle{Unified Point-of-Interest Recommendation with
  Temporal Interval Assessment}. In \bibinfo{booktitle}{\emph{{KDD}}}.
  \bibinfo{publisher}{{ACM}}, \bibinfo{pages}{1015--1024}.
\newblock


\bibitem[\protect\citeauthoryear{Liu, Pham, Cong, and Yuan}{Liu
  et~al\mbox{.}}{2017}]%
        {DBLP:journals/pvldb/LiuPCY17}
\bibfield{author}{\bibinfo{person}{Yiding Liu}, \bibinfo{person}{Tuan{-}Anh
  Pham}, \bibinfo{person}{Gao Cong}, {and} \bibinfo{person}{Quan Yuan}.}
  \bibinfo{year}{2017}\natexlab{}.
\newblock \showarticletitle{An Experimental Evaluation of Point-of-interest
  Recommendation in Location-based Social Networks}.
\newblock \bibinfo{journal}{\emph{{PVLDB}}} \bibinfo{volume}{10},
  \bibinfo{number}{10} (\bibinfo{year}{2017}), \bibinfo{pages}{1010--1021}.
\newblock


\bibitem[\protect\citeauthoryear{Liu, Wei, Sun, and Miao}{Liu
  et~al\mbox{.}}{2014}]%
        {DBLP:conf/cikm/LiuWSM14}
\bibfield{author}{\bibinfo{person}{Yong Liu}, \bibinfo{person}{Wei Wei},
  \bibinfo{person}{Aixin Sun}, {and} \bibinfo{person}{Chunyan Miao}.}
  \bibinfo{year}{2014}\natexlab{}.
\newblock \showarticletitle{Exploiting Geographical Neighborhood
  Characteristics for Location Recommendation}. In
  \bibinfo{booktitle}{\emph{{CIKM}}}. \bibinfo{publisher}{{ACM}},
  \bibinfo{pages}{739--748}.
\newblock


\bibitem[\protect\citeauthoryear{Lops, Jannach, Musto, Bogers, and Koolen}{Lops
  et~al\mbox{.}}{2019}]%
        {DBLP:journals/umuai/LopsJMBK19}
\bibfield{author}{\bibinfo{person}{Pasquale Lops}, \bibinfo{person}{Dietmar
  Jannach}, \bibinfo{person}{Cataldo Musto}, \bibinfo{person}{Toine Bogers},
  {and} \bibinfo{person}{Marijn Koolen}.} \bibinfo{year}{2019}\natexlab{}.
\newblock \showarticletitle{Trends in content-based recommendation - Preface to
  the special issue on Recommender systems based on rich item descriptions}.
\newblock \bibinfo{journal}{\emph{User Model. User Adapt. Interact.}}
  \bibinfo{volume}{29}, \bibinfo{number}{2} (\bibinfo{year}{2019}),
  \bibinfo{pages}{239--249}.
\newblock


\bibitem[\protect\citeauthoryear{Ma, Zhang, Wang, and Liu}{Ma
  et~al\mbox{.}}{2018}]%
        {DBLP:conf/cikm/MaZWL18}
\bibfield{author}{\bibinfo{person}{Chen Ma}, \bibinfo{person}{Yingxue Zhang},
  \bibinfo{person}{Qinglong Wang}, {and} \bibinfo{person}{Xue Liu}.}
  \bibinfo{year}{2018}\natexlab{}.
\newblock \showarticletitle{Point-of-Interest Recommendation: Exploiting
  Self-Attentive Autoencoders with Neighbor-Aware Influence}. In
  \bibinfo{booktitle}{\emph{{CIKM}}}. \bibinfo{publisher}{{ACM}},
  \bibinfo{pages}{697--706}.
\newblock


\bibitem[\protect\citeauthoryear{Manotumruksa, Macdonald, and
  Ounis}{Manotumruksa et~al\mbox{.}}{2018}]%
        {DBLP:conf/sigir/ManotumruksaMO18}
\bibfield{author}{\bibinfo{person}{Jarana Manotumruksa}, \bibinfo{person}{Craig
  Macdonald}, {and} \bibinfo{person}{Iadh Ounis}.}
  \bibinfo{year}{2018}\natexlab{}.
\newblock \showarticletitle{A Contextual Attention Recurrent Architecture for
  Context-Aware Venue Recommendation}. In \bibinfo{booktitle}{\emph{{SIGIR}}}.
  \bibinfo{publisher}{{ACM}}, \bibinfo{pages}{555--564}.
\newblock


\bibitem[\protect\citeauthoryear{Mazumdar, Patra, and Babu}{Mazumdar
  et~al\mbox{.}}{2020}]%
        {DBLP:journals/tweb/MazumdarPB20}
\bibfield{author}{\bibinfo{person}{Pramit Mazumdar},
  \bibinfo{person}{Bidyut~Kr. Patra}, {and} \bibinfo{person}{Korra~Sathya
  Babu}.} \bibinfo{year}{2020}\natexlab{}.
\newblock \showarticletitle{Cold-start Point-of-interest Recommendation through
  Crowdsourcing}.
\newblock \bibinfo{journal}{\emph{{ACM} Trans. Web}} \bibinfo{volume}{14},
  \bibinfo{number}{4} (\bibinfo{year}{2020}), \bibinfo{pages}{19:1--19:36}.
\newblock


\bibitem[\protect\citeauthoryear{McNee, Riedl, and Konstan}{McNee
  et~al\mbox{.}}{2006}]%
        {DBLP:conf/chi/McNeeRK06}
\bibfield{author}{\bibinfo{person}{Sean~M. McNee}, \bibinfo{person}{John
  Riedl}, {and} \bibinfo{person}{Joseph~A. Konstan}.}
  \bibinfo{year}{2006}\natexlab{}.
\newblock \showarticletitle{Being accurate is not enough: how accuracy metrics
  have hurt recommender systems}. In \bibinfo{booktitle}{\emph{{CHI} Extended
  Abstracts}}. \bibinfo{publisher}{{ACM}}, \bibinfo{pages}{1097--1101}.
\newblock


\bibitem[\protect\citeauthoryear{Miller}{Miller}{2004}]%
        {MHJ2004}
\bibfield{author}{\bibinfo{person}{Harvey~J. Miller}.}
  \bibinfo{year}{2004}\natexlab{}.
\newblock \showarticletitle{Tobler's First Law and Spatial Analysis}.
\newblock \bibinfo{journal}{\emph{Annals of the Association of American
  Geographers}} \bibinfo{volume}{94}, \bibinfo{number}{2}
  (\bibinfo{year}{2004}), \bibinfo{pages}{284--289}.
\newblock
\urldef\tempurl%
\url{https://doi.org/10.1111/j.1467-8306.2004.09402005.x}
\showDOI{\tempurl}


\bibitem[\protect\citeauthoryear{Naili, Cha{\"{\i}}bi, and Ghezala}{Naili
  et~al\mbox{.}}{2017}]%
        {DBLP:conf/kes/NailiCG17}
\bibfield{author}{\bibinfo{person}{Marwa Naili}, \bibinfo{person}{Anja~Habacha
  Cha{\"{\i}}bi}, {and} \bibinfo{person}{Henda Hajjami~Ben Ghezala}.}
  \bibinfo{year}{2017}\natexlab{}.
\newblock \showarticletitle{Comparative study of word embedding methods in
  topic segmentation}. In \bibinfo{booktitle}{\emph{{KES}}}
  \emph{(\bibinfo{series}{Procedia Computer Science},
  Vol.~\bibinfo{volume}{112})}. \bibinfo{publisher}{Elsevier},
  \bibinfo{pages}{340--349}.
\newblock


\bibitem[\protect\citeauthoryear{Nie, Liu, Zhu, and Su}{Nie
  et~al\mbox{.}}{2016}]%
        {DBLP:journals/mta/NieLZS16}
\bibfield{author}{\bibinfo{person}{Weizhi Nie}, \bibinfo{person}{Anan Liu},
  \bibinfo{person}{Xiaorong Zhu}, {and} \bibinfo{person}{Yuting Su}.}
  \bibinfo{year}{2016}\natexlab{}.
\newblock \showarticletitle{Quality models for venue recommendation in
  location-based social network}.
\newblock \bibinfo{journal}{\emph{Multimedia Tools Appl.}}
  \bibinfo{volume}{75}, \bibinfo{number}{20} (\bibinfo{year}{2016}),
  \bibinfo{pages}{12521--12534}.
\newblock


\bibitem[\protect\citeauthoryear{Ning, Desrosiers, and Karypis}{Ning
  et~al\mbox{.}}{2015}]%
        {DBLP:reference/sp/NingDK15}
\bibfield{author}{\bibinfo{person}{Xia Ning}, \bibinfo{person}{Christian
  Desrosiers}, {and} \bibinfo{person}{George Karypis}.}
  \bibinfo{year}{2015}\natexlab{}.
\newblock \showarticletitle{A Comprehensive Survey of Neighborhood-Based
  Recommendation Methods}.
\newblock In \bibinfo{booktitle}{\emph{Recommender Systems Handbook}}.
  \bibinfo{publisher}{Springer}, \bibinfo{pages}{37--76}.
\newblock


\bibitem[\protect\citeauthoryear{Noulas, Scellato, Lathia, and Mascolo}{Noulas
  et~al\mbox{.}}{2012}]%
        {DBLP:conf/socialcom/NoulasSLM12}
\bibfield{author}{\bibinfo{person}{Anastasios Noulas},
  \bibinfo{person}{Salvatore Scellato}, \bibinfo{person}{Neal Lathia}, {and}
  \bibinfo{person}{Cecilia Mascolo}.} \bibinfo{year}{2012}\natexlab{}.
\newblock \showarticletitle{A Random Walk around the City: New Venue
  Recommendation in Location-Based Social Networks}. In
  \bibinfo{booktitle}{\emph{SocialCom/PASSAT}}. \bibinfo{publisher}{{IEEE}
  Computer Society}, \bibinfo{pages}{144--153}.
\newblock


\bibitem[\protect\citeauthoryear{Palumbo, Rizzo, Troncy, and Baralis}{Palumbo
  et~al\mbox{.}}{2017}]%
        {DBLP:conf/recsys/Palumbo0TB17}
\bibfield{author}{\bibinfo{person}{Enrico Palumbo}, \bibinfo{person}{Giuseppe
  Rizzo}, \bibinfo{person}{Rapha{\"{e}}l Troncy}, {and} \bibinfo{person}{Elena
  Baralis}.} \bibinfo{year}{2017}\natexlab{}.
\newblock \showarticletitle{Predicting Your Next Stop-over from Location-based
  Social Network Data with Recurrent Neural Networks}. In
  \bibinfo{booktitle}{\emph{RecTour@RecSys}} \emph{(\bibinfo{series}{{CEUR}
  Workshop Proceedings}, Vol.~\bibinfo{volume}{1906})}.
  \bibinfo{publisher}{CEUR-WS.org}, \bibinfo{pages}{1--8}.
\newblock


\bibitem[\protect\citeauthoryear{Papalexakis, Pelechrinis, and
  Faloutsos}{Papalexakis et~al\mbox{.}}{2014}]%
        {DBLP:conf/www/PapalexakisPF14}
\bibfield{author}{\bibinfo{person}{Evangelos~E. Papalexakis},
  \bibinfo{person}{Konstantinos Pelechrinis}, {and} \bibinfo{person}{Christos
  Faloutsos}.} \bibinfo{year}{2014}\natexlab{}.
\newblock \showarticletitle{Spotting misbehaviors in location-based social
  networks using tensors}. In \bibinfo{booktitle}{\emph{{WWW} (Companion
  Volume)}}. \bibinfo{publisher}{{ACM}}, \bibinfo{pages}{551--552}.
\newblock


\bibitem[\protect\citeauthoryear{Perozzi, Al{-}Rfou, and Skiena}{Perozzi
  et~al\mbox{.}}{2014}]%
        {DBLP:conf/kdd/PerozziAS14}
\bibfield{author}{\bibinfo{person}{Bryan Perozzi}, \bibinfo{person}{Rami
  Al{-}Rfou}, {and} \bibinfo{person}{Steven Skiena}.}
  \bibinfo{year}{2014}\natexlab{}.
\newblock \showarticletitle{DeepWalk: online learning of social
  representations}. In \bibinfo{booktitle}{\emph{{KDD}}}.
  \bibinfo{publisher}{{ACM}}, \bibinfo{pages}{701--710}.
\newblock


\bibitem[\protect\citeauthoryear{Qian, Liu, Nguyen, and Yin}{Qian
  et~al\mbox{.}}{2019}]%
        {DBLP:journals/tois/QianLNY19}
\bibfield{author}{\bibinfo{person}{Tieyun Qian}, \bibinfo{person}{Bei Liu},
  \bibinfo{person}{Quoc Viet~Hung Nguyen}, {and} \bibinfo{person}{Hongzhi
  Yin}.} \bibinfo{year}{2019}\natexlab{}.
\newblock \showarticletitle{Spatiotemporal Representation Learning for
  Translation-Based {POI} Recommendation}.
\newblock \bibinfo{journal}{\emph{{ACM} Trans. Inf. Syst.}}
  \bibinfo{volume}{37}, \bibinfo{number}{2} (\bibinfo{year}{2019}),
  \bibinfo{pages}{18:1--18:24}.
\newblock


\bibitem[\protect\citeauthoryear{Rahmani, Aliannejadi, Baratchi, and
  Crestani}{Rahmani et~al\mbox{.}}{2020}]%
        {DBLP:conf/ecir/RahmaniABC20}
\bibfield{author}{\bibinfo{person}{Hossein~A. Rahmani},
  \bibinfo{person}{Mohammad Aliannejadi}, \bibinfo{person}{Mitra Baratchi},
  {and} \bibinfo{person}{Fabio Crestani}.} \bibinfo{year}{2020}\natexlab{}.
\newblock \showarticletitle{Joint Geographical and Temporal Modeling Based on
  Matrix Factorization for Point-of-Interest Recommendation}. In
  \bibinfo{booktitle}{\emph{{ECIR} {(1)}}} \emph{(\bibinfo{series}{Lecture
  Notes in Computer Science}, Vol.~\bibinfo{volume}{12035})}.
  \bibinfo{publisher}{Springer}, \bibinfo{pages}{205--219}.
\newblock


\bibitem[\protect\citeauthoryear{Ravi and Subramaniyaswamy}{Ravi and
  Subramaniyaswamy}{2017}]%
        {DBLP:journals/wpc/RaviS17}
\bibfield{author}{\bibinfo{person}{Logesh Ravi} {and} \bibinfo{person}{V.
  Subramaniyaswamy}.} \bibinfo{year}{2017}\natexlab{}.
\newblock \showarticletitle{A Reliable Point of Interest Recommendation based
  on Trust Relevancy between Users}.
\newblock \bibinfo{journal}{\emph{Wireless Personal Communications}}
  \bibinfo{volume}{97}, \bibinfo{number}{2} (\bibinfo{year}{2017}),
  \bibinfo{pages}{2751--2780}.
\newblock


\bibitem[\protect\citeauthoryear{Ren, Song, E, and Song}{Ren
  et~al\mbox{.}}{2017}]%
        {DBLP:journals/ijon/RenSES17}
\bibfield{author}{\bibinfo{person}{Xingyi Ren}, \bibinfo{person}{Meina Song},
  \bibinfo{person}{Haihong E}, {and} \bibinfo{person}{Junde Song}.}
  \bibinfo{year}{2017}\natexlab{}.
\newblock \showarticletitle{Context-aware probabilistic matrix factorization
  modeling for point-of-interest recommendation}.
\newblock \bibinfo{journal}{\emph{Neurocomputing}}  \bibinfo{volume}{241}
  (\bibinfo{year}{2017}), \bibinfo{pages}{38--55}.
\newblock


\bibitem[\protect\citeauthoryear{Rendle, Freudenthaler, Gantner, and
  Schmidt{-}Thieme}{Rendle et~al\mbox{.}}{2009}]%
        {DBLP:conf/uai/RendleFGS09}
\bibfield{author}{\bibinfo{person}{Steffen Rendle}, \bibinfo{person}{Christoph
  Freudenthaler}, \bibinfo{person}{Zeno Gantner}, {and} \bibinfo{person}{Lars
  Schmidt{-}Thieme}.} \bibinfo{year}{2009}\natexlab{}.
\newblock \showarticletitle{{BPR:} Bayesian Personalized Ranking from Implicit
  Feedback}. In \bibinfo{booktitle}{\emph{{UAI}}}. \bibinfo{publisher}{{AUAI}
  Press}, \bibinfo{pages}{452--461}.
\newblock


\bibitem[\protect\citeauthoryear{Rendle, Freudenthaler, and
  Schmidt{-}Thieme}{Rendle et~al\mbox{.}}{2010}]%
        {DBLP:conf/www/RendleFS10}
\bibfield{author}{\bibinfo{person}{Steffen Rendle}, \bibinfo{person}{Christoph
  Freudenthaler}, {and} \bibinfo{person}{Lars Schmidt{-}Thieme}.}
  \bibinfo{year}{2010}\natexlab{}.
\newblock \showarticletitle{Factorizing personalized Markov chains for
  next-basket recommendation}. In \bibinfo{booktitle}{\emph{{WWW}}}.
  \bibinfo{publisher}{{ACM}}, \bibinfo{pages}{811--820}.
\newblock


\bibitem[\protect\citeauthoryear{Ricci, Rokach, and Shapira}{Ricci
  et~al\mbox{.}}{2015}]%
        {DBLP:reference/sp/RicciRS15}
\bibfield{author}{\bibinfo{person}{Francesco Ricci}, \bibinfo{person}{Lior
  Rokach}, {and} \bibinfo{person}{Bracha Shapira}.}
  \bibinfo{year}{2015}\natexlab{}.
\newblock \showarticletitle{Recommender Systems: Introduction and Challenges}.
\newblock In \bibinfo{booktitle}{\emph{Recommender Systems Handbook}}.
  \bibinfo{publisher}{Springer}, \bibinfo{pages}{1--34}.
\newblock


\bibitem[\protect\citeauthoryear{Said and Bellog{\'{\i}}n}{Said and
  Bellog{\'{\i}}n}{2014}]%
        {DBLP:conf/recsys/SaidB14}
\bibfield{author}{\bibinfo{person}{Alan Said} {and} \bibinfo{person}{Alejandro
  Bellog{\'{\i}}n}.} \bibinfo{year}{2014}\natexlab{}.
\newblock \showarticletitle{Comparative recommender system evaluation:
  benchmarking recommendation frameworks}. In
  \bibinfo{booktitle}{\emph{RecSys}}. \bibinfo{publisher}{{ACM}},
  \bibinfo{pages}{129--136}.
\newblock


\bibitem[\protect\citeauthoryear{S{\'{a}}nchez and
  Bellog{\'{\i}}n}{S{\'{a}}nchez and Bellog{\'{\i}}n}{2020}]%
        {DBLP:journals/umuai/SanchezB20}
\bibfield{author}{\bibinfo{person}{Pablo S{\'{a}}nchez} {and}
  \bibinfo{person}{Alejandro Bellog{\'{\i}}n}.}
  \bibinfo{year}{2020}\natexlab{}.
\newblock \showarticletitle{Applying reranking strategies to route
  recommendation using sequence-aware evaluation}.
\newblock \bibinfo{journal}{\emph{User Model. User Adapt. Interact.}}
  \bibinfo{volume}{30}, \bibinfo{number}{4} (\bibinfo{year}{2020}),
  \bibinfo{pages}{659--725}.
\newblock


\bibitem[\protect\citeauthoryear{Shani, Heckerman, and Brafman}{Shani
  et~al\mbox{.}}{2005}]%
        {DBLP:journals/jmlr/ShaniHB05}
\bibfield{author}{\bibinfo{person}{Guy Shani}, \bibinfo{person}{David
  Heckerman}, {and} \bibinfo{person}{Ronen~I. Brafman}.}
  \bibinfo{year}{2005}\natexlab{}.
\newblock \showarticletitle{An MDP-Based Recommender System}.
\newblock \bibinfo{journal}{\emph{Journal of Machine Learning Research}}
  \bibinfo{volume}{6} (\bibinfo{year}{2005}), \bibinfo{pages}{1265--1295}.
\newblock


\bibitem[\protect\citeauthoryear{Shen, Deng, and Gao}{Shen
  et~al\mbox{.}}{2016}]%
        {DBLP:journals/ijon/ShenDG16}
\bibfield{author}{\bibinfo{person}{Junge Shen}, \bibinfo{person}{Cheng Deng},
  {and} \bibinfo{person}{Xinbo Gao}.} \bibinfo{year}{2016}\natexlab{}.
\newblock \showarticletitle{Attraction recommendation: Towards personalized
  tourism via collective intelligence}.
\newblock \bibinfo{journal}{\emph{Neurocomputing}}  \bibinfo{volume}{173}
  (\bibinfo{year}{2016}), \bibinfo{pages}{789--798}.
\newblock


\bibitem[\protect\citeauthoryear{Si, Zhang, and Liu}{Si et~al\mbox{.}}{2019}]%
        {DBLP:journals/kbs/SiZL19}
\bibfield{author}{\bibinfo{person}{Yali Si}, \bibinfo{person}{Fuzhi Zhang},
  {and} \bibinfo{person}{Wenyuan Liu}.} \bibinfo{year}{2019}\natexlab{}.
\newblock \showarticletitle{An adaptive point-of-interest recommendation method
  for location-based social networks based on user activity and spatial
  features}.
\newblock \bibinfo{journal}{\emph{Knowl.-Based Syst.}}  \bibinfo{volume}{163}
  (\bibinfo{year}{2019}), \bibinfo{pages}{267--282}.
\newblock


\bibitem[\protect\citeauthoryear{Staab, Werthner, Ricci, Zipf, Gretzel,
  Fesenmaier, Paris, and Knoblock}{Staab et~al\mbox{.}}{2002}]%
        {DBLP:journals/expert/StaabWRZGFPK02}
\bibfield{author}{\bibinfo{person}{Steffen Staab}, \bibinfo{person}{Hannes
  Werthner}, \bibinfo{person}{Francesco Ricci}, \bibinfo{person}{Alexander
  Zipf}, \bibinfo{person}{Ulrike Gretzel}, \bibinfo{person}{Daniel~R.
  Fesenmaier}, \bibinfo{person}{C{\'{e}}cile Paris}, {and}
  \bibinfo{person}{Craig~A. Knoblock}.} \bibinfo{year}{2002}\natexlab{}.
\newblock \showarticletitle{Intelligent Systems for Tourism}.
\newblock \bibinfo{journal}{\emph{{IEEE} Intell. Syst.}} \bibinfo{volume}{17},
  \bibinfo{number}{6} (\bibinfo{year}{2002}), \bibinfo{pages}{53--64}.
\newblock


\bibitem[\protect\citeauthoryear{Stepan, Morawski, Dick, and Miller}{Stepan
  et~al\mbox{.}}{2016}]%
        {DBLP:journals/tcss/StepanMDM16}
\bibfield{author}{\bibinfo{person}{Torin Stepan}, \bibinfo{person}{Jason~M.
  Morawski}, \bibinfo{person}{Scott Dick}, {and} \bibinfo{person}{James
  Miller}.} \bibinfo{year}{2016}\natexlab{}.
\newblock \showarticletitle{Incorporating Spatial, Temporal, and Social Context
  in Recommendations for Location-Based Social Networks}.
\newblock \bibinfo{journal}{\emph{{IEEE} Trans. Comput. Social Systems}}
  \bibinfo{volume}{3}, \bibinfo{number}{4} (\bibinfo{year}{2016}),
  \bibinfo{pages}{164--175}.
\newblock


\bibitem[\protect\citeauthoryear{Symeonidis, Ntempos, and
  Manolopoulos}{Symeonidis et~al\mbox{.}}{2014}]%
        {DBLP:series/sbece/SymeonidisNM14}
\bibfield{author}{\bibinfo{person}{Panagiotis Symeonidis},
  \bibinfo{person}{Dimitrios Ntempos}, {and} \bibinfo{person}{Yannis
  Manolopoulos}.} \bibinfo{year}{2014}\natexlab{}.
\newblock \bibinfo{booktitle}{\emph{Recommender Systems for Location-based
  Social Networks}}.
\newblock \bibinfo{publisher}{Springer}.
\newblock


\bibitem[\protect\citeauthoryear{Symeonidis, Papadimitriou, Manolopoulos,
  Senkul, and Toroslu}{Symeonidis et~al\mbox{.}}{2011}]%
        {DBLP:conf/gis/SymeonidisPMST11}
\bibfield{author}{\bibinfo{person}{Panagiotis Symeonidis},
  \bibinfo{person}{Alexis Papadimitriou}, \bibinfo{person}{Yannis
  Manolopoulos}, \bibinfo{person}{Pinar Senkul}, {and}
  \bibinfo{person}{Ismail~Hakki Toroslu}.} \bibinfo{year}{2011}\natexlab{}.
\newblock \showarticletitle{Geo-social recommendations based on incremental
  tensor reduction and local path traversal}. In
  \bibinfo{booktitle}{\emph{{GIS-LBSN}}}. \bibinfo{publisher}{{ACM}},
  \bibinfo{pages}{89--96}.
\newblock


\bibitem[\protect\citeauthoryear{Trattner, Oberegger, Marinho, and
  Parra}{Trattner et~al\mbox{.}}{2018}]%
        {DBLP:journals/jitt/TrattnerOMP18}
\bibfield{author}{\bibinfo{person}{Christoph Trattner},
  \bibinfo{person}{Alexander Oberegger}, \bibinfo{person}{Leandro~Balby
  Marinho}, {and} \bibinfo{person}{Denis Parra}.}
  \bibinfo{year}{2018}\natexlab{}.
\newblock \showarticletitle{Investigating the utility of the weather context
  for point of interest recommendations}.
\newblock \bibinfo{journal}{\emph{J. of {IT} {\&} Tourism}}
  \bibinfo{volume}{19}, \bibinfo{number}{1-4} (\bibinfo{year}{2018}),
  \bibinfo{pages}{117--150}.
\newblock


\bibitem[\protect\citeauthoryear{Wang, Li, and Feng}{Wang
  et~al\mbox{.}}{2014}]%
        {DBLP:conf/apweb/WangLF14}
\bibfield{author}{\bibinfo{person}{Henan Wang}, \bibinfo{person}{Guoliang Li},
  {and} \bibinfo{person}{Jianhua Feng}.} \bibinfo{year}{2014}\natexlab{}.
\newblock \showarticletitle{Group-Based Personalized Location Recommendation on
  Social Networks}. In \bibinfo{booktitle}{\emph{APWeb}}
  \emph{(\bibinfo{series}{Lecture Notes in Computer Science},
  Vol.~\bibinfo{volume}{8709})}. \bibinfo{publisher}{Springer},
  \bibinfo{pages}{68--80}.
\newblock


\bibitem[\protect\citeauthoryear{Wang, Shen, Ouyang, and Cheng}{Wang
  et~al\mbox{.}}{2018}]%
        {DBLP:conf/ijcai/WangSOC18}
\bibfield{author}{\bibinfo{person}{Hao Wang}, \bibinfo{person}{Huawei Shen},
  \bibinfo{person}{Wentao Ouyang}, {and} \bibinfo{person}{Xueqi Cheng}.}
  \bibinfo{year}{2018}\natexlab{}.
\newblock \showarticletitle{Exploiting POI-Specific Geographical Influence for
  Point-of-Interest Recommendation}. In \bibinfo{booktitle}{\emph{{IJCAI}}}.
  \bibinfo{publisher}{ijcai.org}, \bibinfo{pages}{3877--3883}.
\newblock


\bibitem[\protect\citeauthoryear{Wang, Terrovitis, and Mamoulis}{Wang
  et~al\mbox{.}}{2013}]%
        {DBLP:conf/gis/WangTM13}
\bibfield{author}{\bibinfo{person}{Hao Wang}, \bibinfo{person}{Manolis
  Terrovitis}, {and} \bibinfo{person}{Nikos Mamoulis}.}
  \bibinfo{year}{2013}\natexlab{}.
\newblock \showarticletitle{Location recommendation in location-based social
  networks using user check-in data}. In
  \bibinfo{booktitle}{\emph{{SIGSPATIAL/GIS}}}. \bibinfo{publisher}{{ACM}},
  \bibinfo{pages}{364--373}.
\newblock


\bibitem[\protect\citeauthoryear{Wang, Wang, Tang, Shu, Ranganath, and
  Liu}{Wang et~al\mbox{.}}{2017}]%
        {DBLP:conf/www/WangWTSRL17}
\bibfield{author}{\bibinfo{person}{Suhang Wang}, \bibinfo{person}{Yilin Wang},
  \bibinfo{person}{Jiliang Tang}, \bibinfo{person}{Kai Shu},
  \bibinfo{person}{Suhas Ranganath}, {and} \bibinfo{person}{Huan Liu}.}
  \bibinfo{year}{2017}\natexlab{}.
\newblock \showarticletitle{What Your Images Reveal: Exploiting Visual Contents
  for Point-of-Interest Recommendation}. In \bibinfo{booktitle}{\emph{{WWW}}}.
  \bibinfo{publisher}{{ACM}}, \bibinfo{pages}{391--400}.
\newblock


\bibitem[\protect\citeauthoryear{Wang, Chen, Wang, Chen, and Gong}{Wang
  et~al\mbox{.}}{2020a}]%
        {DBLP:journals/iotj/WangCWCG20}
\bibfield{author}{\bibinfo{person}{Wei Wang}, \bibinfo{person}{Junyang Chen},
  \bibinfo{person}{Jinzhong Wang}, \bibinfo{person}{Junxin Chen}, {and}
  \bibinfo{person}{Zhiguo Gong}.} \bibinfo{year}{2020}\natexlab{a}.
\newblock \showarticletitle{Geography-Aware Inductive Matrix Completion for
  Personalized Point-of-Interest Recommendation in Smart Cities}.
\newblock \bibinfo{journal}{\emph{{IEEE} Internet Things J.}}
  \bibinfo{volume}{7}, \bibinfo{number}{5} (\bibinfo{year}{2020}),
  \bibinfo{pages}{4361--4370}.
\newblock


\bibitem[\protect\citeauthoryear{Wang, Chen, Wang, Chen, Liu, and Gong}{Wang
  et~al\mbox{.}}{2020b}]%
        {DBLP:journals/tii/WangCWCLG20}
\bibfield{author}{\bibinfo{person}{Wei Wang}, \bibinfo{person}{Junyang Chen},
  \bibinfo{person}{Jinzhong Wang}, \bibinfo{person}{Junxin Chen},
  \bibinfo{person}{Jinquan Liu}, {and} \bibinfo{person}{Zhiguo Gong}.}
  \bibinfo{year}{2020}\natexlab{b}.
\newblock \showarticletitle{Trust-Enhanced Collaborative Filtering for
  Personalized Point of Interests Recommendation}.
\newblock \bibinfo{journal}{\emph{{IEEE} Trans. Ind. Informatics}}
  \bibinfo{volume}{16}, \bibinfo{number}{9} (\bibinfo{year}{2020}),
  \bibinfo{pages}{6124--6132}.
\newblock


\bibitem[\protect\citeauthoryear{Xie, Yin, Wang, Xu, Chen, and Wang}{Xie
  et~al\mbox{.}}{2016}]%
        {DBLP:conf/cikm/XieYWXCW16}
\bibfield{author}{\bibinfo{person}{Min Xie}, \bibinfo{person}{Hongzhi Yin},
  \bibinfo{person}{Hao Wang}, \bibinfo{person}{Fanjiang Xu},
  \bibinfo{person}{Weitong Chen}, {and} \bibinfo{person}{Sen Wang}.}
  \bibinfo{year}{2016}\natexlab{}.
\newblock \showarticletitle{Learning Graph-based {POI} Embedding for
  Location-based Recommendation}. In \bibinfo{booktitle}{\emph{{CIKM}}}.
  \bibinfo{publisher}{{ACM}}, \bibinfo{pages}{15--24}.
\newblock


\bibitem[\protect\citeauthoryear{Xiong, Qiao, Han, Xiong, Bu, Li, Yue, and
  Yuan}{Xiong et~al\mbox{.}}{2020}]%
        {DBLP:journals/ijon/XiongQHXBLYY20}
\bibfield{author}{\bibinfo{person}{Xi Xiong}, \bibinfo{person}{Shaojie Qiao},
  \bibinfo{person}{Nan Han}, \bibinfo{person}{Fei Xiong}, \bibinfo{person}{Zhan
  Bu}, \bibinfo{person}{Rong{-}Hua Li}, \bibinfo{person}{Kun Yue}, {and}
  \bibinfo{person}{Guan Yuan}.} \bibinfo{year}{2020}\natexlab{}.
\newblock \showarticletitle{Where to go: An effective point-of-interest
  recommendation framework for heterogeneous social networks}.
\newblock \bibinfo{journal}{\emph{Neurocomputing}}  \bibinfo{volume}{373}
  (\bibinfo{year}{2020}), \bibinfo{pages}{56--69}.
\newblock


\bibitem[\protect\citeauthoryear{Yang, Bai, Zhang, Yuan, and Han}{Yang
  et~al\mbox{.}}{2017}]%
        {DBLP:conf/kdd/YangBZY017}
\bibfield{author}{\bibinfo{person}{Carl Yang}, \bibinfo{person}{Lanxiao Bai},
  \bibinfo{person}{Chao Zhang}, \bibinfo{person}{Quan Yuan}, {and}
  \bibinfo{person}{Jiawei Han}.} \bibinfo{year}{2017}\natexlab{}.
\newblock \showarticletitle{Bridging Collaborative Filtering and
  Semi-Supervised Learning: {A} Neural Approach for {POI} Recommendation}. In
  \bibinfo{booktitle}{\emph{{KDD}}}. \bibinfo{publisher}{{ACM}},
  \bibinfo{pages}{1245--1254}.
\newblock


\bibitem[\protect\citeauthoryear{Yang, Zhang, and Qu}{Yang
  et~al\mbox{.}}{2016}]%
        {DBLP:journals/tist/YangZQ16}
\bibfield{author}{\bibinfo{person}{Dingqi Yang}, \bibinfo{person}{Daqing
  Zhang}, {and} \bibinfo{person}{Bingqing Qu}.}
  \bibinfo{year}{2016}\natexlab{}.
\newblock \showarticletitle{Participatory Cultural Mapping Based on Collective
  Behavior Data in Location-Based Social Networks}.
\newblock \bibinfo{journal}{\emph{{ACM} {TIST}}} \bibinfo{volume}{7},
  \bibinfo{number}{3} (\bibinfo{year}{2016}), \bibinfo{pages}{30:1--30:23}.
\newblock


\bibitem[\protect\citeauthoryear{Yang, Zhang, Yu, and Wang}{Yang
  et~al\mbox{.}}{2013}]%
        {DBLP:conf/ht/YangZYW13}
\bibfield{author}{\bibinfo{person}{Dingqi Yang}, \bibinfo{person}{Daqing
  Zhang}, \bibinfo{person}{Zhiyong Yu}, {and} \bibinfo{person}{Zhu Wang}.}
  \bibinfo{year}{2013}\natexlab{}.
\newblock \showarticletitle{A sentiment-enhanced personalized location
  recommendation system}. In \bibinfo{booktitle}{\emph{{HT}}}.
  \bibinfo{publisher}{{ACM}}, \bibinfo{pages}{119--128}.
\newblock


\bibitem[\protect\citeauthoryear{Yang and Zhang}{Yang and Zhang}{2018}]%
        {DBLP:conf/wsdm/YangZ18}
\bibfield{author}{\bibinfo{person}{Grace~Hui Yang} {and}
  \bibinfo{person}{Sicong Zhang}.} \bibinfo{year}{2018}\natexlab{}.
\newblock \showarticletitle{Differential Privacy for Information Retrieval}. In
  \bibinfo{booktitle}{\emph{{WSDM}}}. \bibinfo{publisher}{{ACM}},
  \bibinfo{pages}{777--778}.
\newblock


\bibitem[\protect\citeauthoryear{Yao, Sheng, Qin, Wang, Shemshadi, and He}{Yao
  et~al\mbox{.}}{2015}]%
        {DBLP:conf/sigir/YaoSQWSH15}
\bibfield{author}{\bibinfo{person}{Lina Yao}, \bibinfo{person}{Quan~Z. Sheng},
  \bibinfo{person}{Yongrui Qin}, \bibinfo{person}{Xianzhi Wang},
  \bibinfo{person}{Ali Shemshadi}, {and} \bibinfo{person}{Qi He}.}
  \bibinfo{year}{2015}\natexlab{}.
\newblock \showarticletitle{Context-aware Point-of-Interest Recommendation
  Using Tensor Factorization with Social Regularization}. In
  \bibinfo{booktitle}{\emph{{SIGIR}}}. \bibinfo{publisher}{{ACM}},
  \bibinfo{pages}{1007--1010}.
\newblock


\bibitem[\protect\citeauthoryear{Yao, Sheng, Wang, Zhang, and Qin}{Yao
  et~al\mbox{.}}{2018}]%
        {DBLP:journals/toit/YaoSWZQ17}
\bibfield{author}{\bibinfo{person}{Lina Yao}, \bibinfo{person}{Quan~Z. Sheng},
  \bibinfo{person}{Xianzhi Wang}, \bibinfo{person}{Wei~Emma Zhang}, {and}
  \bibinfo{person}{Yongrui Qin}.} \bibinfo{year}{2018}\natexlab{}.
\newblock \showarticletitle{Collaborative Location Recommendation by
  Integrating Multi-dimensional Contextual Information}.
\newblock \bibinfo{journal}{\emph{{ACM} Trans. Internet Techn.}}
  \bibinfo{volume}{18}, \bibinfo{number}{3} (\bibinfo{year}{2018}),
  \bibinfo{pages}{32:1--32:24}.
\newblock


\bibitem[\protect\citeauthoryear{Ye, Yin, Lee, and Lee}{Ye
  et~al\mbox{.}}{2011}]%
        {DBLP:conf/sigir/YeYLL11}
\bibfield{author}{\bibinfo{person}{Mao Ye}, \bibinfo{person}{Peifeng Yin},
  \bibinfo{person}{Wang{-}Chien Lee}, {and} \bibinfo{person}{Dik~Lun Lee}.}
  \bibinfo{year}{2011}\natexlab{}.
\newblock \showarticletitle{Exploiting geographical influence for collaborative
  point-of-interest recommendation}. In \bibinfo{booktitle}{\emph{{SIGIR}}}.
  \bibinfo{publisher}{{ACM}}, \bibinfo{pages}{325--334}.
\newblock


\bibitem[\protect\citeauthoryear{Yin, Cui, Chen, Hu, and Zhang}{Yin
  et~al\mbox{.}}{2015}]%
        {DBLP:journals/tkdd/YinCCHZ15}
\bibfield{author}{\bibinfo{person}{Hongzhi Yin}, \bibinfo{person}{Bin Cui},
  \bibinfo{person}{Ling Chen}, \bibinfo{person}{Zhiting Hu}, {and}
  \bibinfo{person}{Chengqi Zhang}.} \bibinfo{year}{2015}\natexlab{}.
\newblock \showarticletitle{Modeling Location-Based User Rating Profiles for
  Personalized Recommendation}.
\newblock \bibinfo{journal}{\emph{{TKDD}}} \bibinfo{volume}{9},
  \bibinfo{number}{3} (\bibinfo{year}{2015}), \bibinfo{pages}{19:1--19:41}.
\newblock


\bibitem[\protect\citeauthoryear{Ying, Wu, Xu, Liu, Liang, Zhang, and
  Xiong}{Ying et~al\mbox{.}}{2019}]%
        {DBLP:journals/www/YingWXLLZX19}
\bibfield{author}{\bibinfo{person}{Haochao Ying}, \bibinfo{person}{Jian Wu},
  \bibinfo{person}{Guandong Xu}, \bibinfo{person}{Yanchi Liu},
  \bibinfo{person}{Tingting Liang}, \bibinfo{person}{Xiao Zhang}, {and}
  \bibinfo{person}{Hui Xiong}.} \bibinfo{year}{2019}\natexlab{}.
\newblock \showarticletitle{Time-aware metric embedding with asymmetric
  projection for successive {POI} recommendation}.
\newblock \bibinfo{journal}{\emph{World Wide Web}} \bibinfo{volume}{22},
  \bibinfo{number}{5} (\bibinfo{year}{2019}), \bibinfo{pages}{2209--2224}.
\newblock


\bibitem[\protect\citeauthoryear{Ying, Kuo, Tseng, and Lu}{Ying
  et~al\mbox{.}}{2014}]%
        {DBLP:journals/tist/YingKTL14}
\bibfield{author}{\bibinfo{person}{Josh~Jia{-}Ching Ying},
  \bibinfo{person}{Wen{-}Ning Kuo}, \bibinfo{person}{Vincent~S. Tseng}, {and}
  \bibinfo{person}{Eric~Hsueh{-}Chan Lu}.} \bibinfo{year}{2014}\natexlab{}.
\newblock \showarticletitle{Mining User Check-In Behavior with a Random Walk
  for Urban Point-of-Interest Recommendations}.
\newblock \bibinfo{journal}{\emph{{ACM} {TIST}}} \bibinfo{volume}{5},
  \bibinfo{number}{3} (\bibinfo{year}{2014}), \bibinfo{pages}{40:1--40:26}.
\newblock


\bibitem[\protect\citeauthoryear{Ying, Lu, Kuo, and Tseng}{Ying
  et~al\mbox{.}}{2012}]%
        {DBLP:conf/kdd/YingLKT12}
\bibfield{author}{\bibinfo{person}{Josh~Jia{-}Ching Ying},
  \bibinfo{person}{Eric~Hsueh{-}Chan Lu}, \bibinfo{person}{Wen{-}Ning Kuo},
  {and} \bibinfo{person}{Vincent~S. Tseng}.} \bibinfo{year}{2012}\natexlab{}.
\newblock \showarticletitle{Urban point-of-interest recommendation by mining
  user check-in behaviors}. In \bibinfo{booktitle}{\emph{UrbComp@KDD}}.
  \bibinfo{publisher}{{ACM}}, \bibinfo{pages}{63--70}.
\newblock


\bibitem[\protect\citeauthoryear{Yu, Xu, and Wang}{Yu et~al\mbox{.}}{2019}]%
        {DBLP:conf/seke/YuXW19}
\bibfield{author}{\bibinfo{person}{Dongjin Yu}, \bibinfo{person}{Kaihui Xu},
  {and} \bibinfo{person}{Dongjing Wang}.} \bibinfo{year}{2019}\natexlab{}.
\newblock \showarticletitle{Modeling User Contextual Behavior Semantics with
  Geographical Influence for Point-Of-Interest Recommendation}. In
  \bibinfo{booktitle}{\emph{{SEKE}}}. \bibinfo{publisher}{{KSI} Research Inc.
  and Knowledge Systems Institute Graduate School}, \bibinfo{pages}{373--484}.
\newblock


\bibitem[\protect\citeauthoryear{Yu and Chen}{Yu and Chen}{2015}]%
        {yu2015survey}
\bibfield{author}{\bibinfo{person}{Yonghong Yu} {and} \bibinfo{person}{Xingguo
  Chen}.} \bibinfo{year}{2015}\natexlab{}.
\newblock \showarticletitle{A survey of point-of-interest recommendation in
  location-based social networks}. In \bibinfo{booktitle}{\emph{Workshops at
  the Twenty-Ninth AAAI Conference on Artificial Intelligence}}.
\newblock


\bibitem[\protect\citeauthoryear{Yuan, Cong, Ma, Sun, and
  Magnenat{-}Thalmann}{Yuan et~al\mbox{.}}{2013}]%
        {DBLP:conf/sigir/YuanCMSM13}
\bibfield{author}{\bibinfo{person}{Quan Yuan}, \bibinfo{person}{Gao Cong},
  \bibinfo{person}{Zongyang Ma}, \bibinfo{person}{Aixin Sun}, {and}
  \bibinfo{person}{Nadia Magnenat{-}Thalmann}.}
  \bibinfo{year}{2013}\natexlab{}.
\newblock \showarticletitle{Time-aware point-of-interest recommendation}. In
  \bibinfo{booktitle}{\emph{{SIGIR}}}. \bibinfo{publisher}{{ACM}},
  \bibinfo{pages}{363--372}.
\newblock


\bibitem[\protect\citeauthoryear{Yuan, Cong, and Sun}{Yuan
  et~al\mbox{.}}{2014}]%
        {DBLP:conf/cikm/YuanCS14}
\bibfield{author}{\bibinfo{person}{Quan Yuan}, \bibinfo{person}{Gao Cong},
  {and} \bibinfo{person}{Aixin Sun}.} \bibinfo{year}{2014}\natexlab{}.
\newblock \showarticletitle{Graph-based Point-of-interest Recommendation with
  Geographical and Temporal Influences}. In \bibinfo{booktitle}{\emph{{CIKM}}}.
  \bibinfo{publisher}{{ACM}}, \bibinfo{pages}{659--668}.
\newblock


\bibitem[\protect\citeauthoryear{Zhang and Chow}{Zhang and Chow}{2015}]%
        {DBLP:conf/sigir/ZhangC15}
\bibfield{author}{\bibinfo{person}{Jia{-}Dong Zhang} {and}
  \bibinfo{person}{Chi{-}Yin Chow}.} \bibinfo{year}{2015}\natexlab{}.
\newblock \showarticletitle{GeoSoCa: Exploiting Geographical, Social and
  Categorical Correlations for Point-of-Interest Recommendations}. In
  \bibinfo{booktitle}{\emph{{SIGIR}}}. \bibinfo{publisher}{{ACM}},
  \bibinfo{pages}{443--452}.
\newblock


\bibitem[\protect\citeauthoryear{Zhang, Chow, and Li}{Zhang
  et~al\mbox{.}}{2014}]%
        {DBLP:conf/gis/ZhangCL14}
\bibfield{author}{\bibinfo{person}{Jia{-}Dong Zhang},
  \bibinfo{person}{Chi{-}Yin Chow}, {and} \bibinfo{person}{Yanhua Li}.}
  \bibinfo{year}{2014}\natexlab{}.
\newblock \showarticletitle{{LORE:} exploiting sequential influence for
  location recommendations}. In \bibinfo{booktitle}{\emph{{SIGSPATIAL/GIS}}}.
  \bibinfo{publisher}{{ACM}}, \bibinfo{pages}{103--112}.
\newblock


\bibitem[\protect\citeauthoryear{Zhang, Chow, and Li}{Zhang
  et~al\mbox{.}}{2015a}]%
        {DBLP:journals/tsc/ZhangCL15}
\bibfield{author}{\bibinfo{person}{Jia{-}Dong Zhang},
  \bibinfo{person}{Chi{-}Yin Chow}, {and} \bibinfo{person}{Yanhua Li}.}
  \bibinfo{year}{2015}\natexlab{a}.
\newblock \showarticletitle{iGeoRec: {A} Personalized and Efficient
  Geographical Location Recommendation Framework}.
\newblock \bibinfo{journal}{\emph{{IEEE} Trans. Services Computing}}
  \bibinfo{volume}{8}, \bibinfo{number}{5} (\bibinfo{year}{2015}),
  \bibinfo{pages}{701--714}.
\newblock


\bibitem[\protect\citeauthoryear{Zhang, Chow, and Zheng}{Zhang
  et~al\mbox{.}}{2015b}]%
        {DBLP:conf/cikm/ZhangCZ15}
\bibfield{author}{\bibinfo{person}{Jia{-}Dong Zhang},
  \bibinfo{person}{Chi{-}Yin Chow}, {and} \bibinfo{person}{Yu Zheng}.}
  \bibinfo{year}{2015}\natexlab{b}.
\newblock \showarticletitle{ORec: An Opinion-Based Point-of-Interest
  Recommendation Framework}. In \bibinfo{booktitle}{\emph{{CIKM}}}.
  \bibinfo{publisher}{{ACM}}, \bibinfo{pages}{1641--1650}.
\newblock


\bibitem[\protect\citeauthoryear{Zhang, Yao, Sun, and Tay}{Zhang
  et~al\mbox{.}}{2019}]%
        {DBLP:journals/csur/ZhangYST19}
\bibfield{author}{\bibinfo{person}{Shuai Zhang}, \bibinfo{person}{Lina Yao},
  \bibinfo{person}{Aixin Sun}, {and} \bibinfo{person}{Yi Tay}.}
  \bibinfo{year}{2019}\natexlab{}.
\newblock \showarticletitle{Deep Learning Based Recommender System: {A} Survey
  and New Perspectives}.
\newblock \bibinfo{journal}{\emph{{ACM} Comput. Surv.}} \bibinfo{volume}{52},
  \bibinfo{number}{1} (\bibinfo{year}{2019}), \bibinfo{pages}{5:1--5:38}.
\newblock


\bibitem[\protect\citeauthoryear{Zhang and Wang}{Zhang and Wang}{2015}]%
        {DBLP:conf/cikm/ZhangW15}
\bibfield{author}{\bibinfo{person}{Wei Zhang} {and} \bibinfo{person}{Jianyong
  Wang}.} \bibinfo{year}{2015}\natexlab{}.
\newblock \showarticletitle{Location and Time Aware Social Collaborative
  Retrieval for New Successive Point-of-Interest Recommendation}. In
  \bibinfo{booktitle}{\emph{{CIKM}}}. \bibinfo{publisher}{{ACM}},
  \bibinfo{pages}{1221--1230}.
\newblock


\bibitem[\protect\citeauthoryear{Zhao, Lou, Qian, and Hou}{Zhao
  et~al\mbox{.}}{2020}]%
        {DBLP:journals/kbs/ZhaoLQH20}
\bibfield{author}{\bibinfo{person}{Guoshuai Zhao}, \bibinfo{person}{Peiliang
  Lou}, \bibinfo{person}{Xueming Qian}, {and} \bibinfo{person}{Xingsong Hou}.}
  \bibinfo{year}{2020}\natexlab{}.
\newblock \showarticletitle{Personalized location recommendation by fusing
  sentimental and spatial context}.
\newblock \bibinfo{journal}{\emph{Knowl. Based Syst.}}  \bibinfo{volume}{196}
  (\bibinfo{year}{2020}), \bibinfo{pages}{105849}.
\newblock


\bibitem[\protect\citeauthoryear{Zhao, Zhu, Liu, Xu, Li, Zhuang, Sheng, and
  Zhou}{Zhao et~al\mbox{.}}{2019}]%
        {DBLP:conf/aaai/ZhaoZLXLZSZ19}
\bibfield{author}{\bibinfo{person}{Pengpeng Zhao}, \bibinfo{person}{Haifeng
  Zhu}, \bibinfo{person}{Yanchi Liu}, \bibinfo{person}{Jiajie Xu},
  \bibinfo{person}{Zhixu Li}, \bibinfo{person}{Fuzhen Zhuang},
  \bibinfo{person}{Victor~S. Sheng}, {and} \bibinfo{person}{Xiaofang Zhou}.}
  \bibinfo{year}{2019}\natexlab{}.
\newblock \showarticletitle{Where to Go Next: {A} Spatio-Temporal Gated Network
  for Next {POI} Recommendation}. In \bibinfo{booktitle}{\emph{{AAAI}}}.
  \bibinfo{publisher}{{AAAI} Press}, \bibinfo{pages}{5877--5884}.
\newblock


\bibitem[\protect\citeauthoryear{Zhao, Lyu, and King}{Zhao
  et~al\mbox{.}}{2018}]%
        {DBLP:series/sbcs/ZhaoLK18}
\bibfield{author}{\bibinfo{person}{Shenglin Zhao}, \bibinfo{person}{Michael~R.
  Lyu}, {and} \bibinfo{person}{Irwin King}.} \bibinfo{year}{2018}\natexlab{}.
\newblock \bibinfo{booktitle}{\emph{Point-of-Interest Recommendation in
  Location-Based Social Networks}}.
\newblock \bibinfo{publisher}{Springer}.
\newblock


\bibitem[\protect\citeauthoryear{Zhao, Zhao, King, and Lyu}{Zhao
  et~al\mbox{.}}{2017}]%
        {DBLP:conf/www/ZhaoZKL17}
\bibfield{author}{\bibinfo{person}{Shenglin Zhao}, \bibinfo{person}{Tong Zhao},
  \bibinfo{person}{Irwin King}, {and} \bibinfo{person}{Michael~R. Lyu}.}
  \bibinfo{year}{2017}\natexlab{}.
\newblock \showarticletitle{Geo-Teaser: Geo-Temporal Sequential Embedding Rank
  for Point-of-interest Recommendation}. In \bibinfo{booktitle}{\emph{{WWW}
  (Companion Volume)}}. \bibinfo{publisher}{{ACM}}, \bibinfo{pages}{153--162}.
\newblock


\bibitem[\protect\citeauthoryear{Zhao, Zhao, Yang, Lyu, and King}{Zhao
  et~al\mbox{.}}{2016}]%
        {DBLP:conf/aaai/ZhaoZYLK16}
\bibfield{author}{\bibinfo{person}{Shenglin Zhao}, \bibinfo{person}{Tong Zhao},
  \bibinfo{person}{Haiqin Yang}, \bibinfo{person}{Michael~R. Lyu}, {and}
  \bibinfo{person}{Irwin King}.} \bibinfo{year}{2016}\natexlab{}.
\newblock \showarticletitle{{STELLAR:} Spatial-Temporal Latent Ranking for
  Successive Point-of-Interest Recommendation}. In
  \bibinfo{booktitle}{\emph{{AAAI}}}. \bibinfo{publisher}{{AAAI} Press},
  \bibinfo{pages}{315--322}.
\newblock


\bibitem[\protect\citeauthoryear{Zheng, Han, and Sun}{Zheng
  et~al\mbox{.}}{2018}]%
        {DBLP:journals/tkde/ZhengHS18}
\bibfield{author}{\bibinfo{person}{Xin Zheng}, \bibinfo{person}{Jialong Han},
  {and} \bibinfo{person}{Aixin Sun}.} \bibinfo{year}{2018}\natexlab{}.
\newblock \showarticletitle{A Survey of Location Prediction on Twitter}.
\newblock \bibinfo{journal}{\emph{{IEEE} Trans. Knowl. Data Eng.}}
  \bibinfo{volume}{30}, \bibinfo{number}{9} (\bibinfo{year}{2018}),
  \bibinfo{pages}{1652--1671}.
\newblock


\bibitem[\protect\citeauthoryear{Zhou and Wang}{Zhou and Wang}{2014}]%
        {DBLP:conf/dsaa/ZhouW14}
\bibfield{author}{\bibinfo{person}{Dequan Zhou} {and} \bibinfo{person}{Xin
  Wang}.} \bibinfo{year}{2014}\natexlab{}.
\newblock \showarticletitle{Probabilistic Category-based Location
  Recommendation Utilizing Temporal Influence and Geographical Influence}. In
  \bibinfo{booktitle}{\emph{{DSAA}}}. \bibinfo{publisher}{{IEEE}},
  \bibinfo{pages}{115--121}.
\newblock


\bibitem[\protect\citeauthoryear{Zhu, Xu, Guan, and Zhang}{Zhu
  et~al\mbox{.}}{2017}]%
        {DBLP:journals/jnca/ZhuXGZ17}
\bibfield{author}{\bibinfo{person}{Liang Zhu}, \bibinfo{person}{Changqiao Xu},
  \bibinfo{person}{Jianfeng Guan}, {and} \bibinfo{person}{Hongke Zhang}.}
  \bibinfo{year}{2017}\natexlab{}.
\newblock \showarticletitle{{SEM-PPA:} {A} semantical pattern and
  preference-aware service mining method for personalized point of interest
  recommendation}.
\newblock \bibinfo{journal}{\emph{J. Network and Computer Applications}}
  \bibinfo{volume}{82} (\bibinfo{year}{2017}), \bibinfo{pages}{35--46}.
\newblock


\end{thebibliography}

\end{document}